\documentclass[preprint,amsmath,amssymb,nofootinbib]{revtex4}
\usepackage[spanish]{babel}
\decimalpoint
\addto\captionsspanish{}
\usepackage{graphicx}
\usepackage{multirow}
\usepackage{mathrsfs}
\allowdisplaybreaks
\usepackage{amsmath}
\usepackage{array}
\usepackage{booktabs}
\makeatletter
\renewcommand{\fnum@figure}{Fig.~ \thefigure}
\makeatother
\usepackage{pstcol}
\usepackage{amsfonts}
\usepackage{bm}
\usepackage{amsmath}
\usepackage{comment}
\usepackage{amssymb}
\usepackage{color}
\usepackage[all]{xy}
\usepackage{comment}
\usepackage{epstopdf,epsfig}
\def\be{\begin{equation}}
\def\ee{\end{equation}}
\def\bea{\begin{eqnarray}}
\def\eea{\end{eqnarray}}
\usepackage{float}
\usepackage{xcolor}
\usepackage{xcolor}
\definecolor{fucsia}{HTML}{FF00FF}
\usepackage{graphicx}
\usepackage{makecell}
\usepackage{multirow}
\usepackage{subcaption}
\usepackage{setspace}
\usepackage{microtype}
\usepackage[
    colorlinks=true,
    linkcolor=blue,     % Color de los enlaces en TOC, ecuaciones, figuras, etc.
    citecolor=red,      % Color de referencias bibliográficas
]{hyperref}
\addto\captionsspanish{}

\usepackage{ragged2e}

\makeatletter
\def\@makecaption#1#2{%
  \vskip 8pt % <-- espacio entre figura y caption (ajusta aquí)
  \begingroup
    \small
    \justifying
    \textbf{#1.} #2\par
  \endgroup
}
\makeatother
\usepackage{scalerel}
\usepackage{tikz}
\usetikzlibrary{svg.path}

\definecolor{orcidlogocol}{HTML}{A6CE39}
\tikzset{
  orcidlogo/.pic={
    \fill[orcidlogocol] svg{M256,128c0,70.7-57.3,128-128,128C57.3,256,0,198.7,0,128C0,57.3,57.3,0,128,0C198.7,0,256,57.3,256,128z};
    \fill[white] svg{M86.3,186.2H70.9V79.1h15.4v48.4V186.2z}
                 svg{M108.9,79.1h41.6c39.6,0,57,28.3,57,53.6c0,27.5-21.5,53.6-56.8,53.6h-41.8V79.1z M124.3,172.4h24.5c34.9,0,42.9-26.5,42.9-39.7c0-21.5-13.7-39.7-43.7-39.7h-23.7V172.4z}
                 svg{M88.7,56.8c0,5.5-4.5,10.1-10.1,10.1c-5.6,0-10.1-4.6-10.1-10.1c0-5.6,4.5-10.1,10.1-10.1C84.2,46.7,88.7,51.3,88.7,56.8z};
  }
}

\newcommand\orcidicon[1]{\href{https://orcid.org/#1}{\mbox{\scalerel*{
\begin{tikzpicture}[yscale=-1,transform shape]
\pic{orcidlogo};
\end{tikzpicture}
}{|}}}}
\begin{document}
\baselineskip=20pt

\title{Dark-Sector Effects on the Phase Structure of Nonlinear Magnetic AdS Black Holes}
\author{
Carlos E. Romero-Figueroa\textsuperscript{\orcidicon{0000-0001-5548-7766}}\textsuperscript{1} 
 and
J.R. Villanueva\textsuperscript{\orcidicon{0000-0002-6726-492X}}\textsuperscript{2}
}\email{\raggedright
carlosed.romero@correo.nucleares.unam.mx; jose.villanueva@uv.cl
}

\affiliation{\textsuperscript{1}Instituto de Ciencias Nucleares, Universidad Nacional Autónoma de México (UNAM), Mexico City, Mexico}
\affiliation{\textsuperscript{2}Instituto de Física y Astronomía, Universidad de Valparaíso, Avenida Gran Bretaña 1111, Valparaíso, Chile}

\begin{abstract}
We investigate the effects of perfect fluid dark matter (PFDM) and a dark-energy field on the phase structure of a nonlinear magnetically charged Anti--de Sitter (NLMC--AdS) black hole. The full thermodynamic system is analyzed numerically, while exact critical points are obtained for the limiting cases. Small--large black hole (SBH/LBH) phase transitions emerge in both the limiting geometries and the full solution within the quintessence regime. Moreover, the critical ratio $\rho_c=P_c v_c/T_c$ differs from the standard Reissner--Nordström--AdS (RN--AdS)/van der Waals (vdW) value, $\rho_c=3/8$. Our results show that the dark-sector parameters significantly influence the strength and persistence of the first-order transition. In the quintessence regime, a stronger dark-energy contribution enhances the swallow-tail structure of the Gibbs free energy and increases the latent heat associated with SBH/LBH coexistence. In contrast, a larger magnitude of the PFDM parameter $|\lambda|$ progressively suppresses the swallow-tail structure and the corresponding first-order transition, eventually leading to a single-phase regime. In the phantom regime, the system instead exhibits spinodal behavior without phase coexistence. We also find that any nonzero $\lambda$ prevents the formation of a regular magnetic core.  Finally, geometrothermodynamics (GTD) reproduces the phase structure through singularities of the thermodynamic curvature, while the associated critical scaling is consistent with results reported for black holes, cosmological horizons, and real fluids, pointing toward a broader universality of thermodynamic critical behavior.

\vspace{0.1in}

\end{abstract}

\maketitle

\tableofcontents

\section{Introduction}\label{sec:1}
The physics of black holes represents one of the greatest areas of interest within the scientific community.
Probably one of the main reasons for this is that its study connects different branches of physics itself and can be approached from various perspectives, such as the study of motion for different kinds of particles in the background generated by black holes \cite{Chandrasekhar:1983,Cruz:2004ts,Hackmann:2009nh,Villanueva:2013gga,Olivares:2013jza,Cruz:2013ufa,Villanueva:2015kua,Villanueva:2018kem,Fathi:2020sfw,Fathi:2022pqv,Fathi:2024kce}, the study of stability and  quasinormal modes \cite{Regge:1957td,Zerilli:1970wzz,Kokkotas:1999bd,Berti:2009kk,Gonzalez:2018xrq,Becar:2019hwk,Tattersall:2018nve,Aragon:2020xtm,Fathi:2025jrk},
thermodynamic approaches \cite{bekenstein1973black,Bekenstein:1972tm,Hawking:1976de,ruppeiner1979thermodynamics,Cruz:2004vp,belgiorno2002notes,belgiorno2003black,belgiorno2003quasi,Appels:2016uha,Astorino:2016ybm,Molina:2021hgx,Fathi:2021liw,Fathi:2021eig,Karmakar:2023mhs,Fathi:2023lau}, among others.
Furthermore, many studies connect theory with observational data intended to test the theory's parameters, providing limits and regions of validity, both in classical \cite{Shapiro:1972zz,Islam:1983rxp,Turyshev:2003ut,Kraniotis:2003ig,Kerr:2003bp,Hackmann:2011wp,Leiva:2012az,Sultana:2012qp,Iorio:2004ee,Rincon:2019zxk} and modern tests \cite{Uniyal:2022vdu,Fathi:2024kda,Porto:2026nva}.

Recently, research on black holes within the framework of nonlinear electrodynamics (NLED), where gravity is coupled to NLED instead of standard Maxwell theory, has garnered increasing interest, as this approach offers a more realistic description of the systems surrounding these compact objects. This is especially relevant today, given that years of theoretical effort have borne fruit, resulting in a comprehensive framework for in-depth study within this theory \cite{Born1933,BornInfeld1934,HeisenbergEuler1936,Plebanski1970,Burinskii:2002pz,Breton:2015cza,Breton:2017hwe,Magos:2020ykt,Breton:2023bwf}.
In this regard, the magnetic interpretation of Bardeen geometry and the work of Ayón-Beato and García showed how NLEDs can support non-singular metrics \cite{AyonBeatoGarcia1998,AyonBeatoGarcia1999,AyonBeatoGarcia2000}.
Thus, it is possible to relate constructions that include electrically and magnetically charged families, de Sitter nuclei, reverse-engineered models, and new power-law or Born-Infeld theories. Recent studies have further explored the thermodynamic, dynamical, and observational properties of black holes in the presence of dark-sector components, quantum corrections, and modified gravitational frameworks. In particular, black holes surrounded by dark-matter distributions, including a perfect fluid dark matter (PFDM) and Dehnen-type dark matter halos, have been investigated through their thermodynamic and observational properties \cite{Ashraf2025PFDM,Ashraf2025Dehnen}. Regular and quantum-corrected black-hole solutions have also attracted attention in connection with thermodynamics, particle dynamics, and gravitational lensing \cite{ditta2024particle,Ditta2025LQBH,Saleem2025deSitter,Ditta2025LimitingCurvature}. Other recent studies have considered black-hole dynamics and observational signatures in modified-gravity frameworks, including Bumblebee, Horndeski, and $F(R)$ gravity \cite{mustafa2025epicyclic,mustafa2025orbital,Ashraf2025Horndeski,Ditta2025FR}, as well as quantum-gravity effects and alternative black-hole geometries \cite{Ashraf2025QuantumGravity,Ashraf2025ShortHairy,Ashraf2025HuSawicki,Bouzenada2026SUN}.
Our interest lies in nonlinear magnetic black holes which are immersed in a background dominated by a negative cosmological constant (AdS case). These developments have also extended to the thermodynamics of AdS black holes in the presence of additional matter fields and modified gravitational sectors \cite{Ditta2025KalbRamond,Ashraf2025DeformedAdS,Bouzenada2025Frolov,Maurya2025Frolov,Bouzenada2025QOSdS}. The advantage of introducing a magnetic charge is that these models can support regular black-hole solutions within NLED, where the nonlinear electromagnetic field can modify the near-origin geometry and, in suitable models, remove the central singularity and yield finite electromagnetic self-energies. Furthermore, they exhibit particularly interesting thermodynamic behaviors,
including small--large black hole (SBH/LBH) phase transitions.

In parallel, differential geometry approaches provide a powerful framework for analyzing thermodynamic systems, and related formalisms have been widely used to study phase transitions, stability, and microstructure in a wide range of black hole solutions \cite{wei2021general,Banerjee,ref20,Zangeneh_2018,Dehyadegari,Sahay,Hazarika,wei2020extended,ruppeiner2008thermodynamic,quevedo2008geometrothermodynamics,Ladino:2024ned,ladino2025phase}. Specifically, the formalism of Geometrothermodynamics (GTD) has provided a Legendre-invariant description of thermodynamics \cite{quevedo2007geometrothermodynamics,quevedo2023unified,bravetti2017zeroth}, characterizing phase structure and microscopic interactions through curvature scalars on the equilibrium manifold. Curvature singularities indicate thermodynamic critical behavior, while the sign of the scalar curvature reflects the dominant microscopic interaction, whether attractive or repulsive. GTD has been successfully applied to systems ranging from ideal gases and van der Waals (vdW) fluids~\cite{quevedo2022geometrothermodynamics,quevedo2011phase}, magnetic materials~\cite{quevedo2024geometrothermodynamic}, econophysics~\cite{2023IJGMM..2050057Q} and the Ising model~\cite{bravetti2014representation}, to black holes in various gravitational theories~\cite{ladino2025phase}

This article is organized as follows:
Sect. \ref{sec2} is dedicated to presenting the background spacetime geometry of a nonlinear magnetically charged anti--de Sitter (NLMC--AdS) black hole with spherical symmetry, immersed in PFDM distribution and surrounded by a dark energy field.
In particular, the metric that governs spacetime is presented,
and the curvature invariants associated with it are also calculated.
In Sect. \ref{sec3} we study the thermodynamic properties of the black hole solution,
such as Smarr's formula, geometric volume variation, critical behavior in extended phase space, and local stability.
In Sect. \ref{sec4}, aspects of GTD and black-hole microstructure are analyzed, allowing us to identify phase transitions through geometric singularities.
Finally, Sect. \ref{sec:5} presents the conclusions of the work, along with future perspectives for further research.

Throughout this work, we adopt natural units where $G = c = 1 = M_{pl}$. The sign convention
is $(-,+,+,+)$, and the primes denote differentiation with respect to the radial coordinate.

\section{Nonlinear Magnetic AdS Black Holes in PFDM and Dark Energy Backgrounds}\label{sec2}
In this section, we establish the background spacetime geometry of a
spherically symmetric NLMC--AdS
black hole immersed in a PFDM distribution and surrounded by a dark-energy field. Following Kiselev \cite{kiselev2003quintessence,kiselev2003quintessential} and Li and Yang \cite{MHL2012}, the energy-momentum tensor of PFDM in the standard orthogonal basis is given by
\begin{equation}
T^{\text{DM}}_{\,\,\mu\nu} = \mathrm{diag}\left(-\mathcal{E}_{DM},\, P_{r\,DM},\, P_{\theta\,DM},\, P_{\phi\,DM}\right),\label{tensor}
\end{equation}
where $\mathcal{E}_{DM}$ is the energy density, $P_{r\,DM}$ is the
radial pressure, and $P_{\theta\,DM}=P_{\phi\,DM}$ are the tangential
pressures, defined as \cite{MHL2012,rahaman2010perfect,kuncewicz2025perfect}
\begin{equation}
\mathcal{E}_{DM}=-P_{r\,DM}=-\frac{\lambda}{8\pi r^3},
\qquad
P_{\theta\,DM}=P_{\phi\,DM}=-\frac{\lambda}{16\pi r^3}.
\label{energy-density}
\end{equation}
Here, $\lambda$ is a constant parameter characterizing the PFDM
distribution and controlling its contribution to the spacetime
geometry. More general PFDM models can also be formulated with
anisotropic pressure profiles and different equations of state
\cite{kuncewicz2025perfect}. In the present work, we adopt the PFDM
model described above. With the convention used here, a positive PFDM
energy density requires
\begin{equation}
\mathcal{E}_{DM}>0
\qquad\Longrightarrow\qquad
\lambda<0.
\end{equation}
Therefore, throughout this work we restrict the PFDM parameter to the
physical branch $\lambda<0$. The black hole is also surrounded by a
dark-energy field characterized by an equation-of-state parameter $w$
and an intensity parameter $N$. For $-1<w<-1/3$, the field is in the
quintessence regime, whereas $w<-1$ corresponds to the phantom regime \cite{ratra1988cosmological,bronnikov2006regular}.
These two regimes will be considered separately in the analysis below.
The resulting spacetime is described by the line element
\begin{equation}
    ds^2 = -f(r)\,dt^2
    + \dfrac{dr^2}{f(r)}
    + r^2(d\theta^2+\sin^2\theta\,d\varphi^2),
\label{metric}
\end{equation}
where the lapse function is given by \cite{DJG2024,ndongmo2023thermodynamics}
\begin{align}
f(r)=1-\frac{2Mr^2}{r^3+Q_m^3}
+\frac{\lambda}{r}\ln\!\frac{r}{|\lambda|}
-\frac{N}{r^{3w+1}}
+\frac{r^2}{\ell^2},
\label{function}
\end{align}
where $M$ is the black hole mass, $Q_m$ is the magnetic charge, and
$\ell$ is the AdS radius. The admissible parameter domain leading to black hole solutions cannot
be determined analytically because the horizon equation associated with
the lapse function, Eq.~\eqref{function}, is transcendental, as in other
black hole solutions surrounded by dark matter
\cite{ahmed2026shadow}. We therefore turn to the near-origin behavior
of the lapse function to characterize the central structure of the
spacetime. Expanding Eq.~\eqref{function} around $r=0$, we obtain
\begin{align}
f(r)\Big|_{r\to0}=
1
+\frac{\lambda}{r}\ln\!\left(\frac{r}{|\lambda|}\right)
-\frac{N}{r^{3w+1}}
+\left(\frac{1}{\ell^2}-\frac{2M}{Q_m^3}\right)r^2
+\mathcal{O}(r^5).
\label{nearorigin}
\end{align}
Eq.~\eqref{nearorigin} shows that the leading behavior near the origin depends on the equation-of-state parameter $w$. The term $-N/r^{3w+1}$ dominates the expansion for $w>0$, whereas the logarithmic PFDM contribution $(\lambda/r)\ln(r/|\lambda|)$ is dominant for $w\le0$. Therefore, in both the quintessence and phantom regimes considered in this work, the central geometry is governed by the logarithmic PFDM term. Consequently, unlike the solution in Ref.~\cite{ahmed2026shadow}, the near-origin expansion contains no Schwarzschild-like $1/r$ contribution, preventing the definition of a constant effective mass. Hence, for any  $\lambda\neq 0$, the logarithmic divergence cannot be removed by tuning $M$, $Q_m$, $N$, or $\ell$, ruling out the existence of a regular black hole core.

\begin{figure}[ht!]
\centering

%---------------- First row ----------------%
\begin{minipage}{0.48\linewidth}
    \centering
    \includegraphics[width=\linewidth]{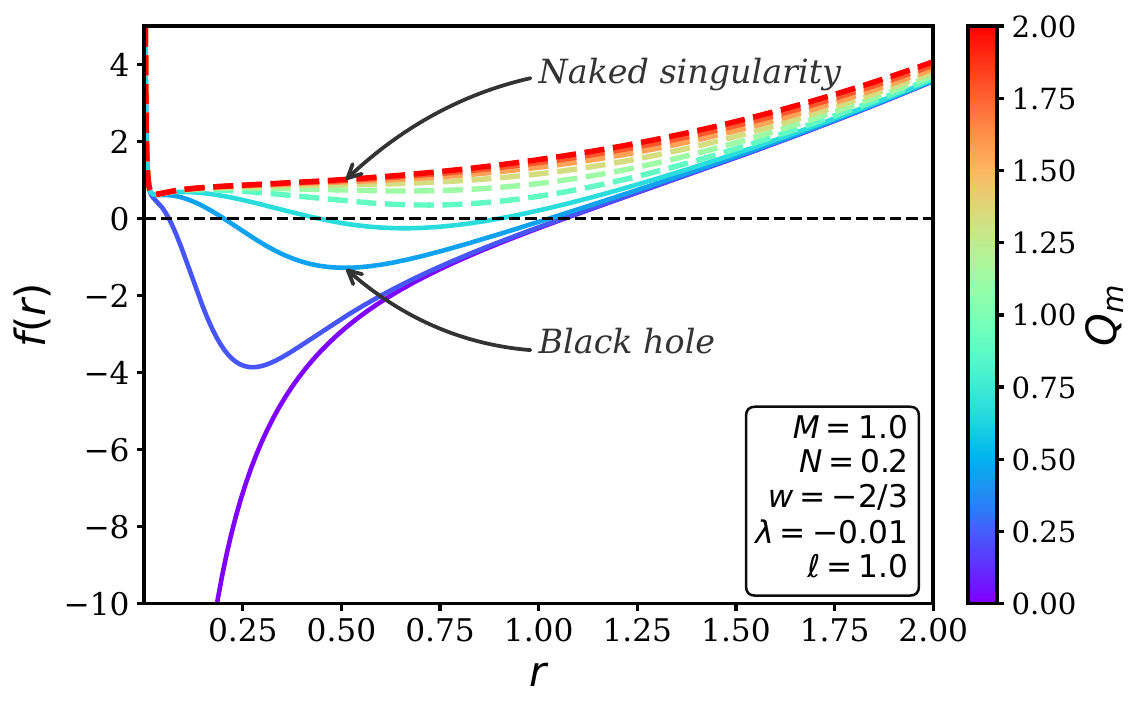}

    (a)
\end{minipage}
\hfill
\begin{minipage}{0.48\linewidth}
    \centering
    \includegraphics[width=\linewidth]{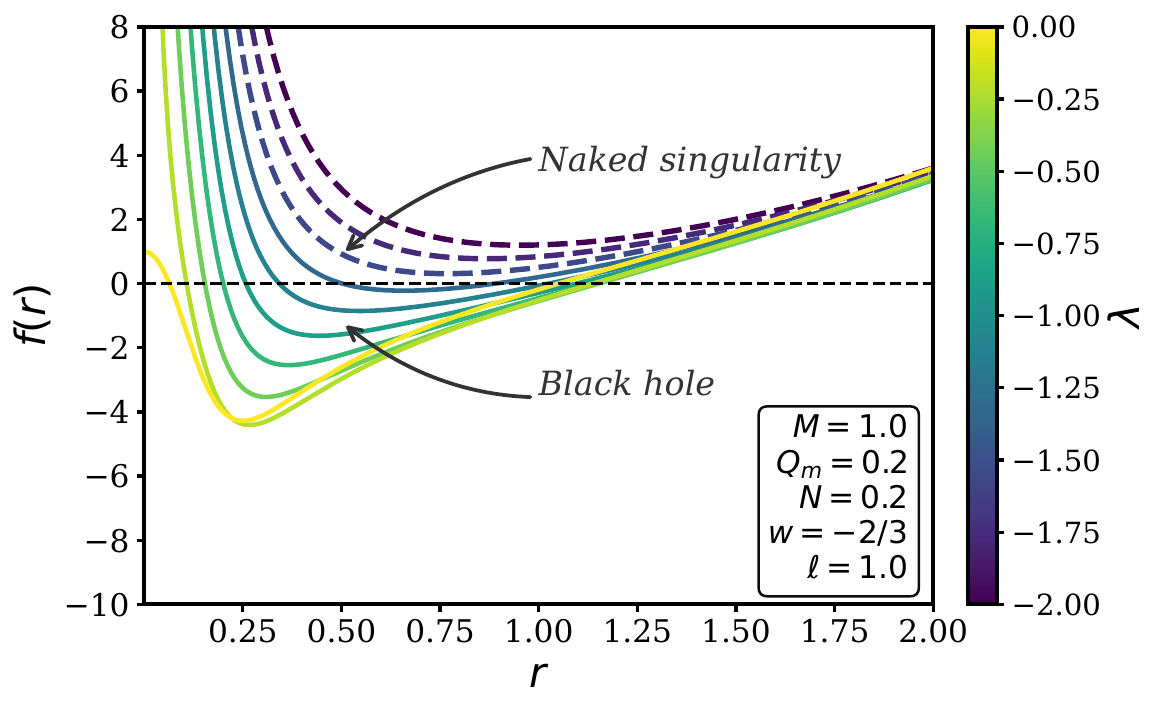}

    (b)
\end{minipage}

\vspace{3mm}

%---------------- Second row ----------------%
\begin{minipage}{0.48\linewidth}
    \centering
    \includegraphics[width=\linewidth]{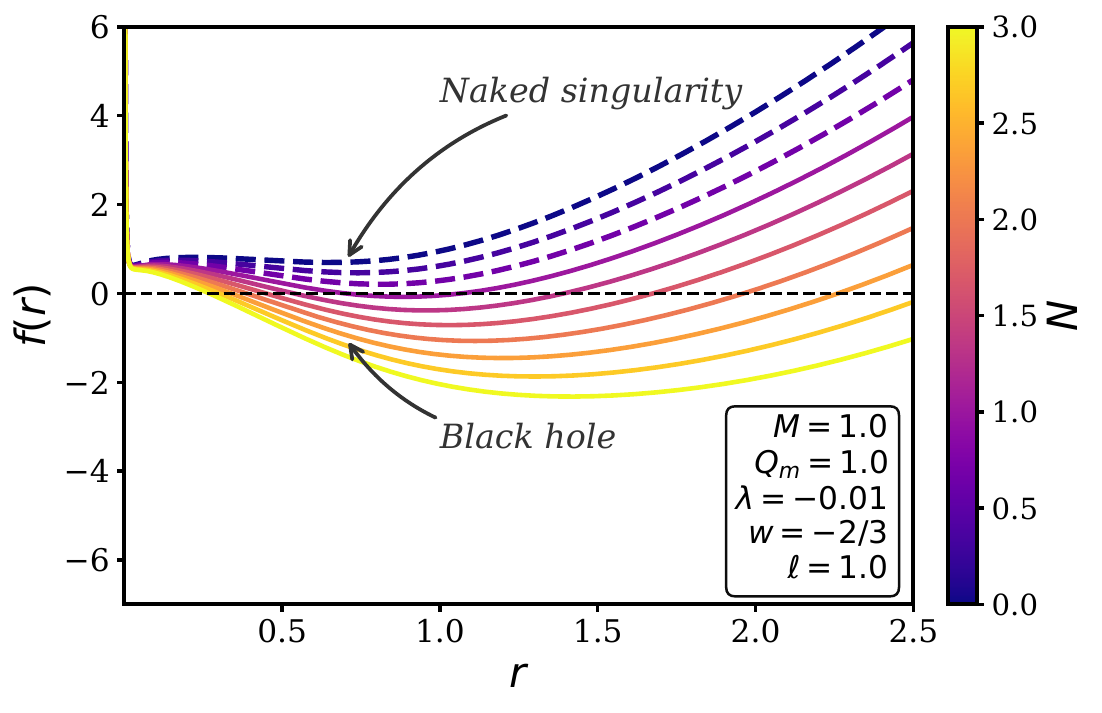}

    (c)
\end{minipage}
\hfill
\begin{minipage}{0.48\linewidth}
    \centering
    \includegraphics[width=\linewidth]{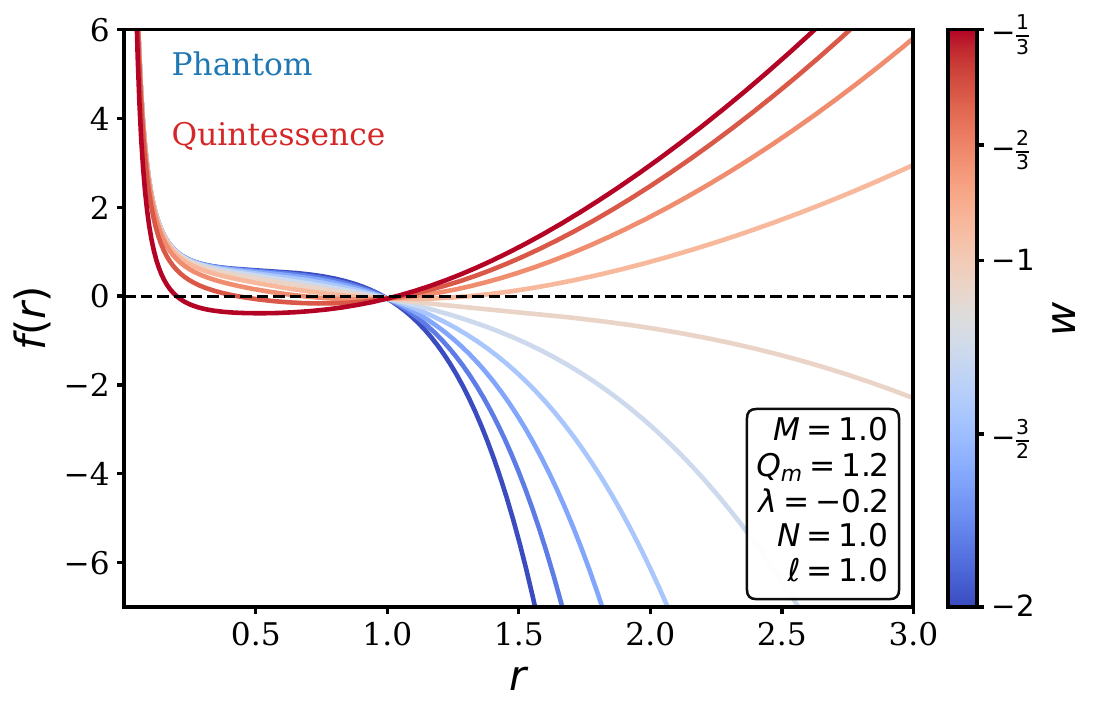}

    (d)
\end{minipage}

\caption{Lapse function $f(r)$ for different values of (a) the magnetic charge $Q_m$, (b) the PFDM coupling $\lambda$, (c) the intensity parameter $N$, and (d) the equation-of-state parameter $w$, with the remaining parameters held fixed. Solid (dashed) curves denote black hole (naked singularity) configurations.}

\label{fig:metric}
\end{figure}
While the near-origin behavior is controlled by the PFDM contribution, the global structure of the spacetime depends on the interplay among the solution parameters. This is illustrated in Fig.~\ref{fig:metric}, which shows the behavior of the lapse function. Depending on the parameter choice, the lapse function may develop one or more horizons, corresponding to black hole configurations, or exhibit no horizons, corresponding to naked singularities. In Fig.~\ref{fig:metric}(a), we observe that increasing the magnetic charge $Q_m$ tends to suppress the formation of horizons, leading to naked singularity configurations. A similar behavior is found in Fig.~\ref{fig:metric}(b), where increasing the magnitude of the PFDM coupling $|\lambda|$ also disfavors black hole formation. This is consistent with our previous results for other black hole solutions, where the dark matter sector was likewise found to hinder the formation of event horizons \cite{ahmed2026shadow}. In contrast, Fig.~\ref{fig:metric}(c) reveals the opposite effect for the dark energy sector. As the intensity parameter $N$ increases, horizons are more easily formed, indicating that the presence of the dark energy fluid favors black hole configurations over naked singularities. Finally, Fig.~\ref{fig:metric}(d) shows that the existence of horizons is relatively insensitive to the equation-of-state parameter $w$. Indeed, black hole solutions are obtained throughout both the quintessence and phantom regimes, indicating that the qualitative horizon structure is primarily controlled by the intensity of the dark energy component rather than by its equation of state. However, as shown in Table~\ref{LimitingCases}, the asymptotic structure does depend on $w$: the spacetime remains asymptotically AdS in the quintessence regime, whereas the phantom contribution dominates at large distances, leading to a non-asymptotically AdS geometry.
\subsection{Curvature Invariants and spacetime regularity}
\label{invariants}
To further characterize the geometry of the solution, we analyze the curvature invariants associated with the metric in Eq.~\eqref{metric}. In particular, we examine whether the spacetime admits a regular core, requiring all independent curvature invariants to remain finite as $r\rightarrow0$. This allows us to identify the parameter sectors in which the magnetic charge and dark-sector contributions lead to a nonsingular center. The independent curvature invariants are 

\begin{align}
\mathcal{R}(r)&\equiv g^{\mu\nu}R_{\mu\nu}
=-\frac{12}{\ell^2}
+\frac{12M Q_m^3\left(2Q_m^3-r^3\right)}
{\Delta(r)^3}
+\frac{3wN(3w-1)}{r^{3(w+1)}}
-\frac{\lambda}{r^3},
\label{RicciScalar}
\\[1cm]
\mathfrak{R}(r)&\equiv R_{\mu\nu}R^{\mu\nu}
=\frac{1}{2r^6}\Bigg[
\frac{72r^6}{\ell^4}
+\frac{144M^2Q_m^6r^6
\left(2Q_m^6-2Q_m^3r^3+5r^6\right)}
{\Delta(r)^6}
\nonumber
\\[0.05cm]&\qquad
+9N^2r^{-6w}w^2\left(5+6w+9w^2\right)
+6Nr^{-3w}w(5+3w)\lambda
+5\lambda^2
\nonumber
\\[0.05cm]&\qquad
+\frac{12r^3}{\ell^2}
\left(
\frac{12MQ_m^3r^3\left(-2Q_m^3+r^3\right)}
{\Delta(r)^3}
+3Nr^{-3w}(1-3w)w+\lambda
\right)
\\[0.05cm]&\qquad
+\frac{24MQ_m^3r^3
\left[
-3Nr^{-3w}w
\left(Q_m^3(1-3w)+2r^3(2+3w)\right)
-\left(Q_m^3+4r^3\right)\lambda
\right]}
{\Delta(r)^3}
\Bigg],
\nonumber
\\[1cm]
\mathcal{K}(r)&\equiv
R_{\mu\nu\rho\sigma}R^{\mu\nu\rho\sigma}
=\frac{1}{r^6}\Bigg[
\left(
\frac{2r^3}{\ell^2}
+\frac{2Mr^3(-2Q_m^3+r^3)}
{\Delta(r)^2}
+Nr^{-3w}(1+3w)
+\lambda
-\lambda\ln\left(\frac{r}{|\lambda|}\right)
\right)^2
\nonumber\\[0.05cm]&\qquad
+4\left(
\frac{r^3}{\ell^2}
-Nr^{-3w}
-\frac{2Mr^3}{\Delta(r)}
+\lambda\ln\left(\frac{r}{|\lambda|}\right)
\right)^2
\nonumber\\[0.05cm]&\qquad
+\frac{3r^{-6w}}{\ell^4\Delta(r)^4}
\Bigg(
2r^{3+3w}\Delta(r)^2
+\ell^2\Big[
N\Delta(r)^2(1+3w)
\label{kretschmann}\\[0.05cm]&\qquad\qquad
+r^{3w}\Big(
M(-4Q_m^3r^3+2r^6)
+\Delta(r)^2\lambda
\Big)\Big]
-\ell^2r^{3w}\Delta(r)^2
\lambda\ln\left(\frac{r}{|\lambda|}\right)
\Bigg)^2
\nonumber\\[0.1cm]&\qquad
+\frac{r^{-6w}}{\ell^4\Delta(r)^6}
\Bigg(
2r^{3+3w}\Delta(r)^3
-\ell^2\Big[
N\Delta(r)^3(2+9w+9w^2)
\nonumber\\[0.05cm]&\qquad\qquad
+r^{3w}\Big(
4Mr^3(Q_m^6-7Q_m^3r^3+r^6)
+3\Delta(r)^3\lambda
\Big)\Big]
+2\ell^2r^{3w}\Delta(r)^3
\lambda\ln\left(\frac{r}{|\lambda|}\right)
\Bigg)^2
\Bigg];
\nonumber
\end{align}

where we have defined $\Delta(r)\equiv Q_m^3+r^3$.
\begin{table}[ht]
\centering
\renewcommand{\arraystretch}{1.5}
\setlength{\tabcolsep}{10pt}
\begin{tabular}{c c c c}
\hline\hline
Limiting case
& $\mathcal{R}$
& $\mathfrak{R}$
& $\mathcal{K}$
\\
\hline

Schwarzschild--AdS$_4$
$(Q_m=N=\lambda=0)$
&
$\displaystyle -\frac{12}{\ell^2}$
&
$\displaystyle \frac{36}{\ell^4}$
&
$\displaystyle \frac{48M^2}{r^6}+\frac{24}{\ell^4}$
\\[0.35cm]

Schwarzschild
$(Q_m=N=\lambda=0,\;\ell\rightarrow\infty)$
&
$0$
&
$0$
&
$\displaystyle \frac{48M^2}{r^6}$
\\[0.35cm]

AdS$_4$
$(M=Q_m=N=\lambda=0)$
&
$\displaystyle -\frac{12}{\ell^2}$
&
$\displaystyle \frac{36}{\ell^4}$
&
$\displaystyle \frac{24}{\ell^4}$
\\[0.35cm]

$r\rightarrow\infty,\;-1<w<-1/3$
&
$\displaystyle -\frac{12}{\ell^2}$
&
$\displaystyle \frac{36}{\ell^4}$
&
$\displaystyle \frac{24}{\ell^4}$
\\[0.35cm]

$r\rightarrow\infty,\;w<-1$
&
$\displaystyle \sim r^{-3(w+1)}$
&
$\displaystyle \sim r^{-6(w+1)}$
&
$\displaystyle \sim r^{-6(w+1)}$
\\[0.35cm]

\hline\hline
\end{tabular}
\caption{
Limiting cases of the curvature invariants. For the quintessence regime, the curvature invariants approach their AdS$_4$ values as $r\rightarrow\infty$, whereas in the phantom regime they diverge at large distances.
}
\label{LimitingCases}
\end{table}
\begin{table*}[ht]
\centering
\small
\renewcommand{\arraystretch}{1.2}
\setlength{\tabcolsep}{3pt}
\begin{tabular}{c c c c c}
\hline\hline
Parameter sector
& $\mathcal{R}( r \to 0)$
& $\mathfrak{R}( r \to 0)$
& $\mathcal{K}( r \to 0)$
& Core
\\
\hline

$\lambda=0,\;Q_m\neq0$
&
$\displaystyle-\frac{12}{\ell^2}+\frac{24M}{Q_m^3}$
&
$\displaystyle\frac{36(-2\ell^2M+Q_m^3)^2}{\ell^4Q_m^6}$
&
$\displaystyle\frac{24(-2\ell^2M+Q_m^3)^2}{\ell^4Q_m^6}$
&
Regular
\\[0.3cm]

$\lambda=0,\;w=-1,\;Q_m\neq0$
&
$\displaystyle-\frac{12}{\ell^2}+\frac{24M}{Q_m^3}+12N$
&
$\displaystyle\frac{36[Q_m^3-\ell^2(2M+NQ_m^3)]^2}{\ell^4Q_m^6}$
&
$\displaystyle\frac{24[Q_m^3-\ell^2(2M+NQ_m^3)]^2}{\ell^4Q_m^6}$
&
Regular
\\[0.3cm]

$\lambda=0,\;-1<w<-1/3$
&
$\displaystyle\sim r^{-3(w+1)}$
&
$\displaystyle\sim r^{-6(w+1)}$
&
$\displaystyle\sim r^{-6(w+1)}$
&
Singular
\\[0.3cm]

$\lambda\neq0$
&
$\displaystyle\sim-\frac{\lambda}{r^3}$
&
$\displaystyle\sim\frac{5\lambda^2}{2r^6}$
&
$\displaystyle\sim\frac{\lambda^2}{r^6}
\ln^2\!\left(\frac{r}{|\lambda|}\right)$
&
Singular
\\

\hline\hline
\end{tabular}

\caption{
Behavior of the curvature invariants near $r=0$ for the relevant parameter sectors. For $Q_m\neq0$ and $\lambda=0$, the phantom regime $w<-1$ and the limiting case $w=-1$ admit a regular core, whereas the quintessence regime $-1<w<-1/3$ is singular. A nonzero $\lambda$ also prevents a regular core.
}
\label{Ricci_table}
\end{table*}
In Table~\ref{LimitingCases}, we summarize the relevant limiting geometries and verify that the known results are consistently recovered. Moreover, in the asymptotic region $r\to\infty$, all curvature invariants in Eqs.~\eqref{RicciScalar}--\eqref{kretschmann} approach their AdS$_4$ values for the quintessence regime, confirming the asymptotically AdS nature of the spacetime. In contrast, in the phantom regime, the dark energy contribution dominates at large distances, and the spacetime is no longer asymptotically AdS. The role of the PFDM parameter $\lambda$ in the central geometry is reflected in the leading behavior of the curvature invariants summarized in Table~\ref{Ricci_table}. For $\lambda\neq0$, $\mathcal{R}\propto-\lambda/r^3$, $\mathfrak{R}\sim5\lambda^2/(2r^6)$, and $\mathcal{K}\propto\lambda^2r^{-6}\ln^2(r/|\lambda|)$ as $r\rightarrow0$. Thus, the invariants diverge and the regular core found for $Q_m\neq0$ and $\lambda=0$ is lost, independently of the regularizing effect of the nonlinear magnetic sector \cite{ahmed2026shadow}. However, since the thermodynamic analysis is formulated locally at the event horizon, the nature of the central geometry does not directly enter the subsequent analysis, and no direct link to the horizon thermodynamics is established within the present framework.

\section{Thermodynamic Properties of the Black Hole Solution}\label{sec3}

We begin this section by examining the thermodynamic properties of the NLMC--AdS black hole. As a first step, we determine the entropy of the system. Since the gravitational sector of the theory is governed by the Einstein--Hilbert action, the entropy is expected to follow the Bekenstein--Hawking area law, $S=A/4$. For the spherically symmetric geometry given by Eq.~\eqref{metric}, the horizon area is $A=4\pi r_h^2$, which leads to
\begin{equation}
S=\pi r_h^2.
\label{entropy}
\end{equation}
However, there exists an alternative way to obtain the entropy by using the standard thermodynamic relation for black holes,
\begin{equation}
\tilde{S}=\int \frac{dM}{T_H},
\label{cc2}
\end{equation}
where $T_H$ is the Hawking temperature, given by
\begin{equation}
\begin{split}
T_{H}\equiv\frac{f'(r_h)}{4\pi}
=\frac{1}{4\pi}\Bigg\{
&\frac{2 r_h}{\ell^2}
+ \frac{N(1+3w)}{r_h^{2+3w}}
+ \frac{\lambda\left[1-\ln\left(\frac{r_h}{|\lambda|}\right)\right]}{r_h^2}
\\
&+ \frac{\left(r_h^3-2 Q_m^3\right)}
{r_h\left(Q_m^3+r_h^3\right)}
\left[
1+\frac{r_h^2}{\ell^2}
-N r_h^{-1-3w}
+\frac{\lambda}{r_h}
\ln\left(\frac{r_h}{|\lambda|}\right)
\right]
\Bigg\}.
\end{split}
\label{cc1}
\end{equation}
Using the horizon condition $f(r_h)=0$ together with the Hawking temperature Eq.~\eqref{cc1}, we obtain
\begin{equation}
\tilde{S}=2\pi \int \left(r_h+\frac{Q_m^3}{r_h^2}\right) dr_h
=\pi r_h^2 \left(1-\frac{2Q_m^3}{r_h^3}\right),
\label{entr2}
\end{equation}
which is the entropy derived in \cite{DJG2024,ndongmo2023thermodynamics}. Clearly, Eq.~\eqref{entropy} and Eq.~\eqref{entr2} coincide only in the limit $Q_m \to 0$. The origin of this discrepancy lies in the fact that the relation in Eq.~\eqref{cc2} determines only the derivative of the entropy with respect to the horizon radius, and implicitly treats the quantities $(Q_m,\lambda,P,N)$ as fixed background parameters. Consequently, the entropy is defined only up to a transformation of the form
\begin{equation}
\tilde{S} \rightarrow \tilde{S} + g(Q_m,\lambda,P,N),
\end{equation}
where $g$ is an arbitrary function of the remaining thermodynamic parameters. In contrast, following our previous works \cite{Ladino:2024ned,ladino2025phase,ahmed2026shadow}, in the present study we adopt the extended thermodynamic framework, where the parameters $(Q_m,\lambda,P,N)$ are regarded as independent thermodynamic variables \cite{kubizvnak2017black}. In this approach, the thermodynamic behavior of the black hole is completely characterized by the fundamental relation $M=M(S,P,Q_m,\lambda,N)$, where the pressure is associated with the AdS radius through $P=3/8\pi \ell^{2}$.
The fundamental relation is obtained by imposing the horizon condition $f(r_h)=0$ and rewriting the mass parameter in terms of the Bekenstein-Hawking entropy Eq.~\eqref{entropy}, read
\begin{equation}
M(S,P,Q_m,\lambda)=
\frac{\pi}{2S}
\left[
\left(\frac{S}{\pi}\right)^{3/2}+Q_m^3
\right]
\left[
1+
\lambda\left(\frac{\pi}{S}\right)^{\frac{1}{2}}
\ln\!\left(\frac{\sqrt{S/\pi}}{|\lambda|}\right)
+
\frac{8PS}{3}-
N \left(\frac{\pi}{S}\right)^{\frac{3w+1}{2}}
\right].
\label{funda}
\end{equation}
The fundamental equation $M(X^i)$ defines a quasi-homogeneous thermodynamic system \cite{quevedo2023unified,quevedo2019quasi}.
This property follows from the dimensional scaling behaviour of the
entropy and coupling constants
$X^i=\{S,P,Q_m,\lambda,N\}$, which transform as
$X^i\to\mu^{\nu_i}X^i$.
Under this generalized scaling, the parameters obey
\begin{equation}
S\to \mu S,
\qquad
P\to \mu^{-1}P,
\qquad
Q_m\to \mu^{1/2}Q_m,
\qquad
\lambda\to \mu^{1/2}\lambda,
\qquad
N\to \mu^{\frac{3w+1}{2}} N , \label{wi}
\end{equation}
where $\mu$ is an arbitrary scaling parameter. As a consequence, the mass transforms according to
\begin{equation}
M(\mu^{w_i}X^i)=\mu^{1/2}M(X^i).
\end{equation}
which is the same scaling behavior obtained in \cite{ahmed2026shadow}. Moreover, using the conjugate thermodynamic parameters 
$\mathcal{Y}_i\equiv\partial M/\partial X^i$, Euler’s theorem yields 
the generalized Smarr relation \cite{romero2024extended1}
\begin{equation}
M=2\sum_i \nu_i X^i\mathcal{Y}_{i}
=2TS-2PV+\Phi_m Q_m+\Pi_\lambda \lambda
+(3w+1)\Psi\,N.
\label{smarr}
\end{equation}
Interestingly, the Smarr relation depends explicitly on the parameter $w$; for example, a similar feature appears in Lifshitz black holes, where it depends on the spacetime scaling parameter \cite{romero2024extended1}. The conjugate thermodynamic variables are defined as

\begin{align}
T&\equiv \left(\frac{\partial M}{\partial S}\right)=\frac{1}{8 \sqrt{\pi} S^{5/2}} \Bigg[
2 \Big(
-2 \pi^{3/2} Q_m^3 \sqrt{S}
+ S^2 (1 + 8 P S)
+ 3 N \, \pi^{\frac{1+3w}{2}} S^{-\frac{3(1-w)}{2}} \, w   \nonumber \\
&+ 3 N \, \pi^{2 + \frac{3w}{2}} Q_m^3 S^{-\frac{3w}{2}} (1+w)
+ \pi^2 Q_m^3 \lambda
+ \sqrt{\pi} S^{3/2} \lambda
\Big) - 3 \pi^2 Q_m^3 \lambda \ln\!\left(\frac{S}{\pi \lambda^2}\right)
\Bigg], \label{eq:Tthermo}  \\[0.6em]
V &\equiv \left(\frac{\partial M}{\partial P}\right)
=
\frac{4\pi}{3}
\left[
\left(\frac{S}{\pi}\right)^{3/2}+Q_m^3
\right],
\label{eq:Vthermo}\\[0.6em]
\Phi_m &\equiv \left(\frac{\partial M}{\partial Q_m}\right)
=
\frac{3\pi Q_m^2}{2S}
\left[
1- N \, \pi^{\frac{1+3w}{2}} \left(\frac{1}{S}\right)^{\frac{1+3w}{2}}
+\frac{\lambda}{\sqrt{S/\pi}}
\ln\!\left(\frac{\sqrt{S/\pi}}{|\lambda|}\right)
+
\frac{8PS}{3}
\right],
\label{eq:Phim}\\[0.6em]
\Pi_\lambda &\equiv
\left(\frac{\partial M}{\partial \lambda}\right)
=
\frac{
(\pi^{3/2}Q_m^3+S^{3/2})
}{
4S^{3/2}
}
\left[
\ln\!\left(\frac{S}{\pi\lambda^2}\right)-2
\right],
\label{eq:Pilambda}\\[0.6em]
\Psi &\equiv
\left(\frac{\partial M}{\partial N}\right)
=
- \frac{1}{2} \pi^{\frac{3w}{2}} \left(\frac{1}{S}\right)^{\frac{3(1+w)}{2}} \left[ \pi^{\frac{3}{2}} Q_m^3 + S^{\frac{3}{2}} \right].\label{eq:Psi}
\end{align}

Therefore, promoting the coupling constants of the theory to thermodynamic
variables \cite{romero2024extended1,quevedo2019quasi}, the extended first law of black
hole thermodynamics becomes
\begin{equation}
dM=T\,dS+V\,dP+\Phi\,dQ_m+\Pi_\lambda\,d\lambda +\Psi dN.
\label{first law}
\end{equation}
Interestingly, when an approach based on scaling arguments is employed, the resulting thermodynamic quantities differ from their geometric counterparts. In particular, the thermodynamic temperature and the thermodynamic volume given in Eq.~\eqref{eq:Tthermo} and Eq.~\eqref{eq:Vthermo} do not coincide with the geometric Hawking temperature in Eq.~\eqref{cc1} and the geometric volume $V=4 \pi r_h^3/3$, respectively. This discrepancy arises because, in the fundamental relation Eq.~\eqref{funda}, the magnetic sector couples directly to the PFDM and to the AdS pressure. Both descriptions coincide only in the limit \( Q_m \to 0 \), where the two entropy expressions in Eq.~\eqref{entropy} and Eq.~\eqref{entr2} become identical.
\begin{figure}[ht!]
\centering

\begin{minipage}[t]{0.48\linewidth}
    \centering
    \includegraphics[width=\linewidth]{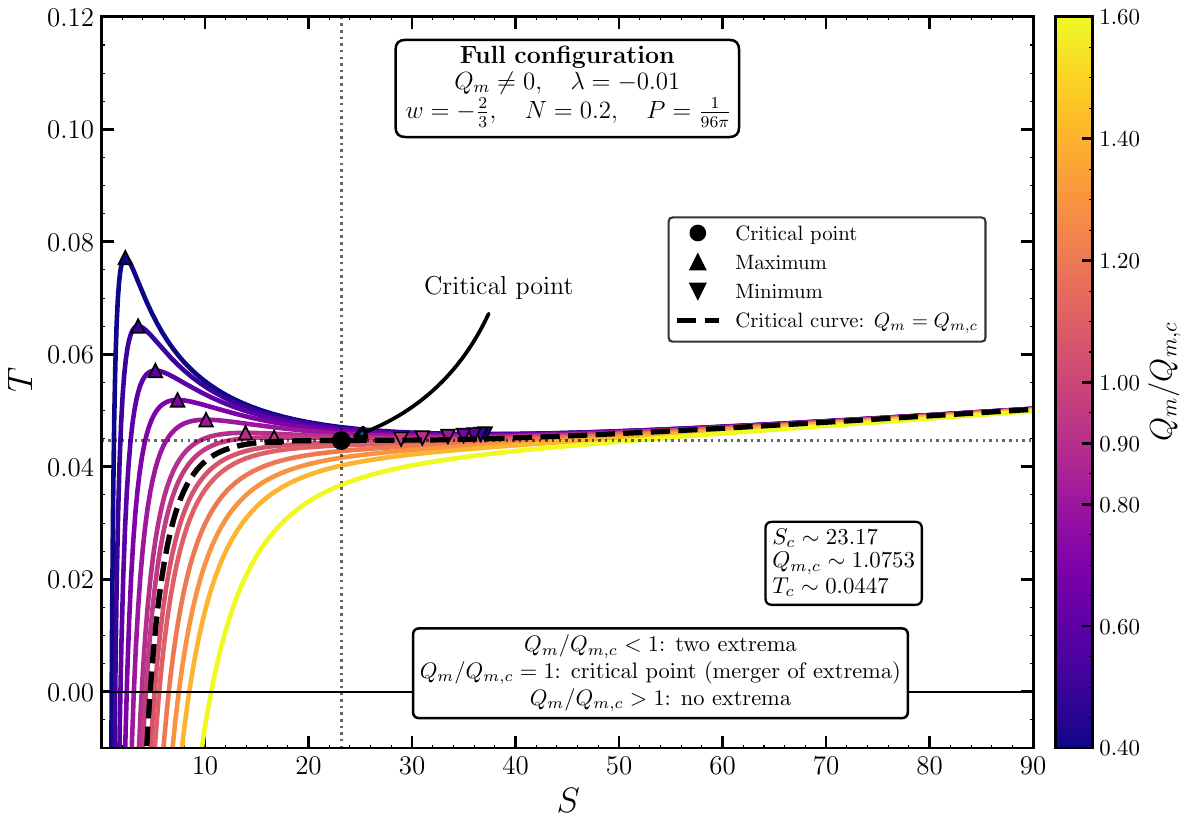}\\ 
    (a)
\end{minipage}
\hfill
\begin{minipage}[t]{0.48\linewidth}
    \centering
    \includegraphics[width=\linewidth]{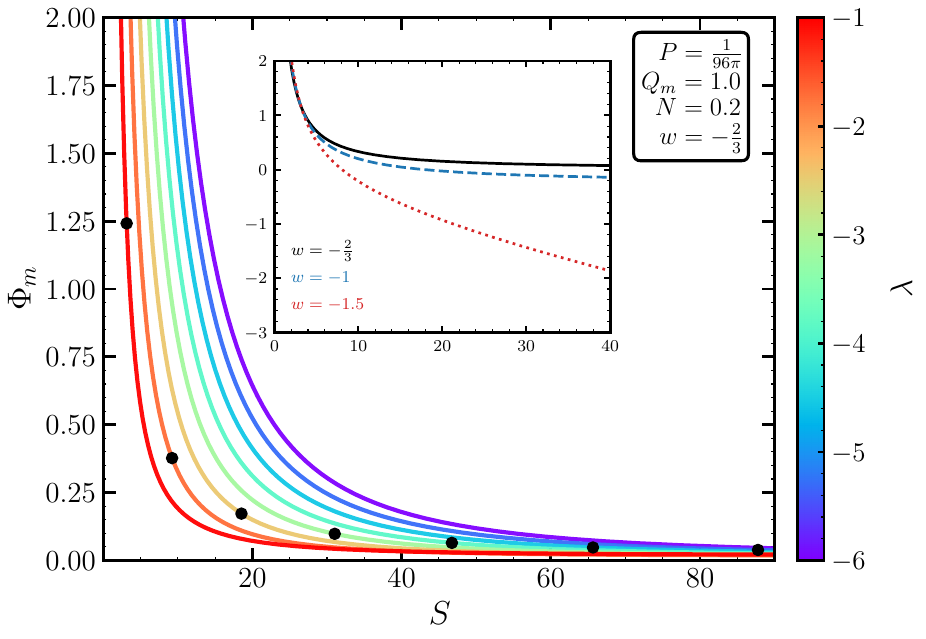}\\
    (b)
\end{minipage}
\hfill\\ \vspace{0.3in}

\begin{minipage}[t]{0.48\linewidth}
    \centering
    \includegraphics[width=\linewidth]{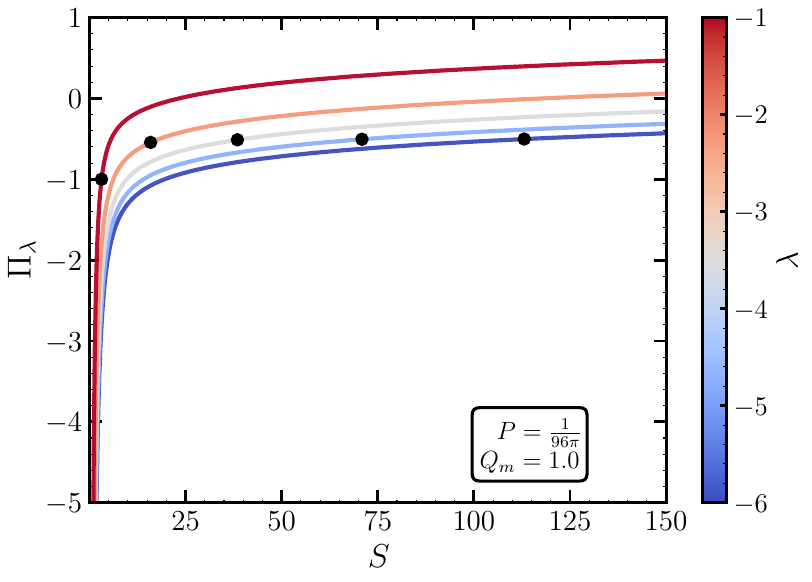}\\ 
    (c)
\end{minipage}
\hfill
\begin{minipage}[t]{0.48\linewidth}
    \centering
    \includegraphics[width=\linewidth]{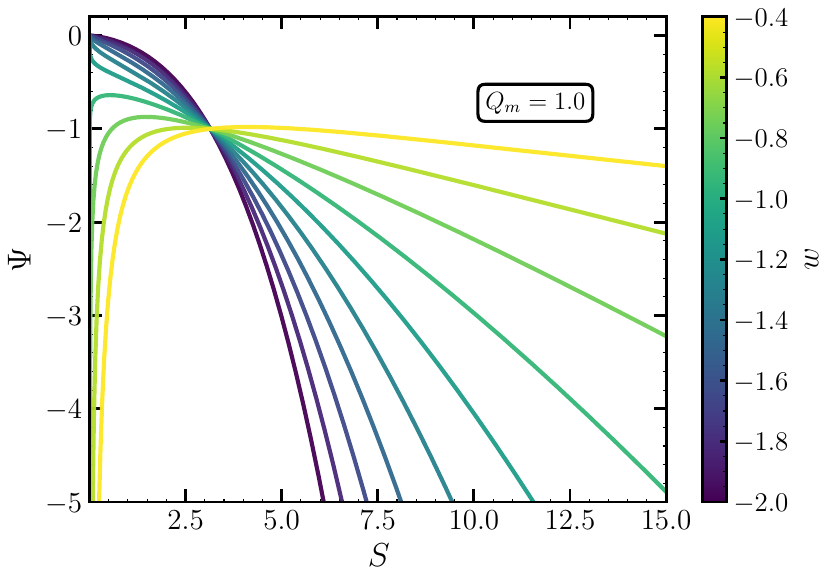}\\ 
    (d)
    \end{minipage}
    \hfill
\caption{Thermodynamic functions for the full black hole configuration with all parameters nonzero: (a) temperature $T(S)$, (b) magnetic potential $\Phi_m(S)$, (c) $\Pi_\lambda(S)$, and (d) $\Psi(S)$.}
\label{temperature_function}
\end{figure}
\begin{figure}[ht!]
\centering

\begin{minipage}[t]{0.48\linewidth}
    \centering
    \includegraphics[width=\linewidth]{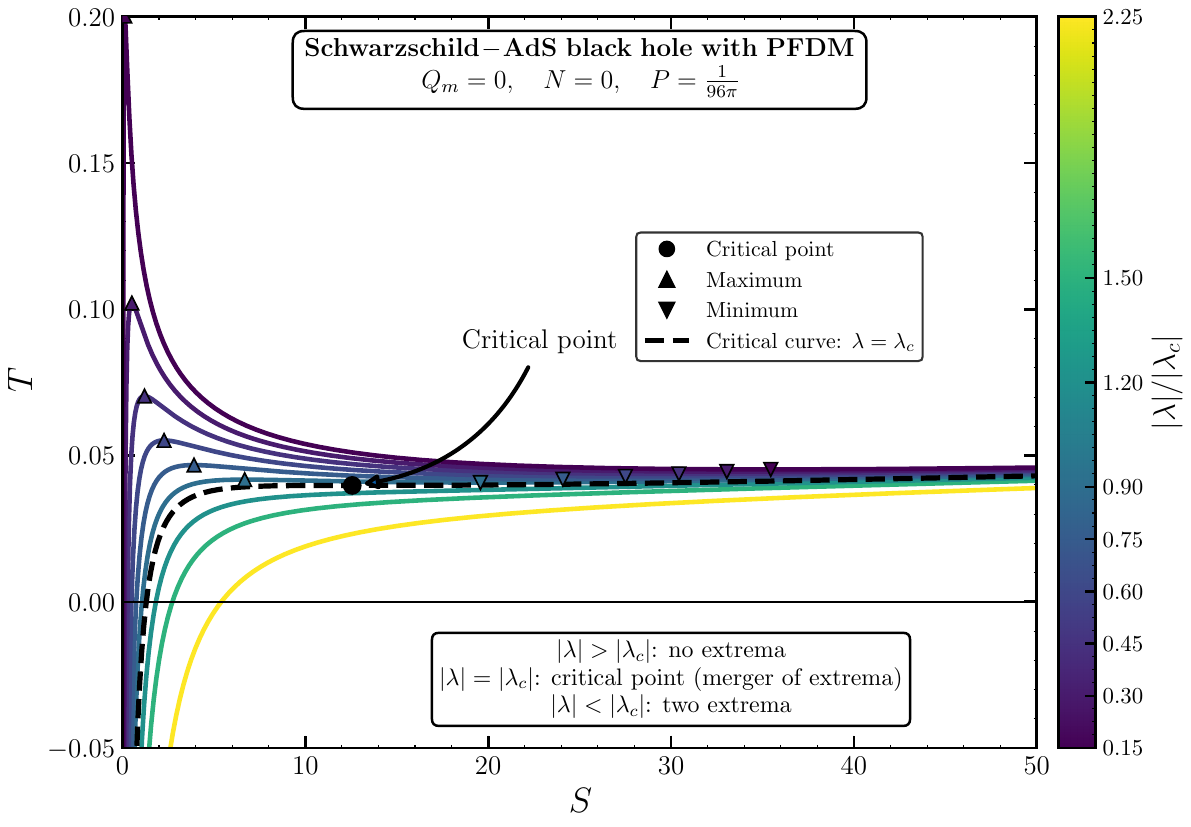}\\ 
    (a)
\end{minipage}
\hfill
\begin{minipage}[t]{0.48\linewidth}
    \centering
    \includegraphics[width=\linewidth]{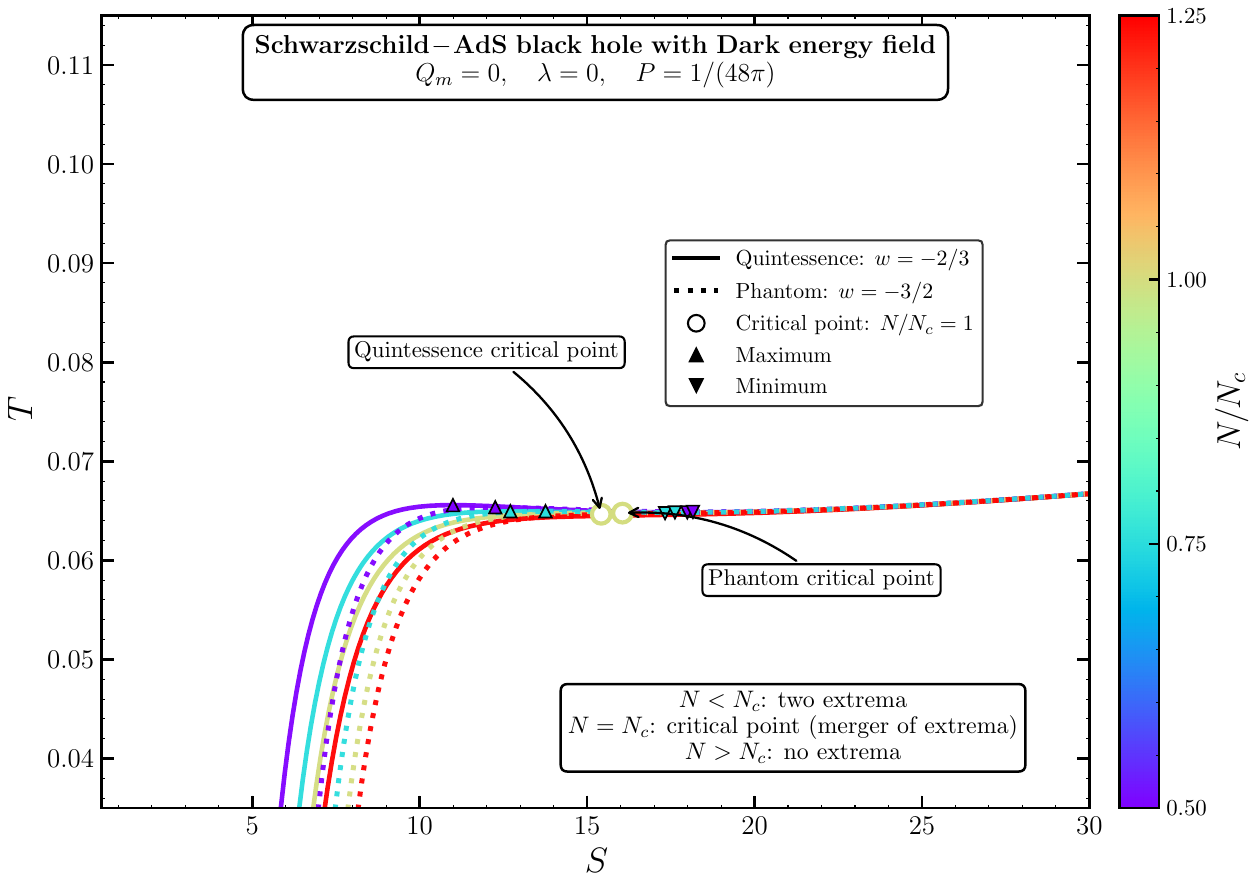}\\
    (b)
\end{minipage}
\hfill\\ \vspace{0.3in}

\begin{minipage}[t]{0.48\linewidth}
    \centering
    \includegraphics[width=\linewidth]{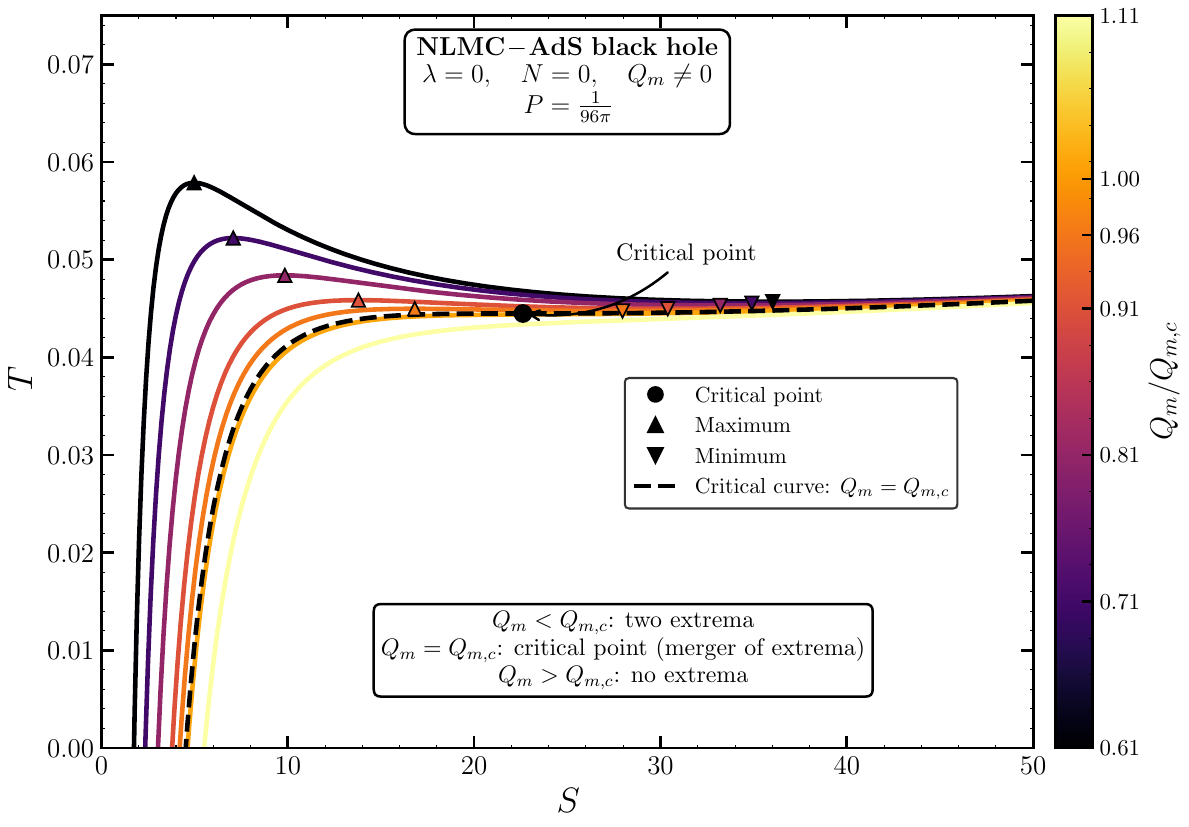}\\ 
    (c)
\end{minipage}
    \hfill
\caption{Temperature--entropy diagrams at fixed pressure for (a) the Schwarzschild--AdS black hole with PFDM, (b) the Schwarzschild--AdS black hole with dark energy, and (c) the NLMC--AdS black hole.}
\label{temperature_function2}
\end{figure}

\begin{table}[ht!]
\centering
\renewcommand{\arraystretch}{1.35}
\setlength{\tabcolsep}{10pt}

\resizebox{\linewidth}{!}{%
\begin{tabular}{lccc}
\hline
Configuration
& $S_c$
& $T_c$
& Critical parameter
\\
\hline

Schwarzschild--AdS + PFDM
&
$\displaystyle \frac{1}{24P}$
&
$\displaystyle \frac{\sqrt{6P}}{2\sqrt{\pi}}$
&
$\displaystyle
\lambda_c=-\frac{1}{\sqrt{216\pi P}}
$
\\[12pt]

Schwarzschild--AdS + dark energy
&
$\displaystyle
\frac{7-3w}{24P(3-w)}
$
&
$\displaystyle
\frac{7-3w}
{2(8-3w)\sqrt{\pi}}
\sqrt{\frac{24P(3-w)}{7-3w}}
$
&
$\displaystyle
N_c=
-\frac{2}
{9w(3-w)(8-3w)}
\pi^{-\frac{1+3w}{2}}
\left[
\frac{7-3w}{24P(3-w)}
\right]^{\frac{7-3w}{2}}
$
\\[14pt]

NLMC--AdS
&
$\displaystyle
\frac{3}{40P}
$
&
$\displaystyle
\frac{\sqrt{30P}}{4\sqrt{\pi}}
$
&
$\displaystyle
Q_{m,c}
=
\frac{2^{5/6}3^{1/2}5^{1/6}}
{40\sqrt{\pi P}}
$
\\

\hline
\end{tabular}%
}

\caption{Critical parameters in the $T$--$S$ plane for the three limiting black-hole configurations at fixed pressure $P$.}
\label{tab:critical_TS}
\end{table}
In Fig.~\ref{temperature_function}, we show the temperature and conjugate potentials of the full black hole solution as functions of the entropy. In the $T$--$S$ plane, Fig.~\ref{temperature_function} (a), the temperature can develop a local maximum and minimum that may merge into an inflection point signaling criticality. Although $Q_m$ is used as a reference parameter to illustrate this behavior, the critical structure is determined by the interplay of all thermodynamic parameters entering the equation of state, namely $P$, $Q_m$, $\lambda$, $N$, and $w$. The extrema are determined numerically for the full solution and analytically for the limiting cases, as shown in Fig.~\ref{temperature_function2}, with the corresponding critical points summarized in Table~\ref{tab:critical_TS}. The magnetic potential provides a complementary view of the thermodynamic behavior. In Fig.~\ref{temperature_function}(b), we show $\Phi_m$, defined in Eq.~\eqref{eq:Phim} as the potential conjugate to $Q_m$. In the large--entropy regime $S \gg 1$, it approaches the constant value $\Phi_m \to 4\pi P Q_m^2$, indicating that the magnetic potential saturates for a LBH. In contrast, in the small--entropy limit $\Phi_m \to \infty$, reflecting the strong interaction between the magnetic charge and the near--horizon geometry. The PFDM parameter introduces a characteristic entropy scale $S_c = \pi \lambda^2$ (black dot), as illustrated in Fig.~\ref{temperature_function}(b). At this scale, the logarithmic PFDM correction vanishes and changes sign, revealing a characteristic length scale $r_h \sim |\lambda|$ associated with the PFDM distribution. An inset in the same figure Fig.~\ref{temperature_function}(b) shows the behavior of $\Phi_m$ for different values of the state parameter $w$. Notably, in the phantom regime ($w < -1$), the magnetic potential becomes negative in the LBH branch and eventually diverges, signaling a drastic modification of the thermodynamic response induced by the exotic nature of the surrounding fluid. Additionally, the PFDM conjugate potential $\Pi_\lambda$ in Eq.~\eqref{eq:Pilambda} quantifies the response of the black hole mass to variations of the parameter $\lambda$. As shown in Fig.~\ref{temperature_function}(c), it displays a behavior similar to that of the magnetic potential, where the same characteristic scale marks the transition of the logarithmic contribution. The quantity $\Psi$ represents the thermodynamic potential conjugate to the intensity $N$ of the dark energy field, measuring the response of the black hole mass to variations in the surrounding fluid in extended black hole thermodynamics. Notably, $\Psi$ is negative, indicating that it reduces the effective mass of the black hole, consistent with the interpretation of the surrounding fluid as contributing negative energy. As shown in Fig.~\ref{temperature_function}(d), in the phantom regime $w<-1$, $\Psi$ vanishes in the SBH limit, indicating a negligible contribution of the fluid to the energy, while it diverges for a LBH, where the impact of the surrounding fluid becomes dominant. In contrast, in the quintessence regime $-1<w<-1/3$, $\Psi$ diverges in both the SBH and LBH limits, showing that the effect of the fluid remains relevant across all scales. It is worth emphasizing that $w$ is used here as a phenomenological parameter characterizing the surrounding dark-energy field, rather than as a microscopic description of its interaction with the black hole. The quintessence and phantom regimes therefore represent distinct parameter sectors of the solution, leading to different contributions to the black-hole thermodynamic quantities. Any microscopic mechanism underlying a direct connection between the dark-energy properties and the horizon thermodynamics lies beyond the scope of this work.

\subsection{Reverse Isoperimetric Inequality}
Due to the presence of the magnetic field, the geometric volume differs from the thermodynamic volume. This deviation opens the possibility for the black hole to exhibit super--entropic behavior \cite{kubizvnak2017black}. Additionally, following the standard approach of black hole chemistry
\cite{kubizvnak2017black}, and motivated by the molecular interpretation
of the horizon degrees of freedom $\mathscr{N}$ the effective specific volume is
defined as $v \sim V/\mathscr{N}$. Since the number of horizon degrees of freedom
scales with the horizon area, $\mathscr{N} \sim A \sim r_h^2$, the specific volume
for a spherically symmetric spacetime can be written as
\begin{equation}
v=2\left(\frac{3V}{4\pi}\right)^{1/3},
\qquad
v=2(r_h^3+Q_m^3)^{1/3}.
\label{vespe}
\end{equation}
For $Q_m \to 0$ the standard result $v=2r_h$ is recovered. Hence the
magnetic charge introduces a correction proportional to $Q_m^3$,
effectively enlarging the microscopic volume associated with the
horizon degrees of freedom. The quantity that measures this deviation from the geometric volume is the reverse isoperimetric ratio $\mathcal{R}_{\mathrm{iso}}$, which provides a geometric bound relating the thermodynamic volume and the entropy of a black hole. In four dimensions, for a spherically symmetric black hole, this quantity is defined through the ratio \cite{romero2024extended1}
\begin{equation}
\mathcal{R}_{\mathrm{iso}}=
\left(\frac{3V}{4\pi}\right)^{1/3}
\left(\frac{4\pi}{A}\right)^{1/2}\ge 1 ,
\end{equation}
where $A$ denotes the horizon area. Using $A=4S$ and the thermodynamic volume Eq.~\eqref{eq:Vthermo}, the isoperimetric ratio becomes
\begin{equation}
\mathcal{R}_{\mathrm{iso}}=
\left(1+\frac{\pi^{3/2}Q_m^3}{S^{3/2}}\right)^{1/3}.
\end{equation}
\begin{figure}[ht!]
\centering
\begin{minipage}{0.48\textwidth}
\centering
\includegraphics[width=\linewidth]{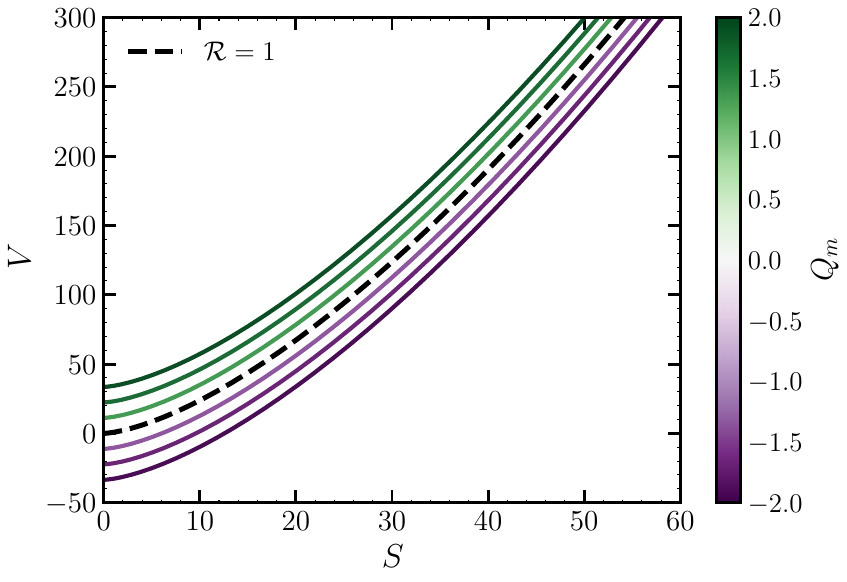}\\
(a)
\end{minipage}
\hfill
\begin{minipage}{0.48\textwidth}
\centering
\includegraphics[width=\linewidth]{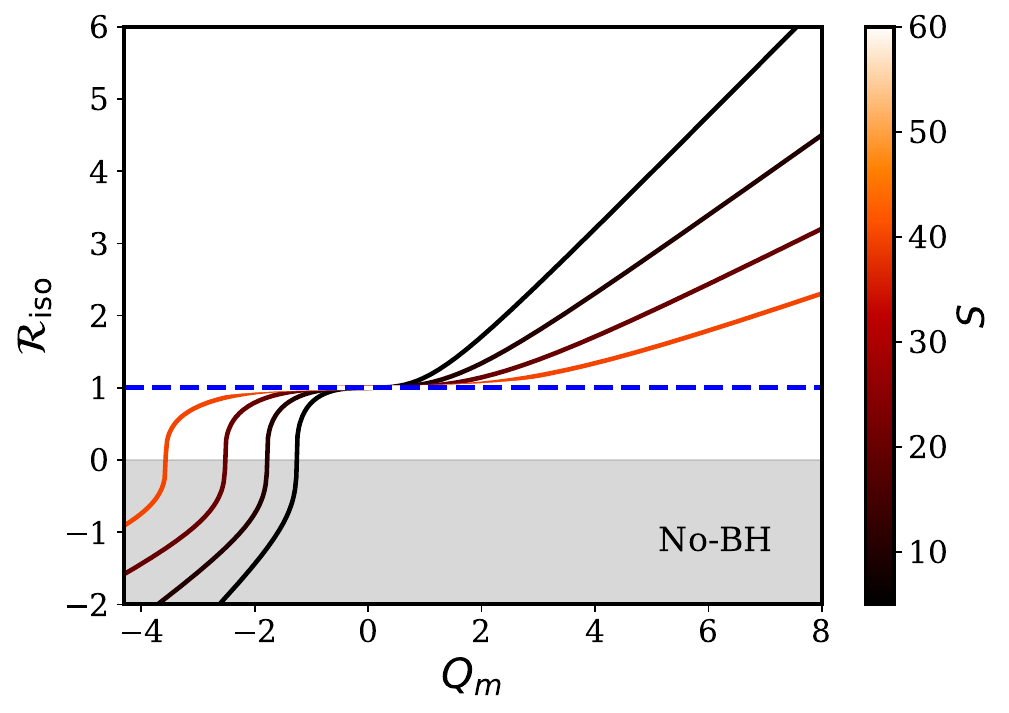}\\
(b)
\end{minipage}
\hfill
\caption{(a) Thermodynamic volume $V$ as a function of the entropy $S$ for different values of $Q_m$. 
(b) Reverse isoperimetric ratio as a function of $Q_m$ for different values of $S$. 
The shaded region corresponds to $\mathcal{R}_{\mathrm{iso}}<0$, which does not represent a physical black hole configuration.}
\label{fig:volume_R}
\end{figure}
The thermodynamic volume is illustrated in Fig.~\ref{fig:volume_R}(a), where the shift induced by the magnetic core can be clearly observed. Notice that for $Q_m>0$, the inequality $\mathcal{R}_{\mathrm{iso}}\ge 1$ is always satisfied, as shown in Fig.~\ref{fig:volume_R}(b), and it is saturated in the limit $Q_m=0$, corresponding to the Schwarzschild--AdS case. In the limit of large black holes, $S\gg Q_m$, the isoperimetric ratio approaches unity, $\mathcal{R}_{\mathrm{iso}}\to1$, recovering the standard geometric behaviour. Conversely, for $Q_m<0$ there always exists a region where $\mathcal{R}_{\mathrm{iso}}<1$, signaling a violation of the reverse isoperimetric conjecture. In this regime, the black hole exhibits super--entropic behavior, meaning that its entropy exceeds the value expected for a given thermodynamic volume. Similar behavior has also been reported for black holes with negative coupling parameters; see, for instance, \cite{romero2024extended1,feng2017horndeski}. Furthermore, for $Q_m<-\sqrt{S/\pi}$ one finds $\mathcal{R}_{\mathrm{iso}}<0$, indicating that no physical black hole solution exists in this regime.

\subsection{$P$--$V$ Criticality}
\label{PV section}
\begin{figure*}[ht!]
\centering

\begin{minipage}[t]{0.48\textwidth}
\centering
\includegraphics[width=\linewidth]{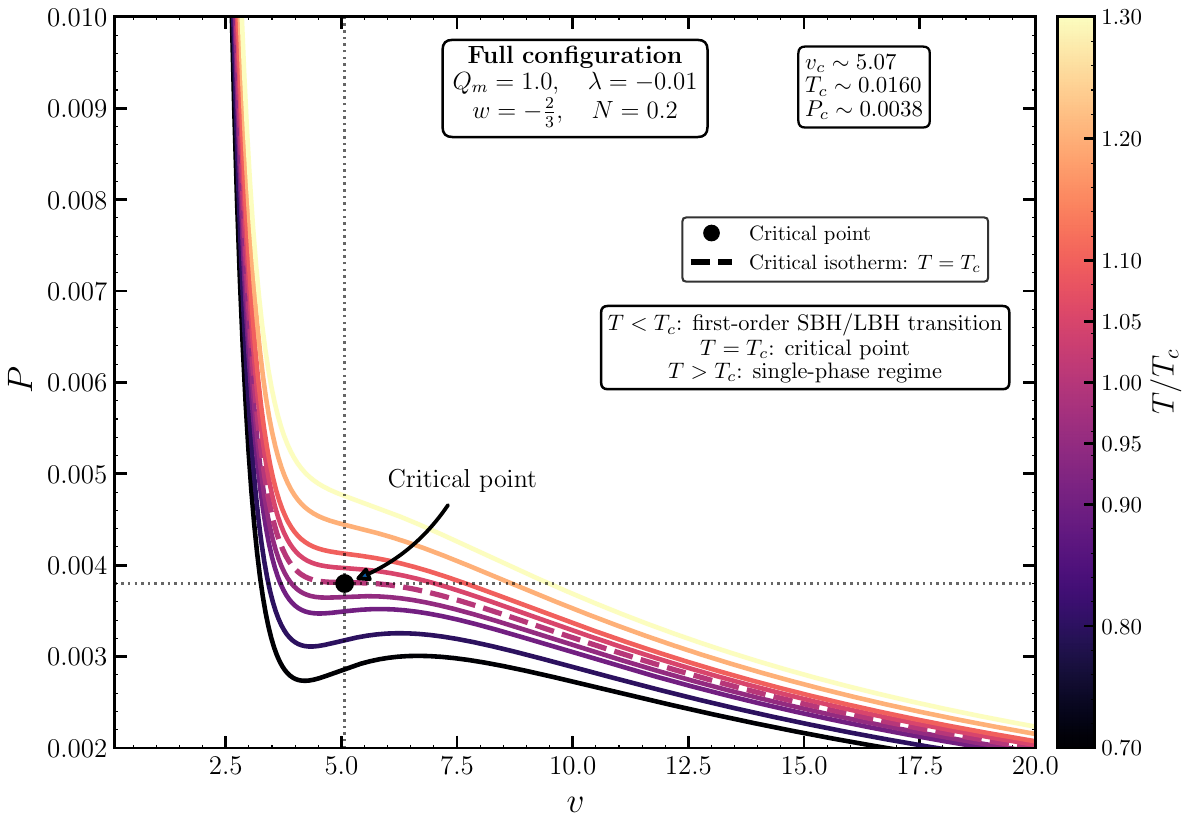}

(a)
\end{minipage}
\hfill
\begin{minipage}[t]{0.48\textwidth}
\centering
\includegraphics[width=\linewidth]{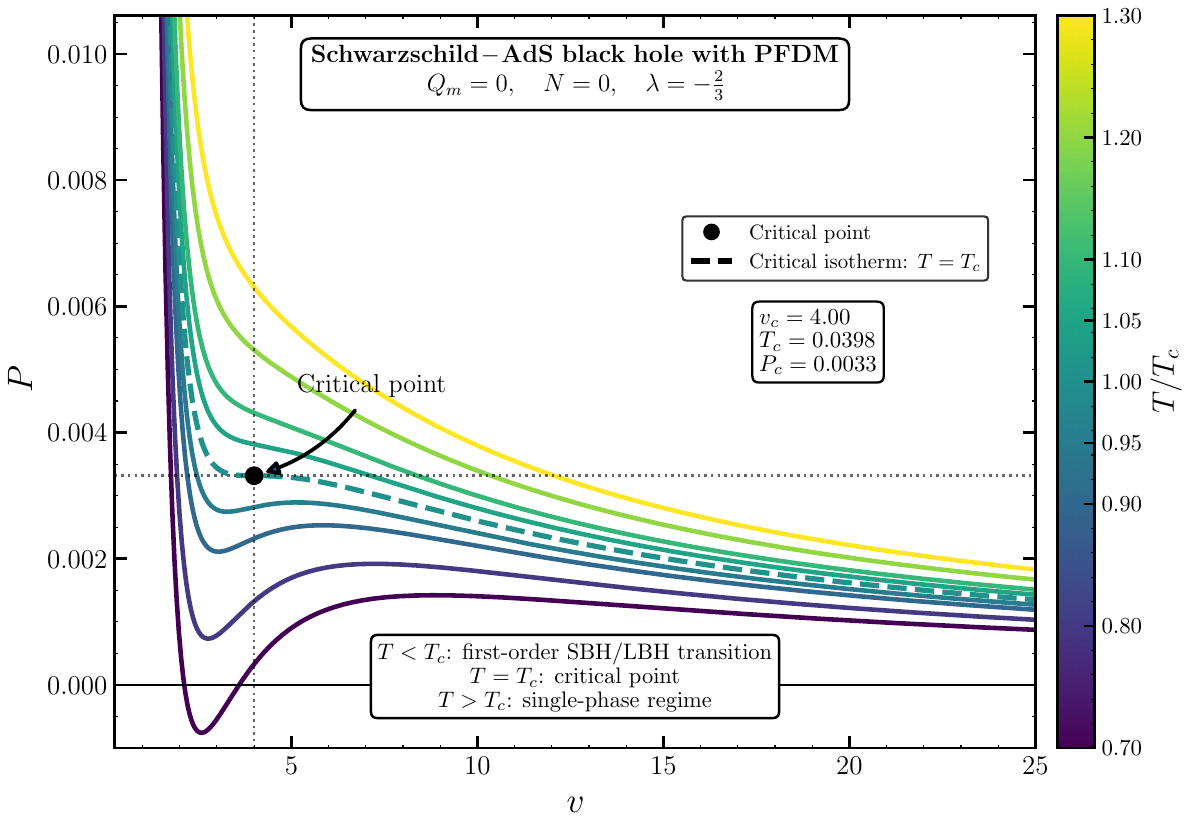}

(b)
\end{minipage}

\vspace{0.3cm}

\begin{minipage}[t]{0.48\textwidth}
\centering
\includegraphics[width=\linewidth]{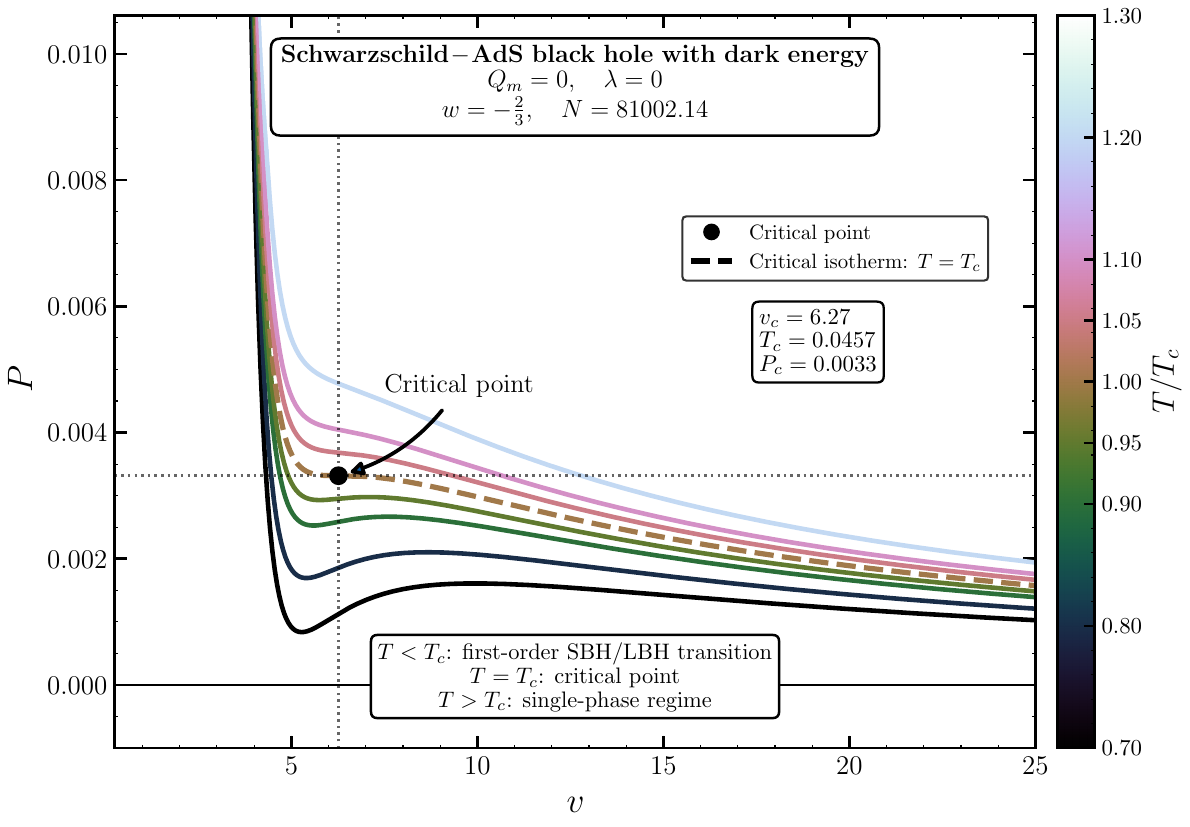}

(c)
\end{minipage}
\hfill
\begin{minipage}[t]{0.48\textwidth}
\centering
\includegraphics[width=\linewidth]{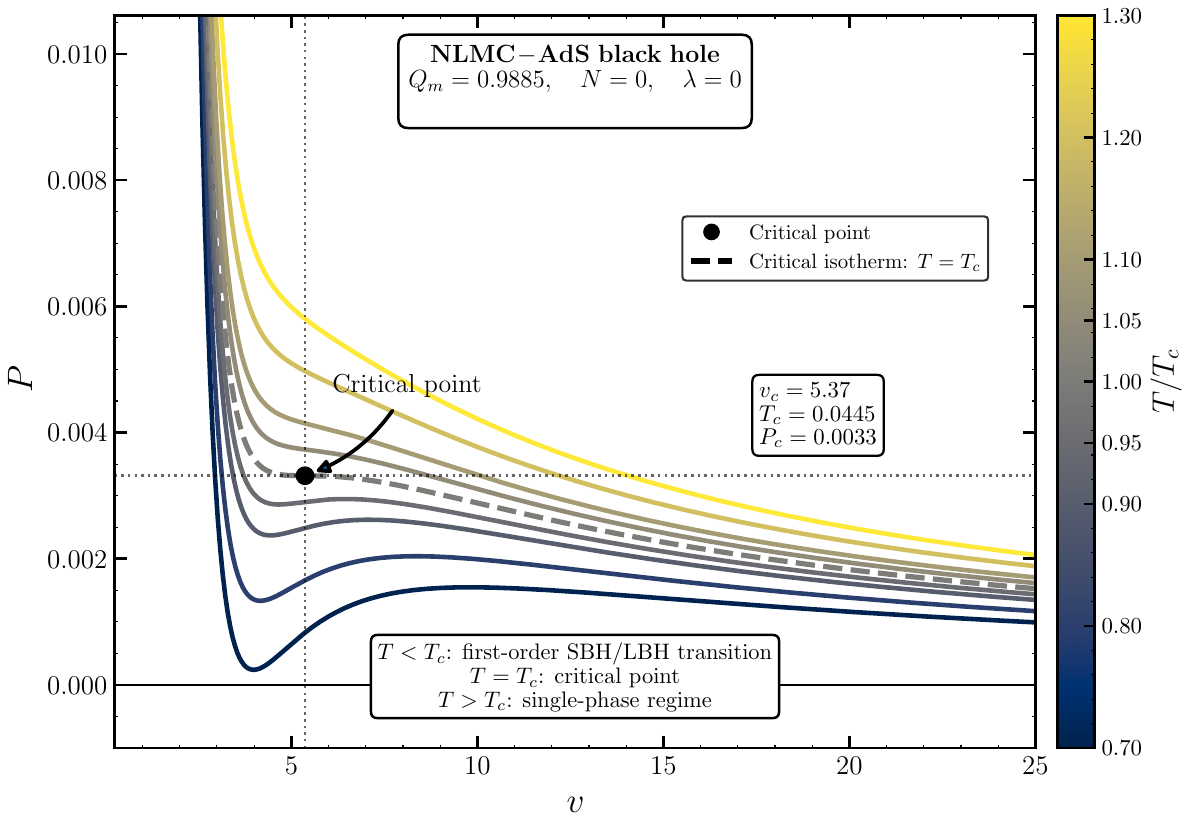}

(d)
\end{minipage}

\caption{
Equation-of-state $P(v,T)$ diagrams for:
(a) the full black hole configuration,
(b) the Schwarzschild--AdS black hole with PFDM,
(c) the Schwarzschild--AdS black hole with dark energy, and
(d) the NLMC--AdS black hole.
}
\label{fig:PV_profiles}
\end{figure*}

We now analyze the critical behavior of the black hole in the $P-v$ plane. Using the equation of state obtained from
Eq.~\eqref{eq:Tthermo} and the specific volume defined in
Eq.~\eqref{vespe}, the pressure can be written as

\begin{equation}
\begin{aligned}
P(v,T)=\;& \frac{T}{v}
-\frac{1}{2\pi v^2}
-\frac{\lambda}{\pi v^3}
-3\frac{ 8^{w} N\, w}{\pi\, v^{3(w+1)}}+\frac{8 Q_m^3}{\pi v^5}-3\frac{\,8^{w+1} N (1+w)\, Q_m^3}{\pi\, v^{3(w+2)}} \\
&+\frac{4 Q_m^3 \lambda}{\pi v^6}
\left[
3 \ln\!\left(\frac{v^2}{4\lambda^2}\right) - 2
\right].
\end{aligned}
\end{equation}
Here we have used the approximation $Q_m\ll r_h$, so that
$v\simeq2r_h$. Unlike the standard Reissner--Nordström AdS (RN--AdS) black hole \cite{Ladino:2024ned}, where the charge term scales as $Q^2/v^4$, here a cubic magnetic contribution $Q_m^3/v^5$ appears, reminiscent of NLEDs models where higher–order charge terms modify the critical behavior of AdS black holes \cite{hendi2013extended,hendi2016p}. The critical point is determined by the conditions
\begin{equation}
\left( \frac{\partial P}{\partial v} \right)_T= 0,
\qquad \left(\frac{\partial^2 P}{\partial v^2}\right)_T = 0. \label{PV}
\end{equation}
The simultaneous fulfillment of Eqs.~\eqref{PV} yields the criticality condition
\begin{align}
-9 \, 8^{w} \, &N(1 + w)\left[
v^{3} w (2 + 3 w) + 8 Q_{m}^{3} (2 + w)(5 + 3 w)
\right]
\nonumber \\
&- v^{3 w}
\left[
v^{3} (v + 6 \lambda)
+ 8 Q_{m}^{3} (-20 v + 63 \lambda)
- 360 Q_{m}^{3} \lambda \ln\left(\frac{v^{2}}{4 \lambda^{2}}\right)
\right]
= 0.
\label{PV1}
\end{align}
Notice that, for the full black hole configuration, the criticality condition Eq.~\eqref{PV1} cannot be solved analytically. However, following the analysis presented in Section~\ref{sec3}, the criticality conditions can be solved exactly for the different limiting configurations. For example, in the NLMC--AdS case ($\lambda,N\to0$), Eq.~\eqref{PV1} reduces to
\begin{equation}
v\left(160 Q_m^3 - v^3\right)=0,
\qquad \Rightarrow \qquad
v_c^{(NLCM)} = (160)^{1/3} Q_m . \label{vc0}
\end{equation}
Thus, using Eq.~\eqref{vc0}, the critical temperature and pressure are obtained from Eq.~\eqref{PV}
\begin{equation}
T_c^{(NLMC)} = \frac{3}{\pi (10240)^{1/3}Q_m}, \qquad P_c^{(NLMC)} = \frac{3}{80\pi (50)^{1/3}Q_m^2};
\end{equation}
with an universal critical ratio given by
\begin{equation}
\rho_c^{(NLMC)}=\frac{P_c^{(NLMC)} v_c^{(NLCM)}}{T_c^{(NLMC)}}=\frac{2}{5}.
\end{equation}
In the same way, the critical points can be determined exactly for all the limiting solutions, as summarized in Table~\ref{tab:critical_PV1}. Moreover, Fig.~\ref{fig:PV_profiles} displays the $P$--$v$ diagrams for the different black-hole configurations considered. Fig.~\ref{fig:PV_profiles}(a) corresponds to the full configuration, for which the critical points are computed numerically, while Fig.~\ref{fig:PV_profiles}(b), Fig.~\ref{fig:PV_profiles}(c), and Fig.~\ref{fig:PV_profiles}(d) correspond, respectively, to the Schwarzschild--AdS black hole with PFDM, the Schwarzschild--AdS black hole with dark energy, and the NLMC--AdS black hole. Interestingly, all the critical ratios presented in Table~\ref{tab:critical_PV1} differ from the standard RN--AdS/vdW value $\rho_c=3/8$ \cite{kubizvnak2012p}. Such deviations are commonly observed in 
higher--dimensional spacetimes or in scenarios where the equation of state is 
modified by NLED, see for instance 
\cite{gunasekaran2012extended,hendi2013extended,hendi2016p}, as well as in gravitational systems characterized by different asymptotic behaviors, including those with Lifshitz or Schrödinger scaling \cite{romero2024extended1,herrera2021hyperscaling,herrera2023anisotropic}.

\begin{table}[ht!]
\centering
\renewcommand{\arraystretch}{1.5}
\setlength{\tabcolsep}{10pt}

\resizebox{\linewidth}{!}{%
\begin{tabular}{lcccc}
\hline
Configuration
& $v_c$
& $T_c$
& $P_c$
& $\rho_c$
\\
\hline
Schwarzschild--AdS + PFDM
&
$\displaystyle -6\lambda$
&
$\displaystyle -\frac{1}{12\pi\lambda}$
&
$\displaystyle \frac{1}{216\pi\lambda^2}$
&
$\displaystyle \frac{1}{3}$
\\[12pt]

Schwarzschild--AdS + dark energy
&
$\displaystyle
2\sqrt{\frac{7-3w}
{24\pi P_c(3-w)}}
$
&
$\displaystyle
\frac{7-3w}
{2(8-3w)\sqrt{\pi}}
\sqrt{\frac{24P_c(3-w)}
{7-3w}}
$
&
$\displaystyle
\frac{7-3w}{24(3-w)}
\left[
-\frac{2\pi^{-\frac{1+3w}{2}}}
{9w(3-w)(8-3w)N}
\right]^{\frac{2}{7-3w}}
$
&
$\displaystyle
\frac{8-3w}{6(3-w)}
$
\\[16pt]

NLMC--AdS
&
$\displaystyle
(160)^{1/3}Q_m
$
&
$\displaystyle
\frac{3}
{\pi(10240)^{1/3}Q_m}
$
&
$\displaystyle
\frac{3}
{80\pi(50)^{1/3}Q_m^2}
$
&
$\displaystyle \frac{2}{5}
$

\\
\hline
\end{tabular}%
}

\caption{Exact critical points in the $P$--$v$ plane for the three limiting black-hole configurations. When evaluated using the numerical parameter values adopted in Fig.~\ref{fig:PV_profiles}, these expressions reproduce the corresponding numerical values.}
\label{tab:critical_PV1}
\end{table}

\begin{figure}[ht!]
\centering

\begin{minipage}{0.48\textwidth}
\centering
\includegraphics[width=\linewidth]{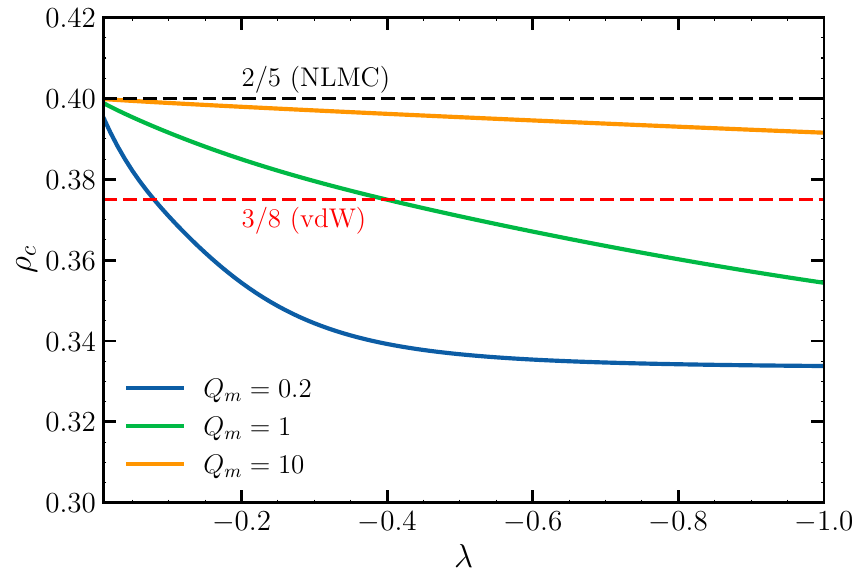}\\
(a)
\end{minipage}
\hfill
\begin{minipage}{0.48\textwidth}
\centering
\includegraphics[width=\linewidth]{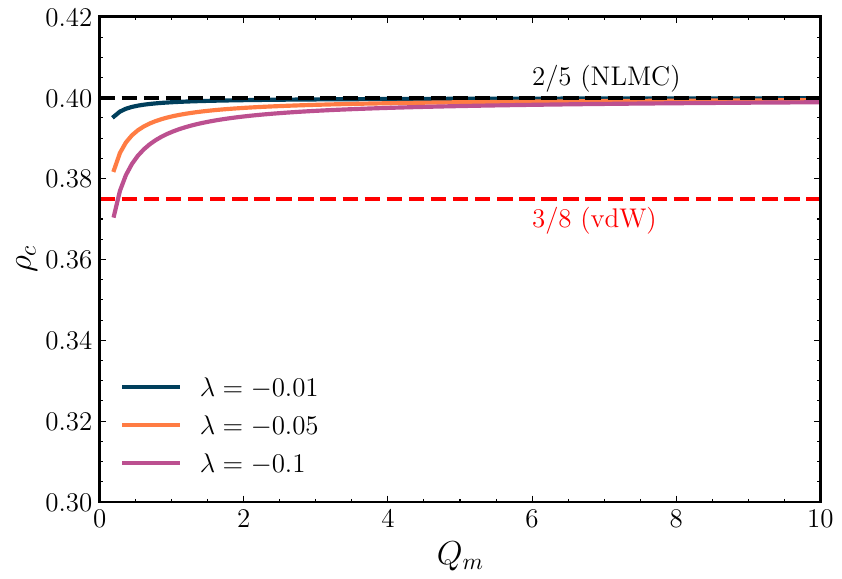}\\
(b)
\end{minipage}
\hfill \\ \vspace{0.3in}

\begin{minipage}{0.48\textwidth}
\centering
\includegraphics[width=\linewidth]{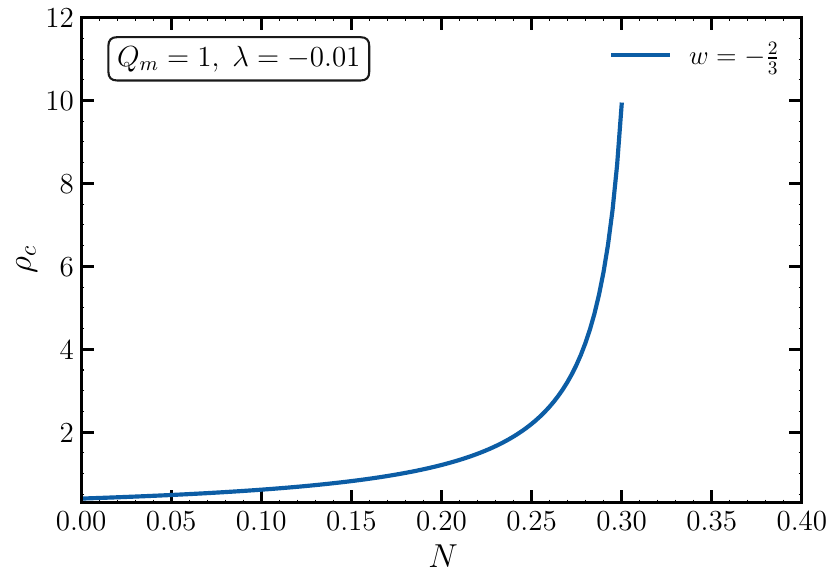}\\
(c)
\end{minipage}
\hfill
\begin{minipage}{0.48\textwidth}
\centering
\includegraphics[width=\linewidth]{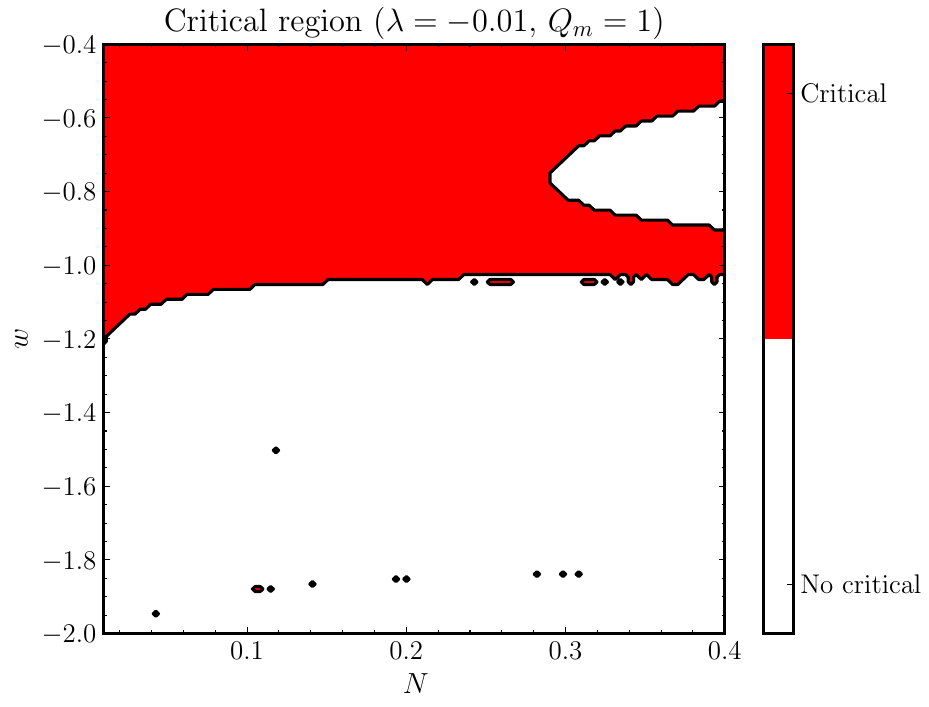}\\
(d)
\end{minipage}
\hfill

\caption{Critical ratio for the NLMC--AdS with PFDM case $(N=0)$: (a) variation with $\lambda$; (b) variation with $Q_m$; (c) variation with $N$ for the full configuration in the quintessence regime ($w=-2/3$); and (d) criticality region in the $(N,w)$ parameter space for $\lambda=-0.01$ and $Q_m=1$.}

\label{ratiolamQ}
\end{figure}
\begin{table}[ht!]
\centering
\setlength{\tabcolsep}{5pt}

% ----------- upper-----------

\begin{minipage}{0.45\textwidth}
\centering
\begin{tabular}{c c c c c c}
\hline
$\lambda$ & $Q_m$ & $v_c$ & $T_c$ & $P_c$ & $\rho_c$ \\
\hline
\multirow{3}{*}{$-0.01$}
& 0.2 & 1.0531 & 0.2206 & 0.0828 & 0.3954 \\
& 1   & 5.3737 & 0.0442 & 0.0033 & 0.3989 \\
& 10  & 54.1999 & 0.0044 & 0.0000 & 0.3999 \\
\hline
\multirow{3}{*}{$-0.05$}
& 0.2 & 1.0501 & 0.2026 & 0.0737 & 0.3821 \\
& 1   & 5.2657 & 0.0441 & 0.0033 & 0.3954 \\
& 10  & 53.9634 & 0.0044 & 0.0000 & 0.3994 \\
\hline
\multirow{3}{*}{$-0.1$}
& 0.2 & 1.1605 & 0.1691 & 0.0541 & 0.3708 \\
& 1   & 5.2028 & 0.0436 & 0.0033 & 0.3915 \\
& 10  & 53.7372 & 0.0044 & 0.0000 & 0.3989 \\
\hline
\multirow{3}{*}{$-1$}
& 0.2 & 6.0247 & 0.0265 & 0.0015 & 0.3338 \\
& 1   & 7.6844 & 0.0230 & 0.0011 & 0.3544 \\
& 10  & 52.0277 & 0.0044 & 0.0000 & 0.3915 \\
\hline
\end{tabular}
\caption{Critical quantities for the NLMC--AdS with PFDM black hole $(N=0)$.}
\label{crit_quan_N0}
\end{minipage}
\hfill
\begin{minipage}{0.45\textwidth}
\centering
\begin{tabular}{c c c c c c}
\hline
$N$ & $\lambda$ & $v_c$ & $T_c$ & $P_c$ & $\rho_c$ \\
\hline
\multirow{3}{*}{0.2}
 & -0.01 & 5.0702 & 0.0159 & 0.0038 & 1.2089 \\
 & -0.05 & 4.9363 & 0.0161 & 0.0039 & 1.1864 \\
 & -0.1  & 4.8470 & 0.0158 & 0.0039 & 1.1920 \\
\hline
\multirow{3}{*}{0.3}
 & -0.01 & 4.9187 & 0.0020 & 0.0041 & 9.8999 \\
 & -0.05 & 4.7709 & 0.0025 & 0.0042 & 8.2111 \\
 & -0.1  & 4.6673 & 0.0022 & 0.0043 & 8.8801 \\
\hline
\multirow{3}{*}{0.5}
 & -0.01 & 4.6178 & -- & 0.0048 & -- \\
 & -0.05 & 4.4404 & -- & 0.0051 & -- \\
 & -0.1  & 4.3059 & -- & 0.0052 & -- \\
\hline
\end{tabular}
\caption{Critical quantities for the full black hole configuration in the quintessence regime ($w=-2/3$) with $Q_m=1$. Dashed entries indicate negative (non-physical) critical quantities.}
\label{tab:critical_values_wminus2/3}
\end{minipage}

\vspace{0.6cm}

% ----------- bottom -----------

\begin{minipage}{0.7\textwidth}
\centering
\begin{tabular}{c c c c c c}
\hline
$N$ & $\lambda$ & $v_c$ & $T_c$ & $P_c$ & $\rho_c$ \\
\hline
\multirow{3}{*}{0.2}
 & -0.01 & 4.4841 & 0.0205 & 0.1218 & 26.5978 \\
 & -0.05 & 4.3949 & 0.0215 & 0.1185 & 24.1917 \\
 & -0.1  & 4.3247 & 0.0196 & 0.1154 & 25.4576 \\
\hline
\multirow{3}{*}{0.3}
 & -0.01 & 4.2623 & 0.0050 & 0.1671 & 141.4085 \\
 & -0.05 & 4.1697 & 0.0018 & 0.1609 & 380.9005 \\
 & -0.1  & 4.0856 & --     & 0.1534 & --       \\
\hline
\multirow{3}{*}{0.5}
 & -0.01 & 3.9660 & -- & 0.2394 & -- \\
 & -0.05 & 3.8730 & -- & 0.2293 & -- \\
 & -0.1  & 3.7967 & -- & 0.2194 & -- \\
\hline
\end{tabular}
\caption{Critical quantities for the full black hole configuration in the phantom regime ($w=-3/2$) with $Q_m=1$.}

\label{tab:critical_values_wminus15}
\end{minipage}

\end{table} Furthermore, For the full equation of state in Eq.~\eqref{PV1}, the criticality conditions were solved numerically for different parameter values. The resulting critical quantities, for representative choices of the parameters, are summarized in Tables~\ref{crit_quan_N0}--\ref{tab:critical_values_wminus15}. The analysis confirms that the black hole admits at most one physically admissible critical point. This result consistently indicates that the system exhibits the standard vdW--type phase structure, characterized by a single SBH/LBH phase transition. For the NLMC--AdS with PFDM black hole case (Table~\ref{crit_quan_N0}), the critical ratio remains close to the analytical value $\rho_c^{(NLMC)}=2/5$, exhibiting only a weak dependence on $\lambda$ and $Q_m$. As illustrated in Figs.~\ref{ratiolamQ}(a) and (b), the ratio is primarily influenced by the PFDM parameter, showing a clear decrease as $\lambda$ becomes more negative. In contrast, its dependence on the magnetic charge is comparatively weak; as $Q_m$ increases, the ratio rapidly approaches the value $2/5$, characteristic of the non-linear magnetic sector. In contrast, for the full black hole configuration $(N\neq 0)$ (Tables~\ref{tab:critical_values_wminus2/3}--\ref{tab:critical_values_wminus15}), significant deviations from this quasi-universal behavior emerge. In these cases, the critical ratio becomes highly sensitive to both $N$ and $\lambda$, and may even diverge in certain parameter regimes. Notably, in the phantom case (Table~\ref{tab:critical_values_wminus15}), one observes large values of $\rho_c$. Overall, while the NLMC--AdS with PFDM black hole solution  preserves a quasi-universal thermodynamic structure, the inclusion of a dark energy field tends to suppress or even destroy criticality. Additionally, in Fig.~\ref{ratiolamQ}(d), we construct the phase diagram in the $(N,w)$ parameter space. The resulting structure identifies the regions where vdW--like phase transitions are allowed. In general, in the phantom regime, criticality is strongly suppressed for the full black hole configuration, and the system effectively behaves as a single-phase thermodynamic system. Nevertheless, spinodal points may still arise, indicating changes in local thermodynamic stability, as shown in Fig.~\ref{fig:heat capacity}(b). In contrast, in the quintessence regime, vdW--type phase transitions can occur, particularly for sufficiently small values of $N\ll1$. As the intensity of the quintessence field increases, however, the critical behavior of the full configuration gradually weakens and eventually disappears.

It is important to emphasize that this suppression is not a generic feature of the dark-energy sector. Criticality and phase transitions can persist in the Schwarzschild--AdS with dark energy and all limiting geometries, as illustrated in Fig.~\ref{fig:PV_profiles}(b)--(d). The suppression observed here arises specifically in the full configuration, where the magnetic, PFDM, and dark-energy contributions interact. A similar effect is found for the PFDM contribution, indicating that the suppression of criticality is a feature of the combined system rather than of any individual sector.

\subsection{Local thermodynamic stability}
In classical thermodynamics, the local stability properties and the phase transition structure of a system are encoded in the behavior of the generalized heat capacities~\cite{romero2024extended1}. In a general setting, the Nambu bracket formalism (see Appendix of~\cite{Ladino:2024ned}) offers a coordinate-independent representation of the heat capacity evaluated at fixed thermodynamic parameters $(x^1,x^2,\ldots,x^{n-1})$ within the equilibrium state space $(y^1,y^2,\ldots,y^{n})$. In this formulation, the heat capacity takes the form
\begin{align}
C_{x^1,\ldots,x^{n-1}}
=
T\left(\frac{\partial S}{\partial T}\right)_{x^1,\ldots,x^{n-1}}
=
T\,
\frac{\{S,x^1,\ldots,x^{n-1}\}_{y^1,\ldots,y^{n}}}
{\{T,x^1,\ldots,x^{n-1}\}_{y^1,\ldots,y^{n}}}.
\label{nambu1}
\end{align}
In the thermodynamic
representation determined by the first law Eq.~\eqref{first law},
the relevant response function for analyzing the local thermodynamic
stability is the heat capacity evaluated at fixed $X \equiv \{P,Q_m,\lambda,N\}$.
It can be written as
\begin{align}
C_{X}
&=T\left(\frac{\partial S}{\partial T}\right)_{X}
=\frac{\mathcal{N_C}}{\mathcal{D_C}}.
\label{heat capacityx}
\end{align}
where the numerator and denominator are
\begin{align}
\mathcal{N_C} &= 
-2 S \Bigg[
-2 \pi^{3/2} Q_m^3 \sqrt{S} + S^2 (1 + 8 P S)
+ 3 N \pi^{\frac{1}{2} + \frac{3w}{2}} S^{-\frac{3}{2}(-1 + w)} w
+ 3 N \pi^{2 + \frac{3w}{2}} Q_m^3 S^{-\frac{1}{2} - \frac{3w}{2}} \sqrt{S} (1 + w)
\\&+ \pi^2 Q_m^3 \lambda
+ \sqrt{\pi} S^{3/2} \lambda
\Bigg]+ 3 \pi^2 Q_m^3 S \lambda \nonumber\ln\left(\frac{S}{\pi \lambda^2}\right),
\\[1em]
\mathcal{D_C} &= 
-8 \pi^{3/2} Q_m^{3} \sqrt{S}
+ S^{2}
- 8 P S^{3}
+ 3 wN  (2 + 3w)\, S^{\frac{3(1-w)}{2}} \left(\frac{\pi}{S}\right)^{\frac{1+3w}{2}} \label{DC} \\
&\qquad
+ 3 N \pi^{3/2} Q_m^{3} (5 + 8w + 3w^{2}) \left(\frac{\pi}{S}\right)^{\frac{1+3w}{2}} \sqrt{S}
+ 8 \pi^{2} Q_m^{3} \lambda
+ 2 \sqrt{\pi} S^{3/2} \lambda - 15 \pi^{2} Q_m^{3} \lambda 
\ln\!\left(\frac{\sqrt{S}}{\sqrt{\pi}\,|\lambda|}\right)\nonumber.
\end{align}

\begin{figure}[H]
\centering
\begin{minipage}{0.48\textwidth}
\centering
\includegraphics[width=\linewidth]{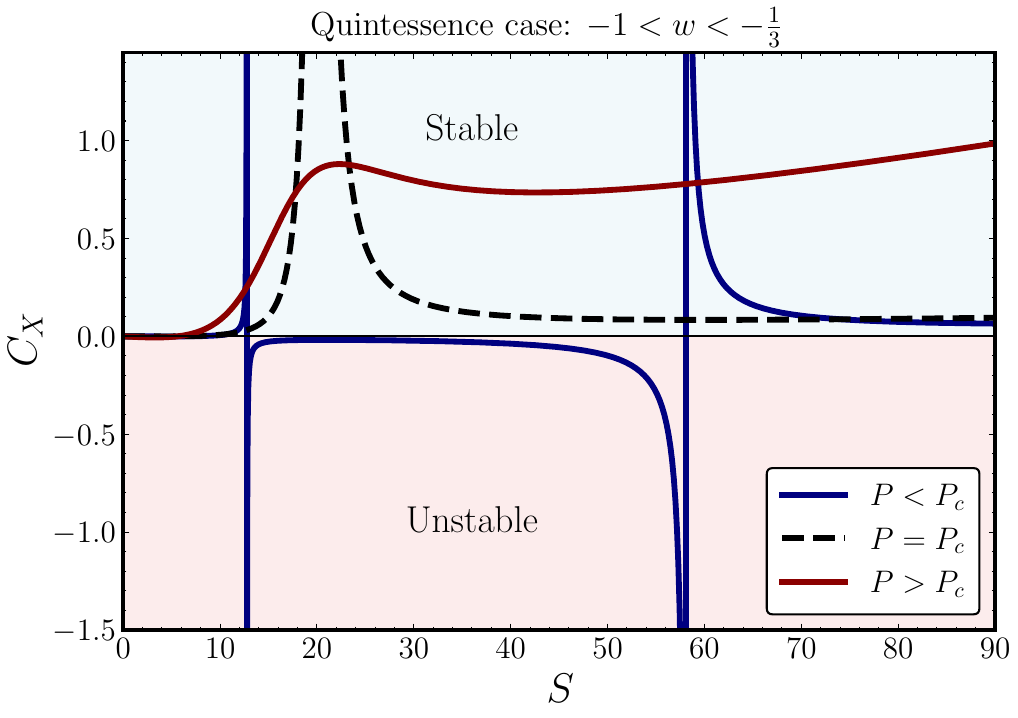}\\
(a)
\end{minipage}
\hfill
\begin{minipage}{0.48\textwidth}
\centering
\includegraphics[width=\linewidth]{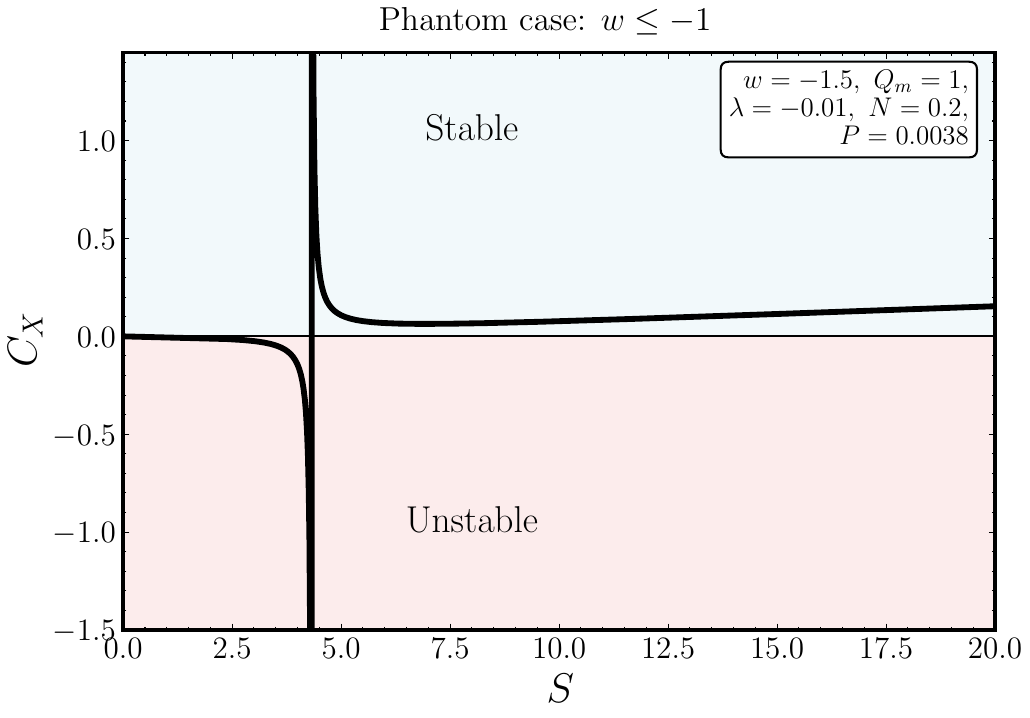}\\
(b)
\end{minipage}
\caption{
Heat capacity of the full black hole solution evaluated at $Q_m=1$, $\lambda=-0.01$, and $N=0.2$. 
(a) Quintessence regime with $w=-2/3$. The dashed curve represents $C_X$ at the critical pressure $P_c \simeq 0.0038$, while the solid curves correspond to $P<P_c$ and $P>P_c$. 
(b) Phantom regime with $w=-1.5$, where a change in the thermodynamic stability is observed.
}
\label{fig:heat capacity}
\end{figure}
As shown in Fig.~\ref{fig:heat capacity}(a), the heat capacity of the black hole coupled to a quintessence fluid exhibits the typical behavior of a thermodynamic system with a single critical point. For $P<P_c$, two divergences appear, separating the solution into three branches. The intermediate branch is thermodynamically unstable, characterized by $C_X<0$, while the small and large black hole phases correspond to regions with positive heat capacity. This structure signals a first-order phase transition between the small and large black hole configurations. 

As the pressure increases to the critical value $P=P_c$, the two singularities merge into a single divergence, indicating the presence of a second-order phase transition. For $P>P_c$, the heat capacity becomes smooth and finite for all values of the entropy, showing that the phase transition disappears and the system enters a supercritical regime. 
\begin{figure}[ht!]
\centering
\begin{minipage}{0.48\textwidth}
    \centering
    \includegraphics[width=\linewidth]{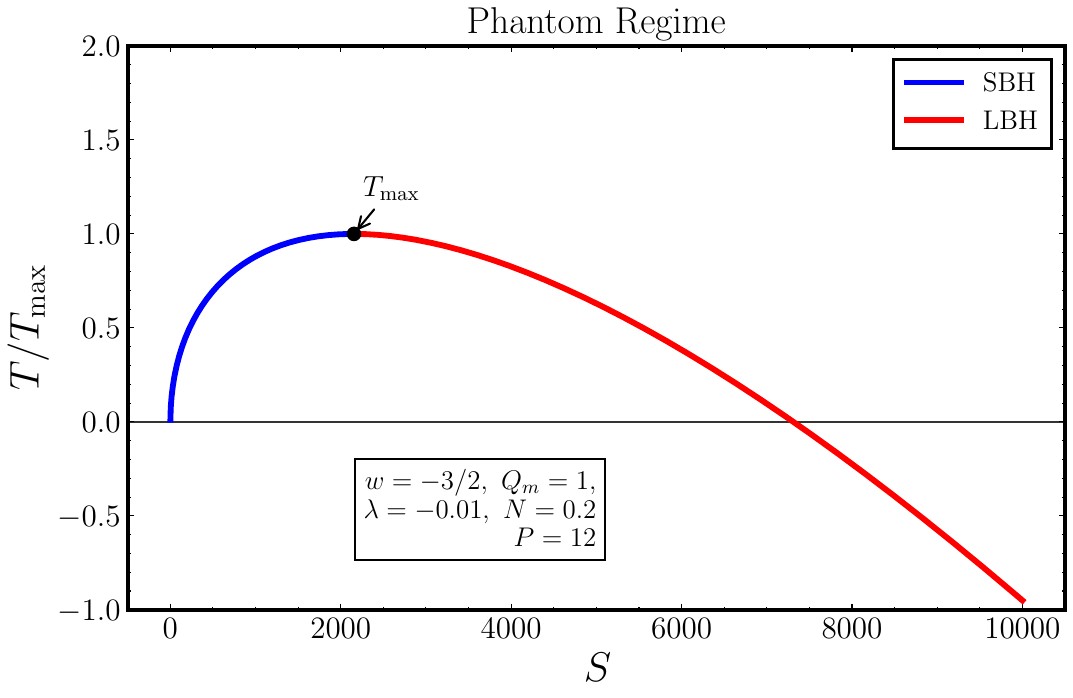}\\
\end{minipage}
\hfill
\caption{
Temperature of the full black hole solution in the phantom regime with $w=-3/2$.
}
\label{fig:T phant}
\end{figure}
Conversely, Fig.~\ref{fig:heat capacity}(b) and Fig.~\ref{fig:T phant} illustrates the behavior of the black hole in the phantom regime. Although divergences in the heat capacity may still occur, they signal changes in local thermodynamic stability rather than genuine critical phenomena. For the full solution, the phantom regime exhibits a thermodynamic structure qualitatively different from the usual vdW behavior. The IBH (intermediate black hole) branch is absent, leaving only the SBH and LBH branches, which join at the spinodal point. This point separates stable and unstable configurations but does not represent phase coexistence; consequently, no swallow-tail structure or first-order phase transition emerges \cite{ladino2026probing}.
\subsection{Global phase structure}
\label{gibbsF}
To characterize the global thermodynamic stability and phase structure, we analyze the Gibbs free energy in the extended phase space, where the black hole mass is interpreted as gravitational enthalpy \cite{kastor2009enthalpy}. It is defined as
\begin{equation}
G\equiv M-T\,S,
\label{Gdef}
\end{equation}
where the mass $M(S,P,Q_m,\lambda)$ is given by
Eq.~\eqref{funda}, and the temperature $T(S,P,Q_m,\lambda)$
is obtained from Eq.~\eqref{eq:Tthermo}.
Substituting these expressions into Eq.~\eqref{Gdef},
we obtain the Gibbs free energy as a function of the entropy
and thermodynamic parameters
\begin{equation}
\begin{aligned}
G=\frac{1}{24 \sqrt{\pi}} \Bigg\{
&32 P \pi^{3/2} Q_m^3
+ \frac{24 \pi^{3/2} Q_m^3}{S}
+ 6 \sqrt{S}
- 16 P S^{3/2}
\\
&- 6 N \pi^{\frac{1}{2}+\frac{3w}{2}}
S^{-\frac{3(1+w)}{2}}
\left[
S^{3/2}(2+3w)
+\pi^{3/2}Q_m^3(5+3w)
\right]
\\
&- 6 \sqrt{\pi}\,\lambda
- \frac{6 \pi^2 Q_m^3 \lambda}{S^{3/2}}
+ \frac{9 \pi^2 Q_m^3 \lambda
\ln\!\left(\frac{S}{\pi \lambda^2}\right)}{S^{3/2}}
\\
&+ \frac{12\left(
\pi^2 Q_m^3 \lambda
+\sqrt{\pi}S^{3/2}\lambda
\right)}
{S^{3/2}}
\ln\!\left(
\frac{\sqrt{S}}
{\sqrt{\pi}\,|\lambda|}
\right)
\Bigg\}.
\end{aligned}
\end{equation}
Since the relation $T(S,P,Q_m,\lambda,N)$ cannot be inverted analytically,
the thermodynamic branches are determined numerically. 
\begin{figure}[ht!]
\centering
\begin{minipage}{0.48\textwidth}
    \centering
    \includegraphics[width=\linewidth]{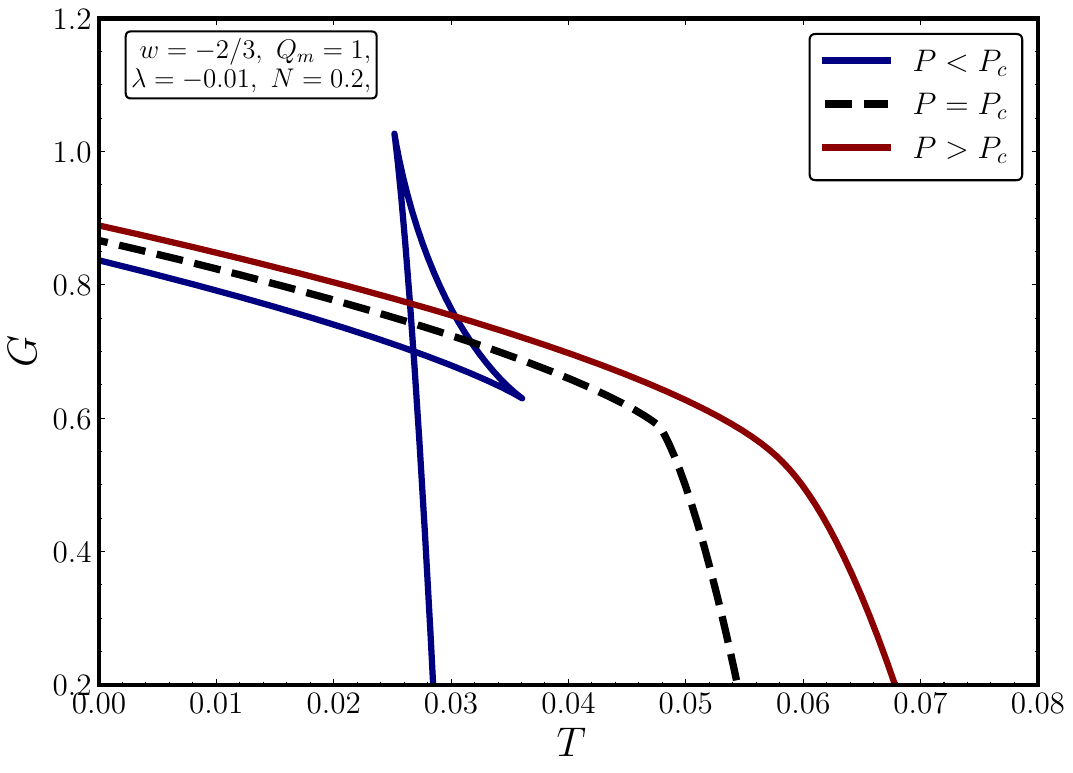}\\ 
(a)
\end{minipage}
\hfill
\begin{minipage}{0.48\textwidth}
    \centering
    \includegraphics[width=\linewidth]{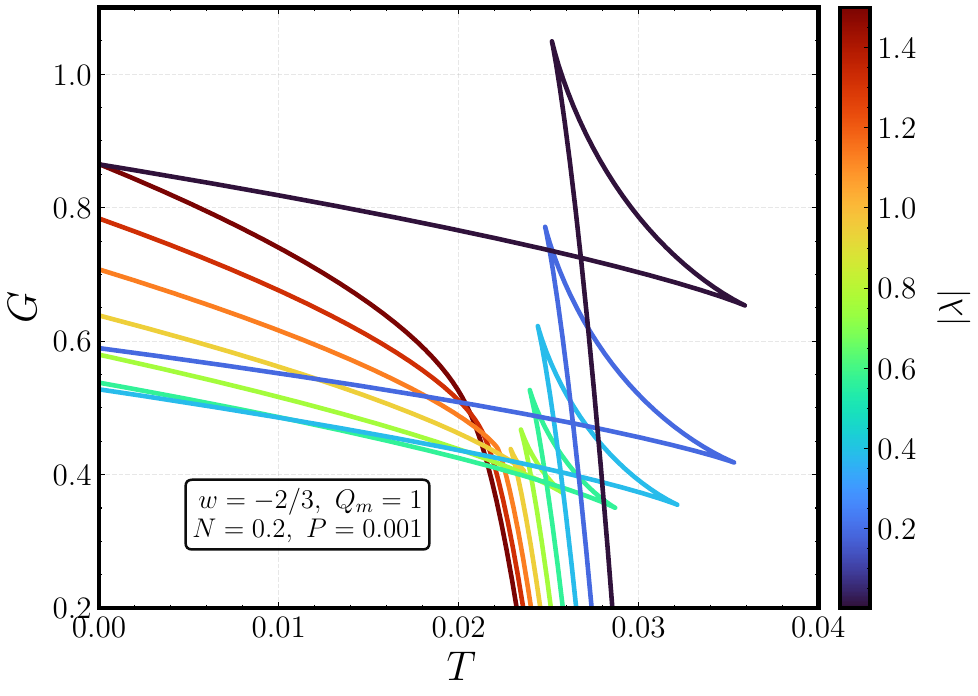}\\
(b)
\end{minipage}
\hfill\\ \vspace{0.3in}
\begin{minipage}{0.48\textwidth}
    \centering
    \includegraphics[width=\linewidth]{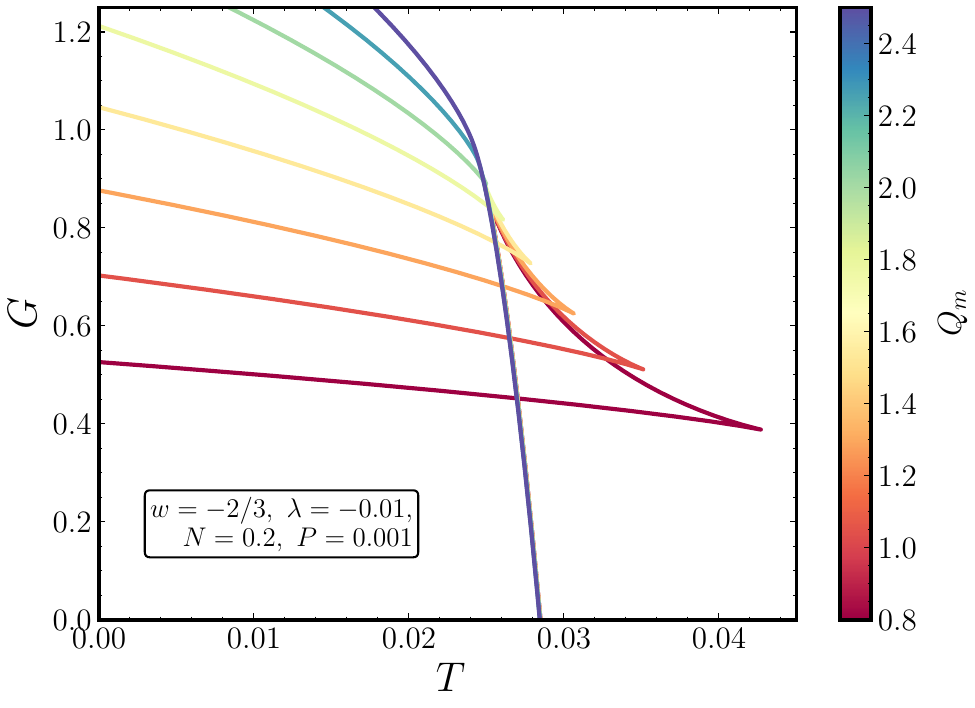}\\
(c)
\end{minipage}
\hfill
\begin{minipage}{0.48\textwidth}
    \centering
    \includegraphics[width=\linewidth]{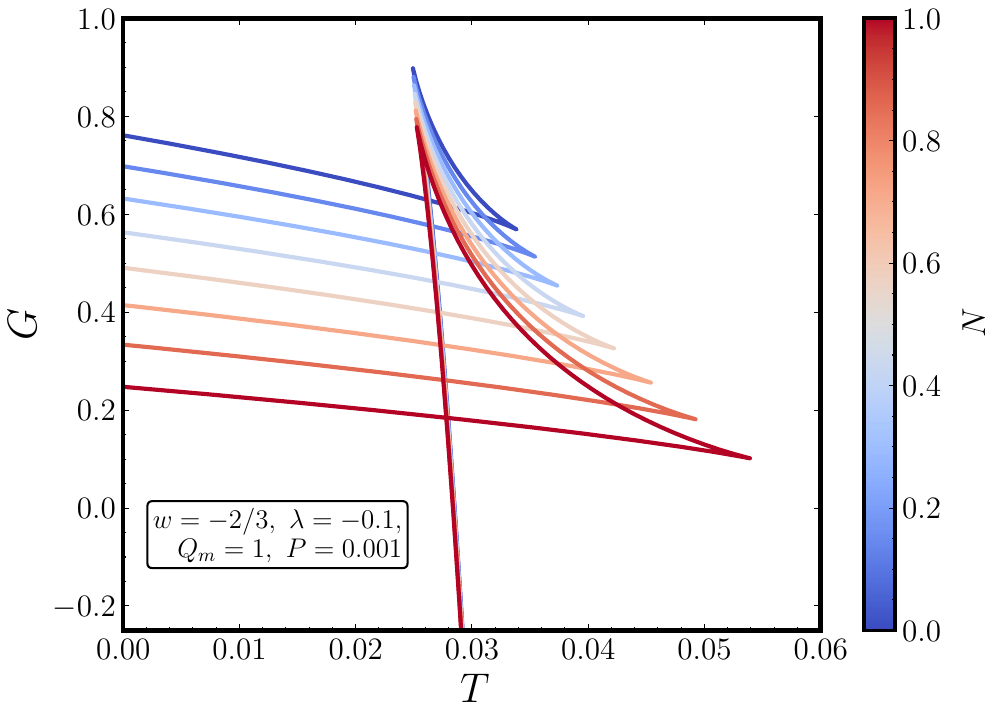}\\
(d)
\end{minipage}
\hfill
\caption{
Gibbs free energy for the full black hole solution.
(a) $G$--$T$ diagram showing the critical behavior at $P_c\simeq0.0038$.
(b) Profiles for different values of $\lambda$ with fixed $Q_m$, $N$, $w$, and $P<P_c$.
(c) Profiles for different values of $Q_m$ with fixed $\lambda$, $N$, $w$, and $P<P_c$.
(d) Profiles for different values of $N$ with fixed $\lambda$, $Q_m$, $w$, and $P<P_c$.}
\label{fig:free_energy}
\end{figure}
\begin{figure}[ht!]
\centering

\begin{minipage}[t]{0.48\linewidth}
    \centering
    \includegraphics[width=\linewidth]{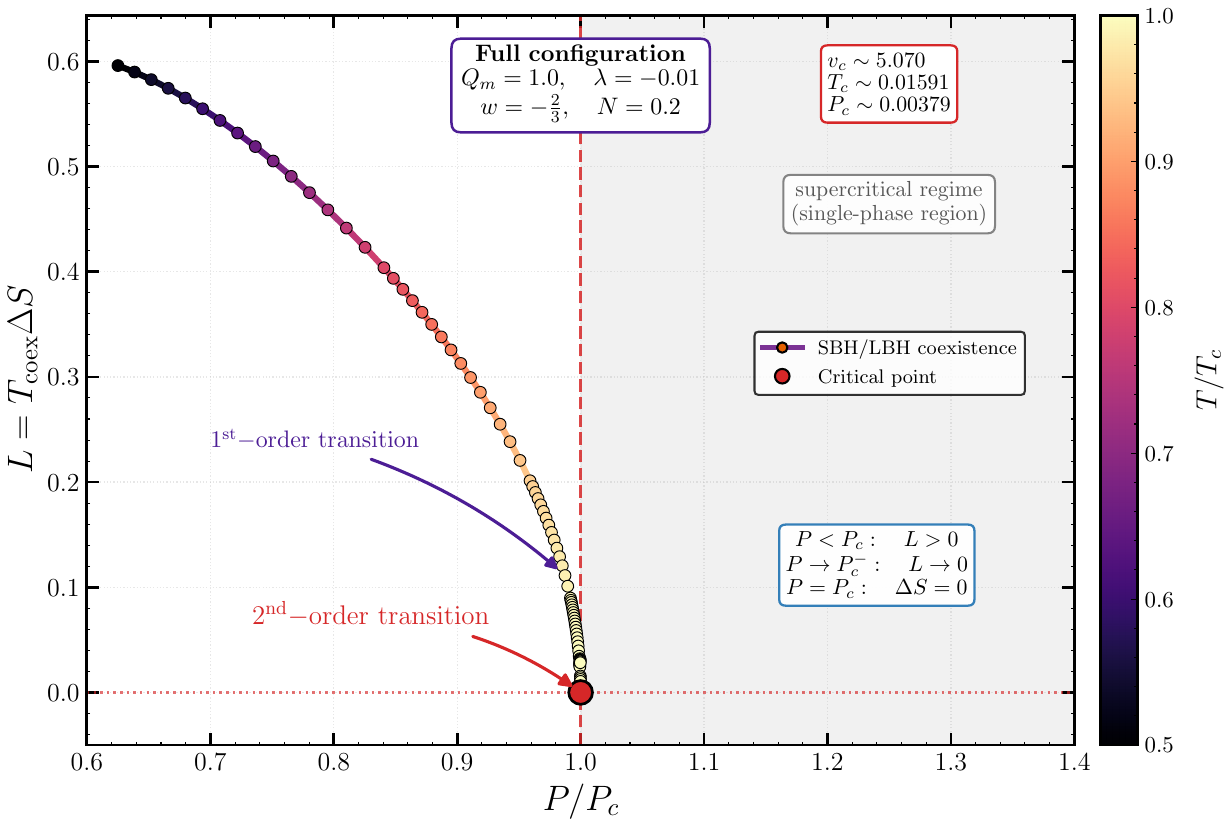}\\
    (a)
\end{minipage}
\hfill
\begin{minipage}[t]{0.48\linewidth}
    \centering
    \includegraphics[width=\linewidth]{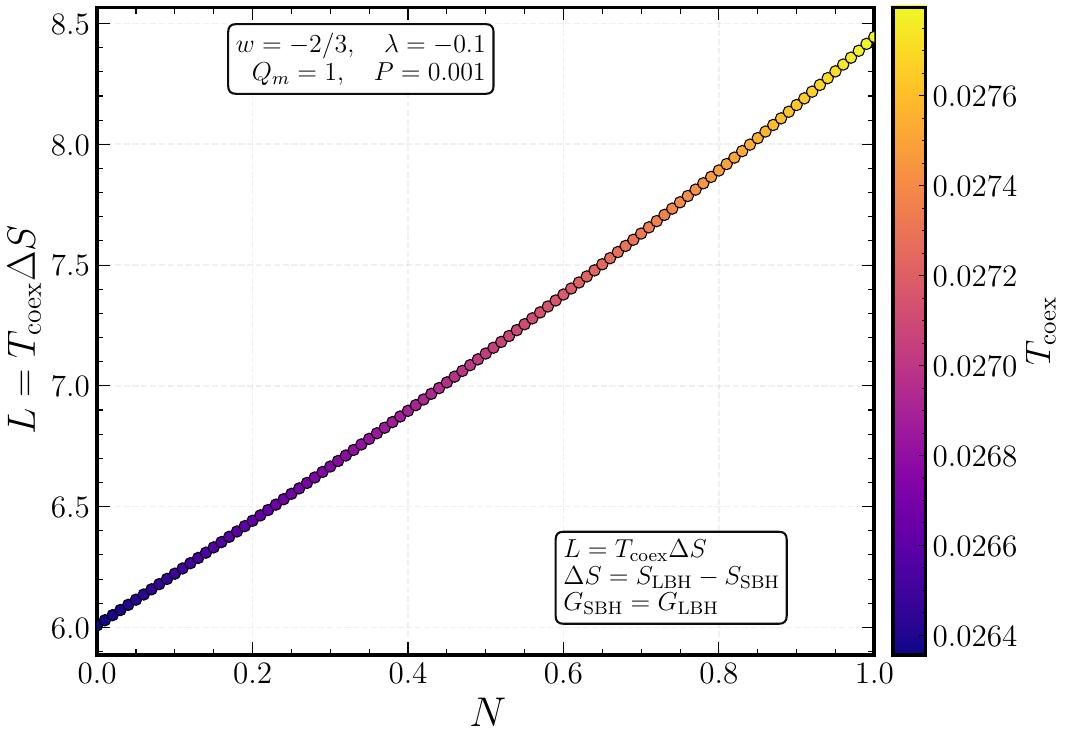}\\
    (b)
\end{minipage}
\hfill\\[0.3in]

\begin{minipage}[t]{0.48\linewidth}
    \centering
    \includegraphics[width=\linewidth]{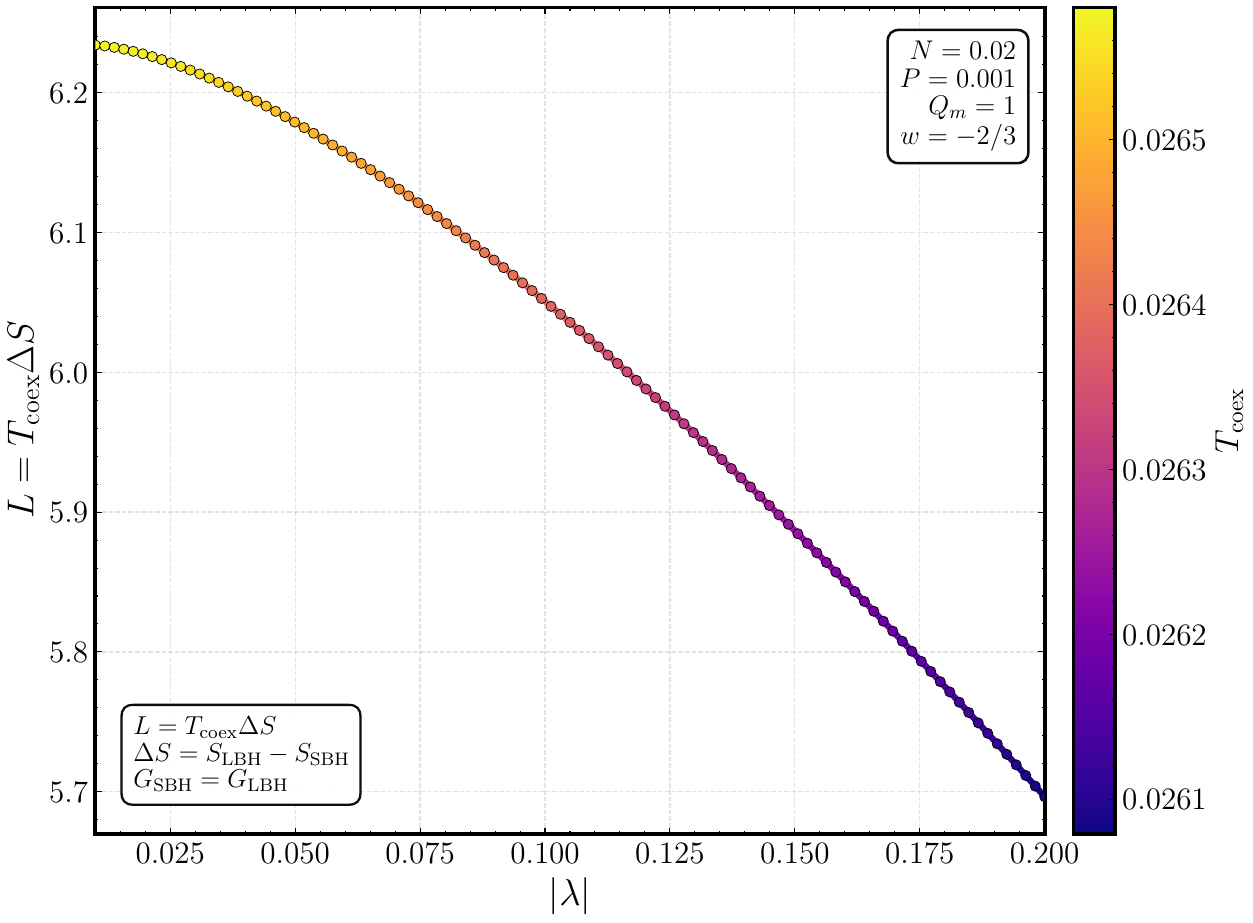}\\
    (c)
\end{minipage}

\caption{Latent heat along the coexistence curve as a function of (a) reduced pressure $P/P_c$, (b) the dark-energy parameter $N$, and (c) the magnitude of the PFDM parameter $|\lambda|$.}

\label{latent_heat_analysis}
\end{figure}
The $G$--$T$ diagram in Fig.~\ref{fig:free_energy}(a) exhibits the typical vdW behavior of AdS black holes \cite{Ladino:2024ned}. For $P<P_c$, three thermodynamic branches, SBH, IBH, and LBH, form a swallow-tail structure, signaling a coexistence curve and a first-order transition between the SBH and LBH phases, as shown in Fig.~\ref{latent_heat_analysis}(a).
At $P=P_c$, the swallow-tail collapses to a single point and the phase
transition becomes second order, whereas for $P>P_c$ the system enters a
supercritical regime with a single stable phase. Figs.~\ref{fig:free_energy}(b)--(d) illustrate the effects of varying the thermodynamic parameters on the free energy. In particular, Fig.~\ref{fig:free_energy}(b) shows that, for the full solution, increasing $|\lambda|$ progressively suppresses the first-order phase transition, leading the system toward single-phase regime. The reduced separation between the thermodynamic branches and the eventual disappearance of the swallow-tail structure reflect a decrease in the entropy difference between the coexisting phases, quantified by the latent heat, $L=T_{coex}\Delta S$. The latent heat vanishes at criticality, where the two phases become indistinguishable; Fig.~\ref{latent_heat_analysis}(a) illustrates this behavior. Moreover, our numerical analysis along the coexistence curve shows that increasing $|\lambda|$ reduces $L$, thereby suppressing the first-order phase transition, as depicted in Fig.~\ref{latent_heat_analysis}(c). This effect arises from the modification of the equation of state by PFDM and its interplay with the other thermodynamic parameters.

These results corroborate our previous findings \cite{ahmed2026shadow}, where we demonstrated that the PFDM fluid does not give rise to novel phase structure\footnote{This conclusion contrasts with that reported in \cite{ndongmo2023thermodynamics}, where the authors claim that for the full solution, the PFDM sector  enhances the phase transition. The origin of this discrepancy is that their analysis considers non-physical values of $\lambda > 0$, which lead to negative energy densities according to Eq.~\eqref{energy-density}.}. Furthermore, the effect of the dark energy field is different. Fig.~\ref{fig:free_energy}(d) reveals that, in the full solution within the quintessence regime, increasing $N$ enhances the swallow-tail structure and increases the latent heat, see Fig.~\ref{latent_heat_analysis}(b). Nevertheless, increasing $N$ narrows the parameter range in which criticality can occur, as discussed in Sec.~\ref{PV section}, with the corresponding behavior summarized in Fig.~\ref{ratiolamQ}(d) and Table~\ref{tab:critical_values_wminus2/3}.
\begin{figure}[ht!]
\centering
\begin{minipage}{0.48\textwidth}
    \centering
    \includegraphics[width=\linewidth]{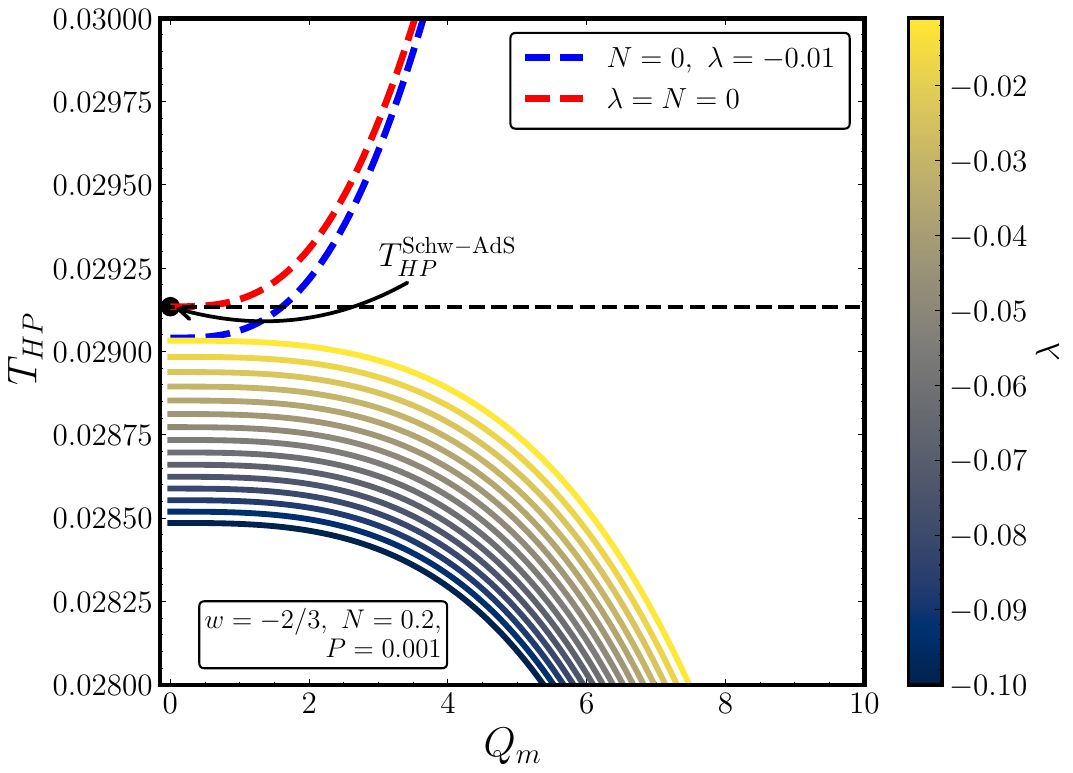}\\
(a)
\end{minipage}
\hfill
\begin{minipage}{0.48\textwidth}
    \centering
    \includegraphics[width=\linewidth]{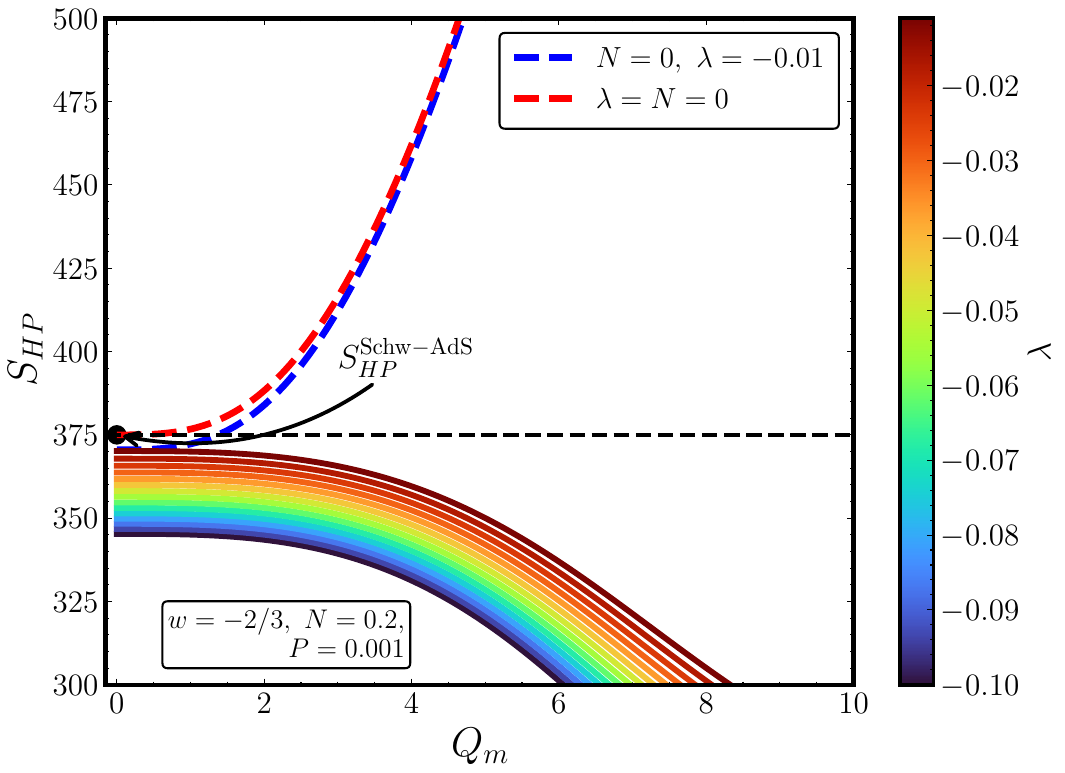}\\
(b)
\end{minipage}
\hfill
\caption{(a) Hawking--Page temperature $T_{HP}$ as a function of the magnetic charge $Q_m$ for different values of the PFDM parameter $\lambda$. (b) Hawking--Page entropy $S_{HP}$ as a function of $Q_m$. 
}
\label{fig:THP}
\end{figure} 
\begin{table}[ht!]
\centering
\begin{tabular}{l@{\hspace{1.2cm}}c@{\hspace{1.2cm}}c@{\hspace{1.2cm}}c}
\hline
Dark-sector parameter & Swallow-tail & Latent heat & Criticality \\
\hline
$|\lambda| \uparrow$ (PFDM)& $\downarrow$ & $\downarrow$ & $\downarrow$ \\
$N \uparrow$ (quintessence) & $\uparrow$ & $\uparrow$ & $\downarrow$ \\
\hline
\end{tabular}
\caption{Summary of the dark-sector effects on the phase structure at fixed $Q_m$.}
\label{tab:dark_sector_effects}
\end{table}

In both the PFDM and dark-energy sectors, these results describe effective macroscopic thermodynamic behavior, while the underlying microscopic mechanisms are not determined within the present framework. The main effects of the dark-sector parameters on the global phase structure are summarized in Table~\ref{tab:dark_sector_effects}. Finally, The Hawking--Page transition can be identified from the Gibbs free
energy \cite{aa19}. The Hawking--Page transition occurs when the Gibbs free energy
vanishes, $G=0$, which corresponds to the point where the thermal AdS
background and the Large black hole phase have the same free energy. For $G>0$ the thermal AdS phase dominates, while for $G<0$ the black
hole configuration becomes thermodynamically preferred. Therefore,
the corresponding temperature defines the Hawking--Page temperature
$T_{HP}$ separating the thermal AdS phase from the stable black hole
phase. The HP transition is sensitive to modifications of the gravitational background. For example,  non-relativistic geometries with anisotropic Lifshitz scaling modify the $T_{HP}$ through a dynamical exponent \cite{herrera2023anisotropic,tarrio2011black}. Similar deviations also appear in higher-dimensional and modified gravity theories \cite{ahmed2026shadow}.  
Fig.~\ref{fig:THP}(a) illustrates the numerical behavior of the Hawking--Page temperature $T_{HP}$ as a function of the magnetic charge $Q_m$, for fixed pressure and several values of the PFDM parameter $\lambda$. The Schwarzschild--AdS limit \cite{Ladino:2024ned} is given by 
$T_{HP}^{\text{Schw-AdS}}=\sqrt{8P/3\pi}$, shown as a horizontal black dashed line, and is recovered in the limit $Q_m=0$ and $\lambda=N=0$. 

The results indicate that increasing both $|\lambda|$ and $Q_m$ shifts the Hawking--Page transition toward lower temperatures. In contrast, when the quintessence field is switched off ($N=0$), the dependence on the magnetic charge is reversed, with larger values of $Q_m$ leading to higher Hawking--Page temperatures. Fig.~\ref{fig:THP}(b) exhibits the same qualitative behavior for the Hawking--Page entropy $S_{HP}$.

\section{Geometrothermodynamics and Black Hole Microstructure}\label{sec4}

In this section, we employ the formalism of geometrothermodynamics (GTD) \cite{quevedo2007geometrothermodynamics} to analyze the thermodynamic microstructure of the black hole solution. GTD provides a Legendre-invariant geometric framework to study thermodynamic systems. In this approach, thermodynamic interactions are encoded in the curvature of the equilibrium manifold, allowing phase transitions to be identified through geometric singularities \cite{quevedo2023unified}. To establish the geometric framework employed in the present work, we
briefly summarize the main ingredients of GTD. Legendre invariance is
implemented by introducing an auxiliary $(2n+1)$-dimensional manifold
$\mathcal{T}$, coordinatized by $Z^A=\{\Phi,E^a,I_a\}$, where $n$ is the
number of thermodynamic degrees of freedom and $\Phi$ denotes the
thermodynamic potential. While $E^a$ and
$I_a$ naturally correspond to extensive and intensive variables in
homogeneous systems, this identification becomes subtler for
quasi-homogeneous systems such as black holes
\cite{quevedo2023unified,quevedo2019quasi}. The space $\mathcal{T}$ is endowed with both a local canonical 1-form \(\Theta_G = d\Phi - I_a \, dE^a\), which satisfies the condition of being non-maximally integrable, defining a contact structure,
and a Riemannian metric $G=G_{AB}dZ^A d Z^B$ where $A,B=0,\ldots,2n$. The triplet \((\mathcal{T}, \Theta_G, G)\) defines a Riemannian contact manifold and is referred to as the thermodynamic phase space in the GTD formalism. Currently, there exist three Legendre-invariant metrics on \(\mathcal{T}\), which are given by \cite{quevedo2023unified}
\begin{equation}
    G^{I/II}=\left(d\Phi-I_adE^a\right)^2+(\xi_{ab}E^aI^b)(\chi_{cd}dE^cdI^d), \label{metrics phase1}
\end{equation}
\begin{equation}
    G^{III}=\left(d\Phi-I_adE^a\right)^2+\sum_{a=1}^{n}\xi_{a}(E_aI_a)^{2k+1}dE^adI^a. \label{metric phase2}
\end{equation}
Here, \(\xi_a\) are \(n\) real constants, \(\xi_{ab}\) is a real diagonal \(n \times n\) matrix, and \(k\) is an integer. Moreover, the matrix \(\chi_{cd}\) is defined as \(\chi_{cd} = \delta_{cd} = \text{diag}(1, 1, \ldots, 1)\) for the metric \(G^I\), and as \(\chi_{cd} = \eta_{cd} = \text{diag}(-1, 1, \ldots, 1)\) for \(G^{II}\). In GTD, thermodynamic states are represented as points in an \( n \)-dimensional subspace of \( \mathcal{T} \), known as the equilibrium space \( \mathcal{E} \). This space is defined by a smooth mapping \( \varphi: \mathcal{E} \rightarrow \mathcal{T} \), for which the condition \( \varphi^*(\Theta_G) = 0 \) holds, where $\varphi^\ast$ represents the pullback. As a consequence, the first law of thermodynamics is naturally satisfied on \( \mathcal{E} \), and the coordinates \( Z^A \) become functions of the variables \( E^a \), that is $ Z^A(E^a) = \{ \Phi(E^a), E^a, I_a(E^a) \}$, where \( \Phi = \Phi(E^a) \)  represents the fundamental equation of the thermodynamic system and $I_a=\partial \Phi/\partial E^a $ the dual variables. Additionally, the line element $G = G_{AB}dZ^AdZ^B$ on $\cal{T}$ induces a line element $g = g_{ab}dE^adE^b$ on $\cal{E}$ by means of the pullback, i.e., $\varphi^\star(G) = g$. Then, from Eqs. (\ref{metrics phase1}) -- \eqref{metric phase2}, we obtain
\begin{align}
g^I &= \sum_{a,b,c=1}^{n} \left( \nu_c E^c \frac{\partial \Phi}{\partial E^c} \right) \frac{\partial^2 \Phi}{\partial E^a \partial E^b} \, dE^a \, dE^b,\label{g111} \\
g^{II} &= \sum_{a,b,c,d=1}^{n} \left( \nu_c E^c \frac{\partial \Phi}{\partial E^c} \right) \eta^d_{\;a} \frac{\partial^2 \Phi}{\partial E^b \partial E^d} \, dE^a \, dE^b, \label{g222}\\
g^{III}&=\sum_{a=1}^{n}\nu_a\left(\delta_{ad}E^d\frac{\partial \Phi}{\partial E^a}\right)^{2k+1}\delta^{ab}\frac{\partial^2 \Phi}{\partial E^b \partial E^c} dE^a dE^c. \label{g333}
\end{align}
The geometry of $\mathcal{E}$ encodes the thermodynamic properties of
the system, with its curvature providing a geometric measure of the underlying thermodynamic interactions. The GTD metrics $g^{I}$, $g^{II}$, and $g^{III}$ defined in Eqs.~(\ref{g111})--(\ref{g333}) give rise to the corresponding Ricci curvature scalars $\mathcal{R}^{I}$, $\mathcal{R}^{II}$, and $\mathcal{R}^{III}$, respectively. In particular, curvature
singularities are generally associated with phase
transitions, and thermodynamic instabilities \cite{quevedo2007geometrothermodynamics,quevedo2023unified}.\begin{figure}[ht!]
\centering
\begin{minipage}{0.48\textwidth}
    \centering
    \includegraphics[width=\linewidth]{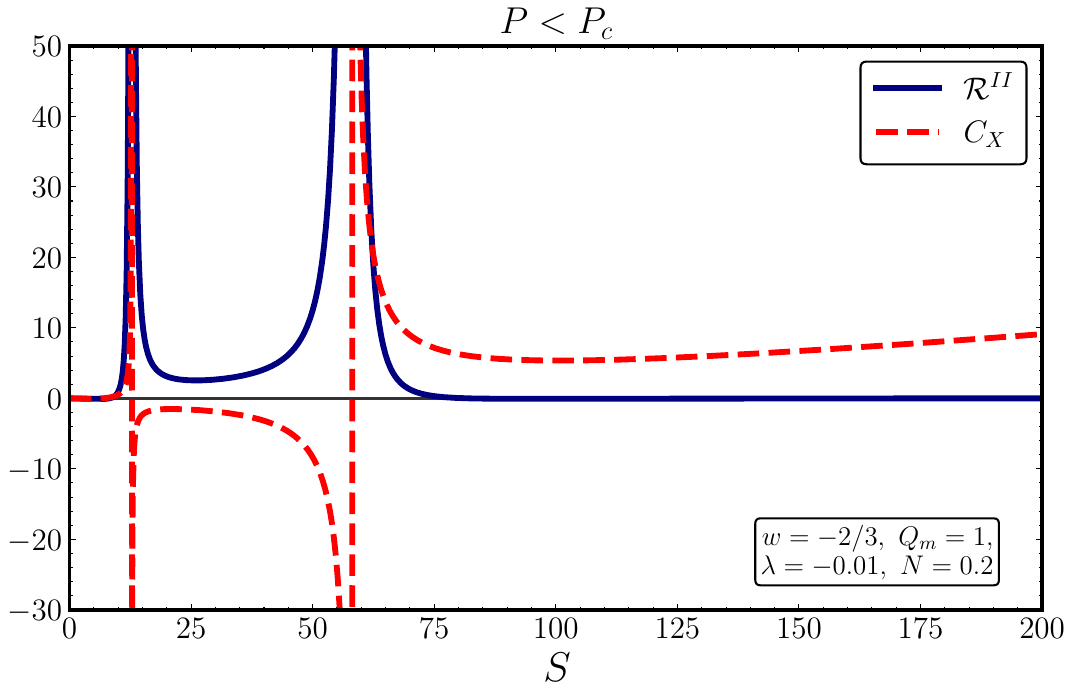}\\
(a)
\end{minipage}
\hfill
\begin{minipage}{0.48\textwidth}
    \centering
    \includegraphics[width=\linewidth]{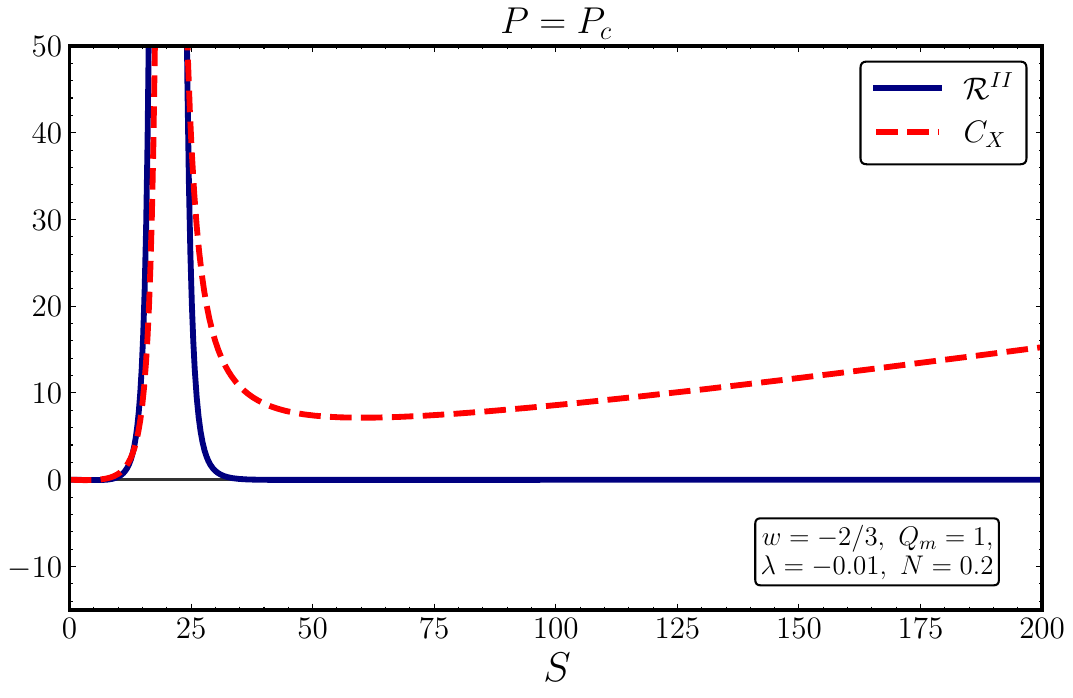}\\
(b)
\end{minipage}
\hfill\\ \vspace{0.3in}
\begin{minipage}{0.48\textwidth}
    \centering
    \includegraphics[width=\linewidth]{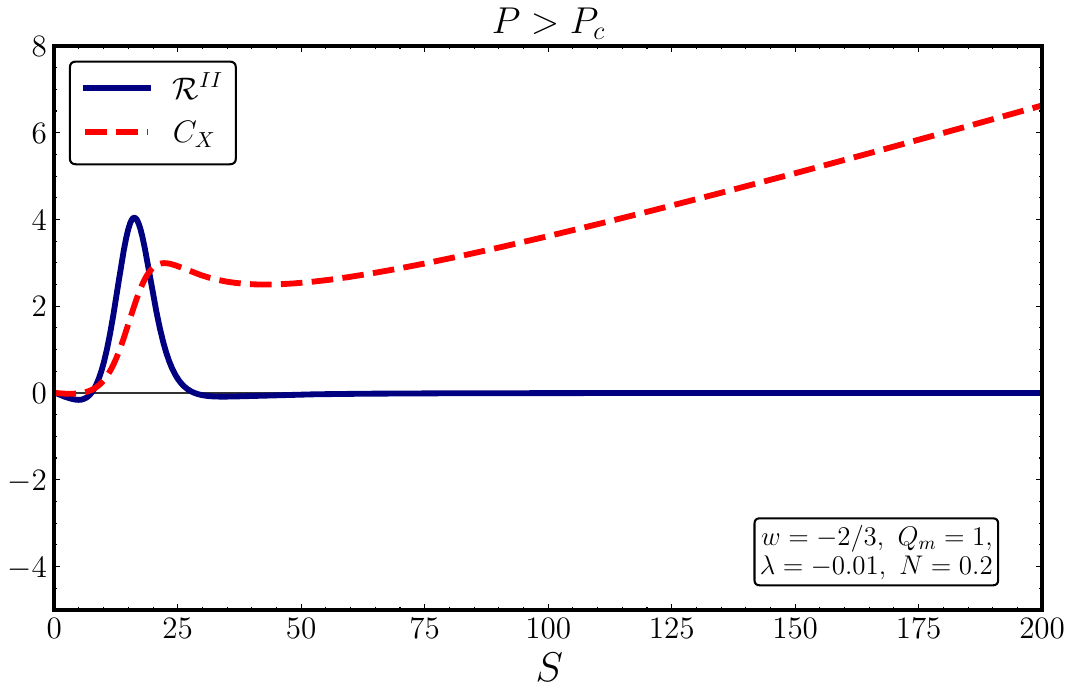}\\
(c)
\end{minipage}
\hfill
\begin{minipage}{0.48\textwidth}
    \centering
    \includegraphics[width=\linewidth]{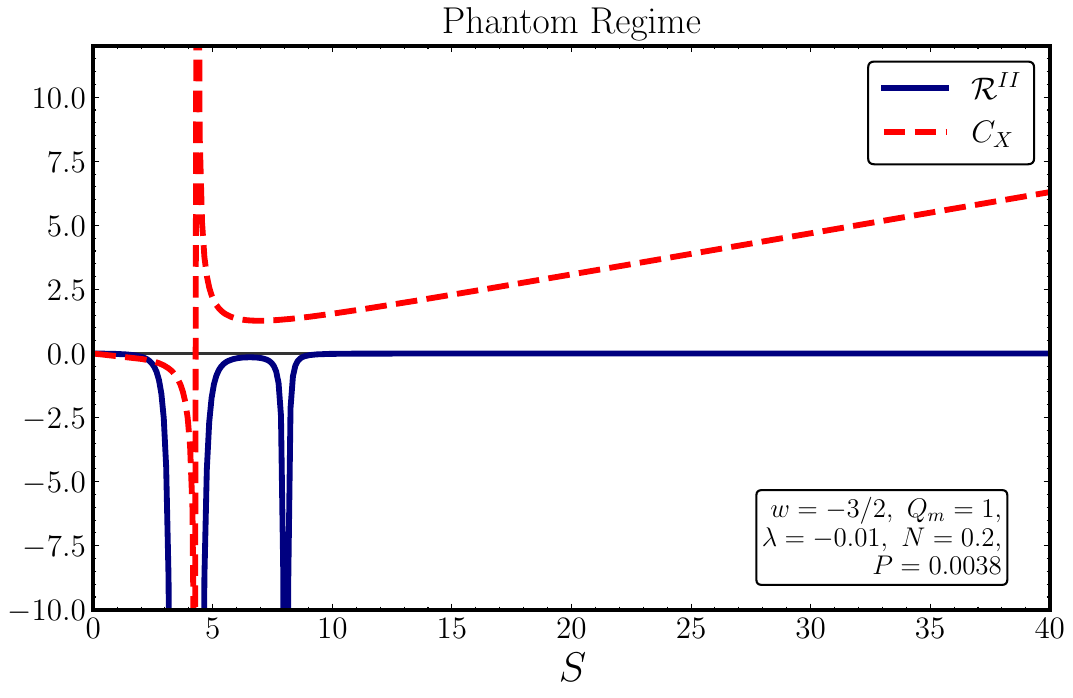}\\
(d)
\end{minipage}
\caption{GTD scalar curvature $\mathcal{R}^{II}$ and heat capacity $C_X$ 
as functions of the entropy $S$ for fixed parameters $Q_M=1$, $\lambda=-0.01$, $N=0.2$. 
The four panels correspond to the (a) subcritical ($P<P_c$), (b) critical ($P=P_c$), (c) supercritical ($P>P_c$), and (d) phantom regimes with $P_c\simeq 0.0038$. Plots have been rescaled for clarity.}
\label{fig:GTD scalars}
\end{figure} Furthermore, according to the interaction hypothesis proposed by
Ruppeiner \cite{ruppeiner1979thermodynamics,ruppeiner1981application},
the sign of the scalar curvature encodes the dominant character of the
effective microscopic interactions: $\mathcal{R}>0$ indicates that
repulsive interactions prevail, whereas $\mathcal{R}<0$ corresponds to
predominantly attractive interactions. The limiting case $\mathcal{R}=0$ is traditionally associated with an ideal-gas-like system, where effective microscopic interactions are absent. However, recent studies have shown that vanishing thermodynamic curvature does not necessarily imply the absence of microscopic interactions \cite{ladino2026probing}, highlighting that the relation between zero curvature and thermodynamic microstructure can be more subtle than the ideal-gas correspondence. 

In this work we focus on the metric $g^{II}$, as it correctly reproduces the phase structure in the appropriate ensemble. This choice is supported by the detailed analysis presented in \cite{ladino2026probing}, and enables a consistent investigation of the thermodynamic microstructure of the black hole.
In general, the black hole solution Eq.~\eqref{function}, regarded as a quasi-homogeneous system, is described by a five-dimensional equilibrium space with coordinates $E^c = \{S, P, Q_m, \lambda,N\}$. For simplicity, we will use a reduced equilibrium space where  only $S$ and the magenitc charge $Q_m$ are fluctuation variables. We expect the qualitative results to remain unchanged in higher-dimensional equilibrium spaces, reflecting the universal thermodynamic behavior of gravitational horizons. Indeed, as shown numerically in \cite{romero2026quasi,ahmed2026shadow}, the behavior near criticality is independent of the dimension of the equilibrium space. Thus, using the weights given in Eq.~\eqref{wi}, the line element in Eq.~\eqref{g222} reduces to
\begin{align}
g^{II}= 
\Sigma
\left(
- \frac{\partial^2 M}{\partial S^2} \, dS^2
+  \frac{\partial^2 M}{\partial Q_m^2} \, dQ_m^2
\right),
\label{g222_reduced}
\end{align}
where the conformal factor is $\Sigma\equiv TS +1/2\Phi_m Q_m $. The corresponding scalar curvature reads
\begin{align}
\mathcal{R}^{II}
&=
\frac{\mathcal{N_R}}
{\mathcal{D_R}},\label{scalar}
\end{align}
where
\begin{align}
\mathcal{D_{\mathcal{R}}} = & \mathcal{D_C}\times
\left(\pi^{3/2} Q_m^{3} + S^{3/2}\right)^{3}
\left[
8 P S^{3/2} 
+ \sqrt{S}\left(1 + 3 wN  \left(\frac{\pi}{S}\right)^{\frac{1+3w}{2}} \right)
+ \sqrt{\pi}\,\lambda
\right]^{3} \label{DR} \\
&\times
\left[
\sqrt{S}\left(-3 + 3 N \left(\frac{\pi}{S}\right)^{\frac{1+3w}{2}} - 8 P S \right)
- 3\sqrt{\pi}\,\lambda 
\ln\!\left(\frac{\sqrt{S}}{\sqrt{\pi}\,|\lambda|}\right)
\right]^{2}. \nonumber
\end{align}
 $\mathcal{N}_R$ is a function of the thermodynamic parameters whose explicit form is too cumbersome to be displayed here. $\mathcal{D}_C$ denotes the denominator of the heat capacity, see Eq.~\eqref{DC}, and therefore encodes the physical singularities of the system as depicted in Fig.~\ref{fig:GTD scalars}. It is worth emphasizing that, in Eq.~\eqref{g222_reduced}, the conformal factor does not coincide with the Smarr relation given in Eq.~\eqref{smarr}, since the analysis is not performed in the full equilibrium space. As a consequence, additional (unphysical) singularities arise from the zeros of Eq.~\eqref{DR}, which are associated with the zeros of the conformal factor $\Sigma$.\begin{figure}[ht!]
\centering
\begin{minipage}{0.48\textwidth}
    \centering
    \includegraphics[width=\linewidth]{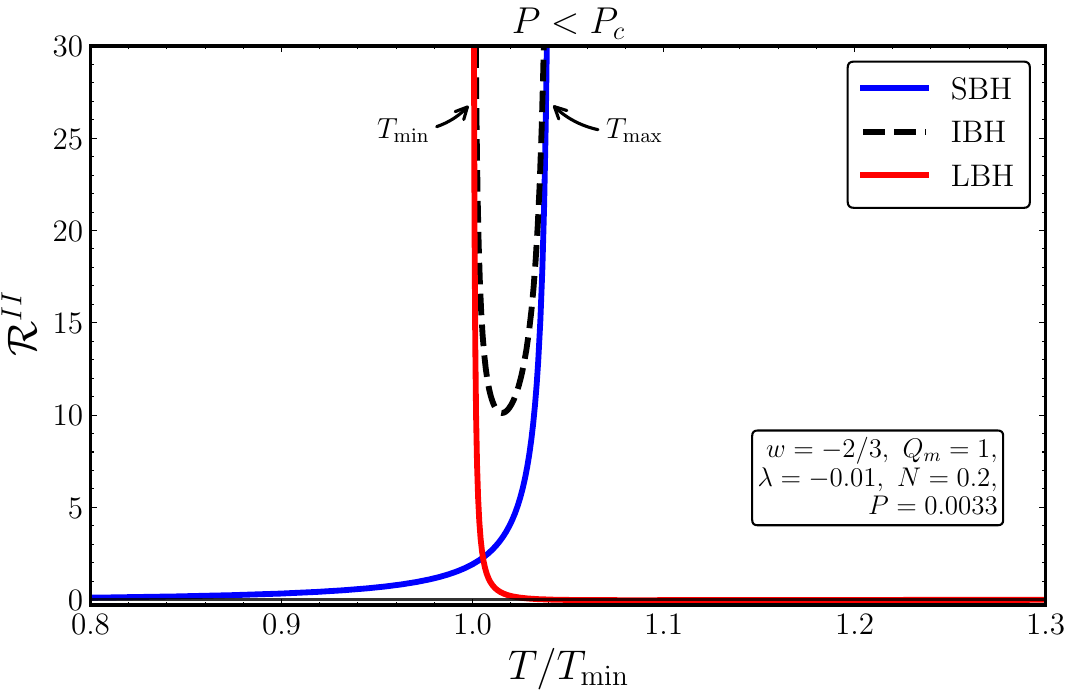}\\
    (a)
\end{minipage}
\hfill
\begin{minipage}{0.48\textwidth}
    \centering
    \includegraphics[width=\linewidth]{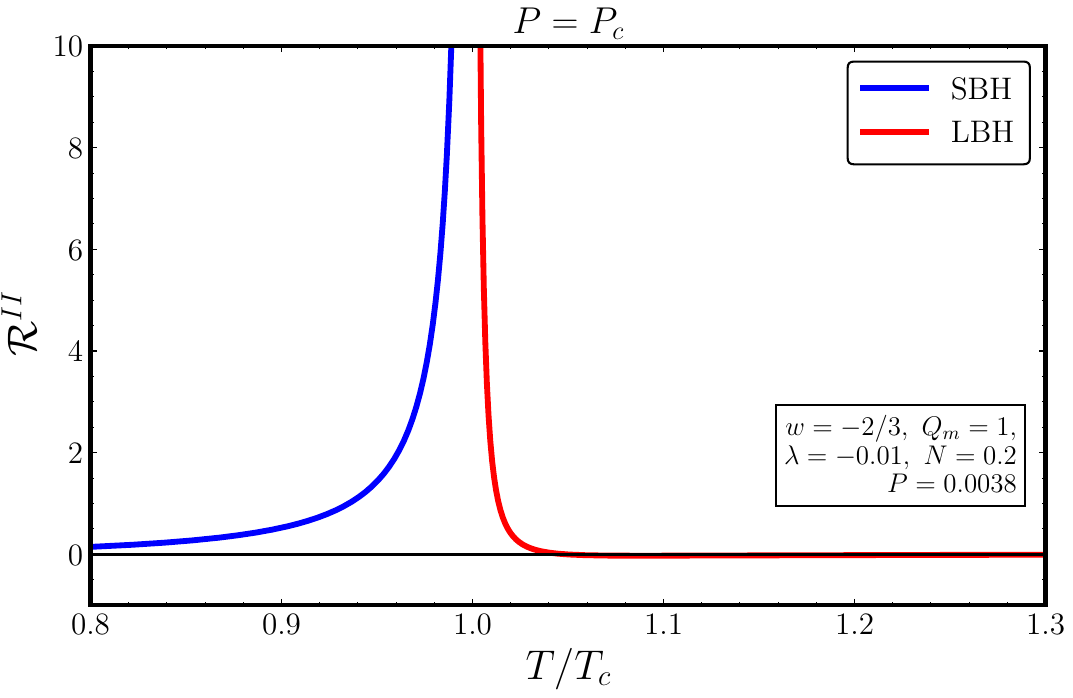}\\
    (b)
\end{minipage}
\hfill\\ \vspace{0.3in}
\begin{minipage}{0.48\textwidth}
    \centering
    \includegraphics[width=\linewidth]{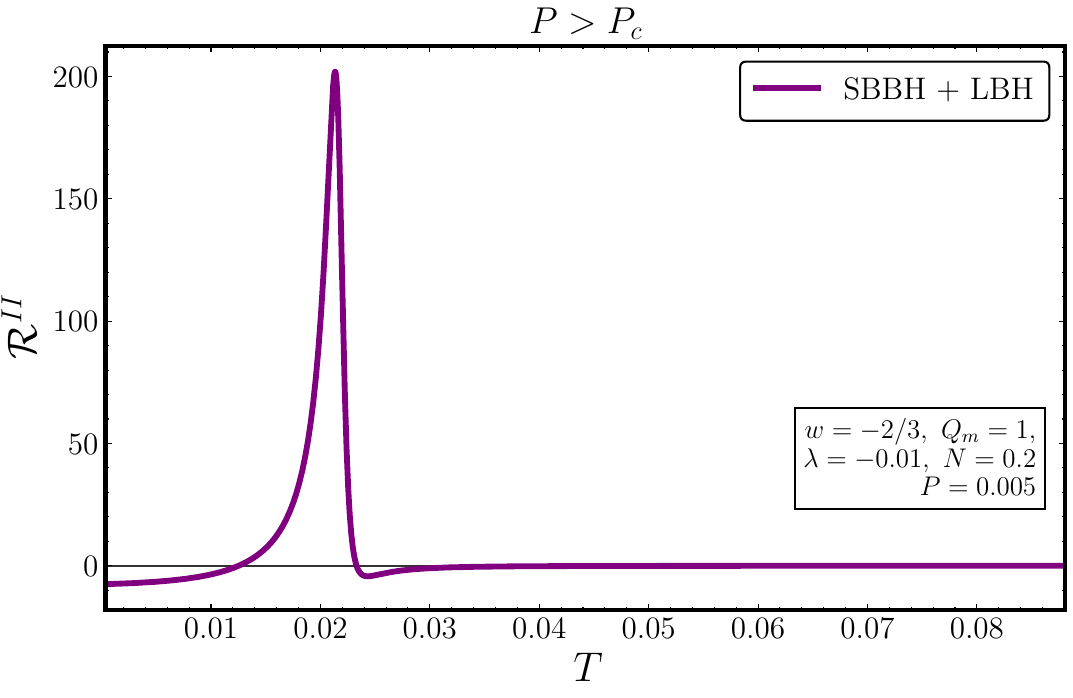}\\
    (c)
\end{minipage}
\hfill
\begin{minipage}{0.48\textwidth}
    \centering
    \includegraphics[width=\linewidth]{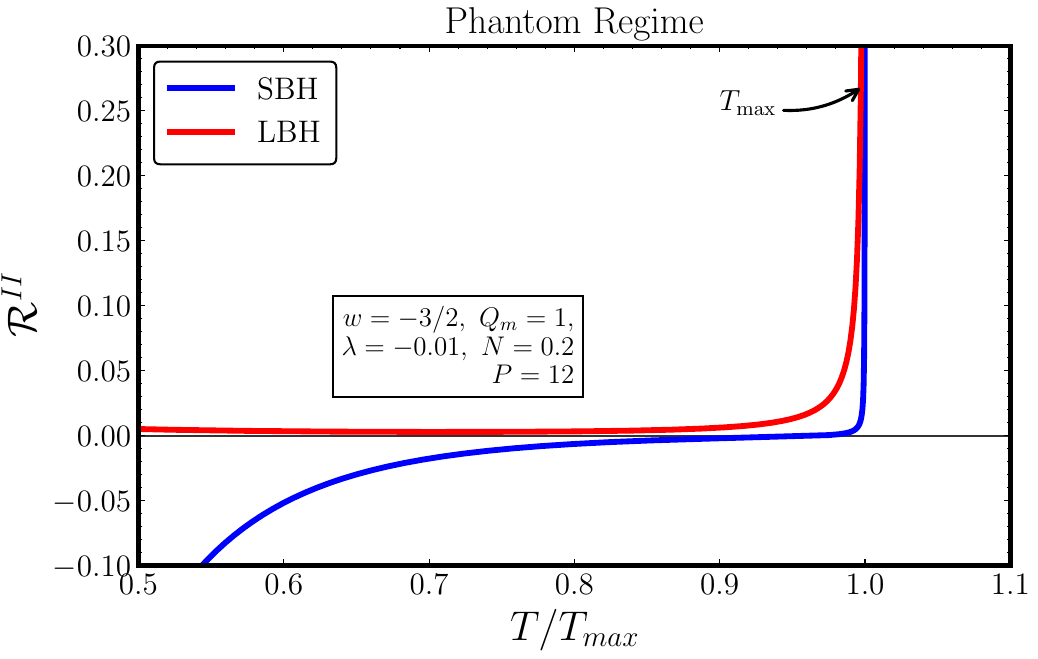}\\
    (d)
\end{minipage}
\caption{
The GTD scalar curvature $\mathcal{R}^{II}$
as a function of the temperature $T$ for fixed parameters $Q_M=1$, $\lambda=-0.01$, $N=0.2$. 
Panels (a)–(c) correspond to the subcritical ($P<P_c$), 
critical ($P=P_c$), and supercritical ($P>P_c$) regimes, 
respectively, with $P_c \simeq 0.0038$ and $w=-2/3$. 
Panel (d) illustrates the phantom regime characterized by $w=-3/2$.
}
\label{fig:GTD scalars tempe}
\end{figure} Nevertheless, for quasi-homogeneous thermodynamic systems defined on the complete equilibrium space, there exists a one-to-one correspondence between geometric singularities and the physical divergences signaling phase transitions, as demonstrated in \cite{quevedo2023unified,romero2026geometric}. From Fig.~\ref{fig:GTD scalars} it is evident that the divergences of the scalar curvature $\mathcal{R}^{II}$ accurately capture the thermodynamic critical behavior of the system. In the subcritical regime (Fig.~\ref{fig:GTD scalars}(a)), two distinct singularities are present; these merge into a single critical divergence at $P = P_c$ (Fig.~\ref{fig:GTD scalars}(b)), and completely disappear in the supercritical region, where no phase transition occurs (Fig.~\ref{fig:GTD scalars}(c)). In the phantom case (Fig.~\ref{fig:GTD scalars}(d)), an additional unphysical singularity emerges as a consequence of working within a reduced equilibrium space. However, the scalar curvature still correctly reproduces the spinodal point associated with the heat capacity.\begin{figure}[ht!]
\centering
\begin{minipage}{0.48\textwidth}
    \centering
    \includegraphics[width=\linewidth]{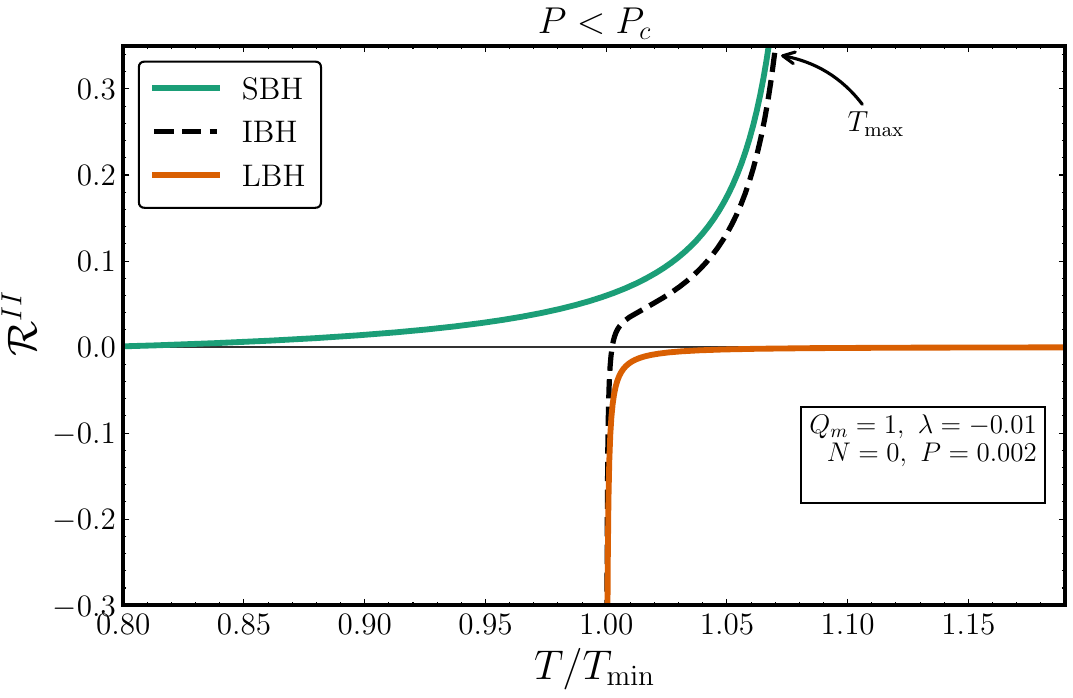}\\
    (a)
\end{minipage}
\hfill
\begin{minipage}{0.48\textwidth}
    \centering
    \includegraphics[width=\linewidth]{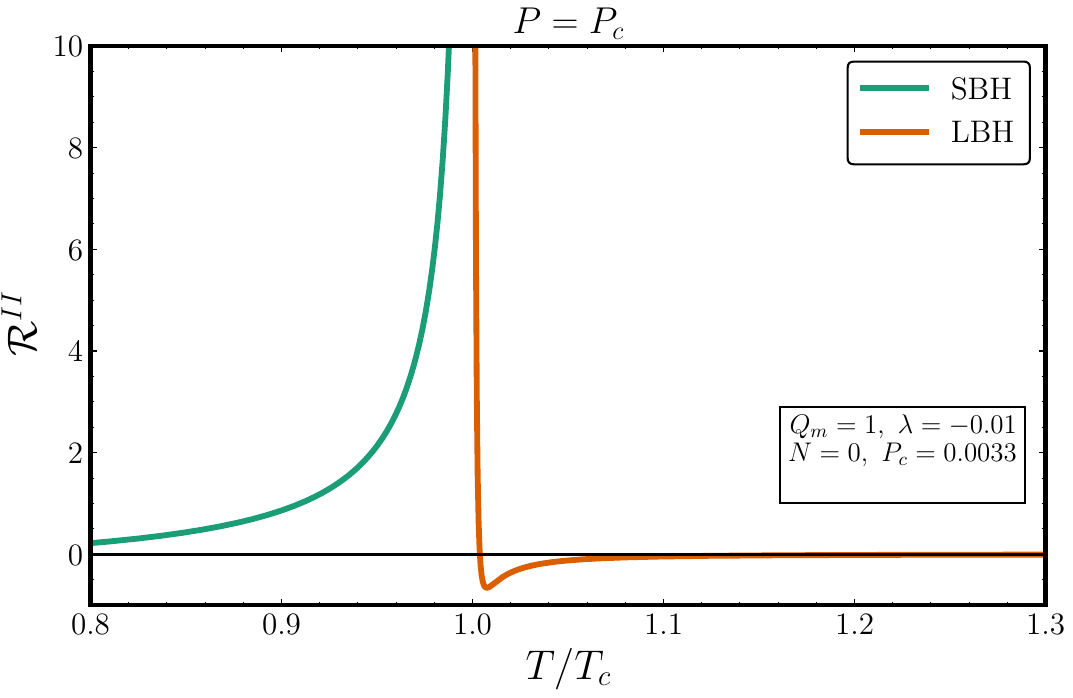}\\
    (b)
\end{minipage}
\hfill \vspace{0.3in}
\begin{minipage}{0.48\textwidth}
    \centering
    \includegraphics[width=\linewidth]{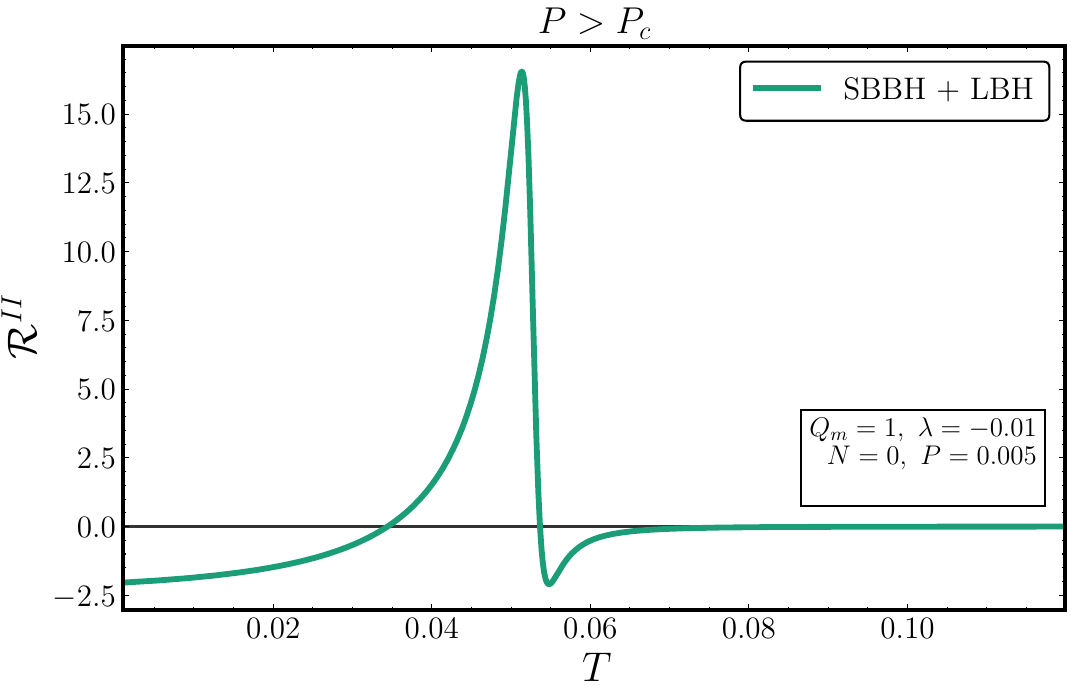}\\ 
    (c)
\end{minipage}
\hfill
\caption{The GTD scalar curvature $\mathcal{R}^{II}$ in the absence of quintessence ($N=0$) 
as a function of the temperature $T$, for fixed parameters 
$Q_M=1$ and $\lambda=-0.01$. 
Panels (a)--(c) correspond to the subcritical ($P<P_c$), 
critical ($P=P_c$), and supercritical ($P>P_c$) regimes, respectively, 
with the critical pressure $P_c \simeq 0.0033$. 
}
\label{fig:GTD scalars tempe quintazero}
\end{figure} Interestingly, in this regime the thermodynamic behavior of the system resembles that of asymptotically flat RN and Kerr black holes \cite{ladino2026probing}.
In particular, in the subcritical regime $P<P_c$ an intermediate IBH branch emerges in the temperature range $T_{\min}<T<T_{\max}$, where the scalar curvature becomes multivalued (see Fig.~\ref{fig:GTD scalars tempe}(a)). This multivalued behavior signals the coexistence region between the SBH and LBH phases. At the critical point $P=P_c$, the two singularities merge, the IBH branch disappears, and the multivalued structure collapses into a single critical configuration (Fig.~\ref{fig:GTD scalars tempe}(b)). For $P>P_c$, the black hole enters the supercritical phase and behaves as a fluid system in thermal equilibrium, with a single thermodynamic branch. In this regime the GTD scalar remains single–valued (Fig.~\ref{fig:GTD scalars tempe}(c)), reflecting the absence of competing phases. In Fig.~\ref{fig:GTD scalars tempe quintazero}, we observe the behavior of the scalar curvature in the absence of quintessence ($N=0$). The overall qualitative behavior remains unchanged; however, the main difference appears in the subcritical regime ($P < P_c$), shown in Fig.~\ref{fig:GTD scalars tempe quintazero}(a), where the LBH branch becomes negative, indicating an attractive thermodynamic interaction.
\noindent
\begin{figure}[ht!]
\centering
\begin{minipage}{0.48\textwidth}
    \centering
    \includegraphics[width=\linewidth]{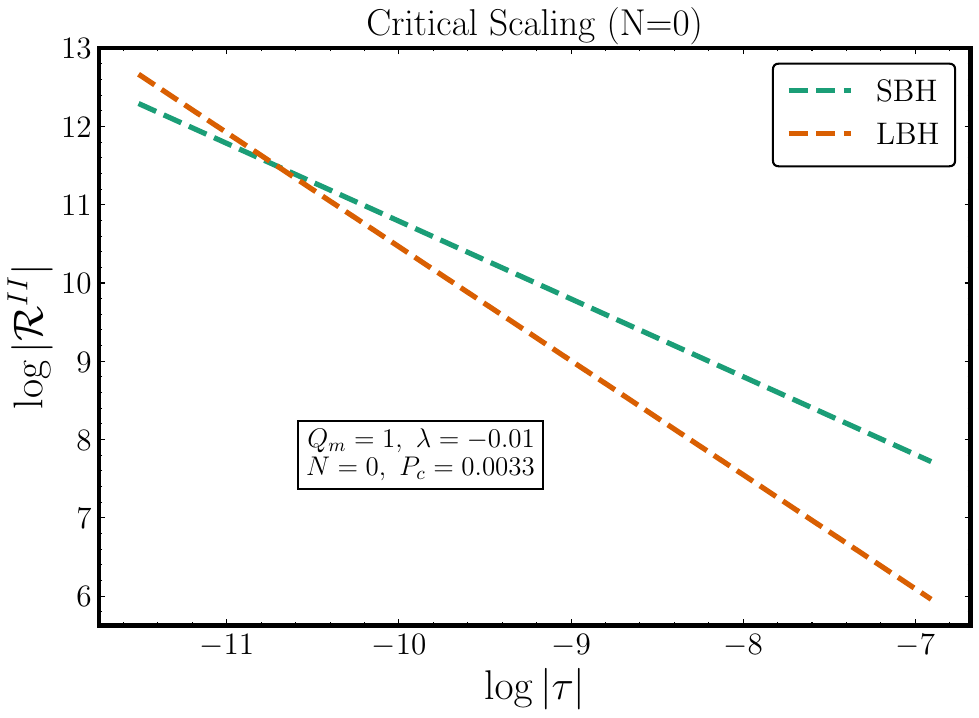}\\
    (a)
\end{minipage}
\hfill
\begin{minipage}{0.48\textwidth}
    \centering
    \includegraphics[width=\linewidth]{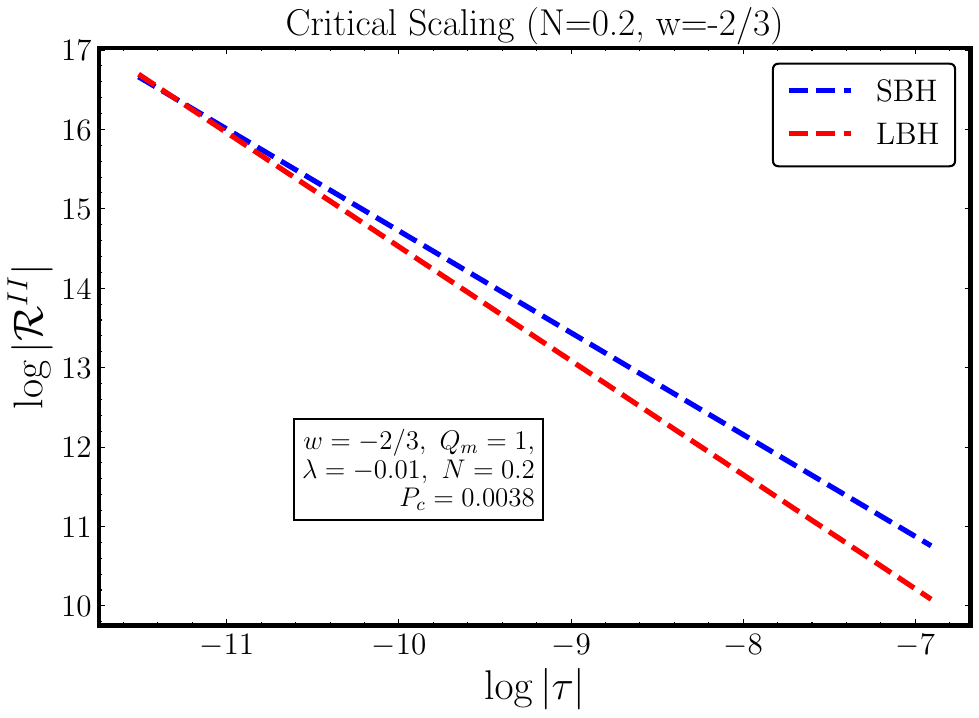}\\
    (b)
\end{minipage}
\hfill\\ 
\caption{
Log--log plots of the GTD scalar curvature $\mathcal{R}^{II}$ as a function of the reduced temperature at the critical pressure. 
Panel (a) corresponds to the case without quintessence ($N=0$), while panel (b) includes quintessence with 
$N=0.2$ and $w=-2/3$. 
The dashed lines represent the linear fits used to extract the critical exponents for the SBH and LBH branches.
}
\label{fig:critical exponent}
\end{figure}
Finally, Fig.~\ref{fig:critical exponent} shows the fitting of the critical behavior of the thermodynamic curvature scalar $\mathcal{R}^{II}$ near the critical point. In this region, the curvature is expected to follow a power-law scaling of the form \cite{romero2026quasi,ladino2026probing,romero2026geometric}
\begin{equation}
\mathcal{R}^{II}(\tau) =
\frac{A_c}{|\tau|^\zeta}
+ \mathcal{R}_0 
+ \mathcal{O}(\tau), \qquad 
\tau \equiv \frac{T - T_c}{T_c}, 
\end{equation}
where $\mathcal{R}_0 $ is a constant. This expression defines the critical amplitude $A_c$ and the critical exponent $\zeta$. From the linear fit in the log–log scale shown in Fig.~\ref{fig:critical exponent}, we extract the critical exponents and their corresponding amplitudes, which are summarized in Table~\ref{tab:critical_exponents}.
\begin{table}[ht!]
\centering
\begin{tabular}{c c c c}
\hline\hline
$N$ & $\zeta_{\text{SBH}}$ & $\zeta_{\text{LBH}}$ & $A_{\text{LBH}}/A_{\text{SBH}}$ \\
\hline
0   & 0.9934 & 1.4563 & 0.0070 \\
0.2 & 1.2825 & 1.4357 & 0.1772 \\
0.3 & 1.3627 & 1.3384 & 1.3389 \\
\hline\hline
\end{tabular}
\caption{Critical exponents and amplitude ratio for SBH and LBH branches 
for different values of $N$ and fixed $w=-2/3$, $Q_m=1, \lambda=-0.01$.}
\label{tab:critical_exponents}
\end{table}
Interestingly, the same critical scaling behavior has also been observed in quantum-corrected cosmological horizons \cite{romero2026quasi}, as well as RN and Kerr black holes \cite{ladino2026probing} and in a variety of real-fluid models \cite{romero2026geometric}. These results show that the GTD curvature singularities consistently capture the thermodynamic critical structure of the systems considered, suggesting a possible broader connection between the geometric description of criticality in gravitational and ordinary thermodynamic systems. However, establishing whether such a correspondence is universal would require a systematic analysis of a wider class of gravitational and conventional thermodynamic systems, which is beyond the scope of the present work.

\section{Conclusions}\label{sec:5}

In this work, we have investigated the thermodynamic phase structure of a NLMC--AdS black hole surrounded by PFDM and a dark-energy field. We find that the entropy obtained from $\int dM/T_H$ does not coincide with the Bekenstein--Hawking entropy for $Q_m\neq0$ when the other parameters are kept fixed. In the quasi-homogeneous extended thermodynamic framework, where $Q_m$, $\lambda$, $P$, and $N$ are treated as independent thermodynamic variables, the entropy is consistently given by the Bekenstein--Hawking area law. This framework allows the effects of the dark sector on the thermodynamic phase structure to be analyzed consistently. We also find that the reverse isoperimetric inequality is satisfied for $Q_m>0$, with $\mathcal{R}_{\mathrm{iso}}\geq1$. It is saturated in the Schwarzschild--AdS limit, $Q_m\to0$, and approaches unity in the LBH limit. Thus, the black hole is not super-entropic in the physical parameter regime.\\

The phase structure was further examined for the limiting geometries, for which the critical points can be obtained analytically. All the cases considered exhibit SBH/LBH phase transitions, with critical ratios that differ from the standard RN--AdS/vdW value. In particular, the NLMC--AdS case  gives $\rho_c=2/5$, while the Schwarzschild--AdS case with PFDM gives $\rho_c=1/3$. For the full solution, where the PFDM, dark-energy, and nonlinear magnetic contributions compete simultaneously, increasing $|\lambda|$ suppresses the SBH/LBH transition and eventually drives the system toward a single-phase regime. Moreover, any nonzero $\lambda$ prevents the formation of a regular magnetic core, as shown by the curvature-invariant analysis. Restricting $\lambda$ to the physical branch $\lambda<0$, required by $\mathcal{E}_{DM}>0$, is also consistent with our previous findings in Ref.~\cite{ahmed2026shadow}, where PFDM was found not to introduce additional phase transitions in the full solution. This result clarifies the apparent discrepancy with Ref.~\cite{ndongmo2023thermodynamics}, whose analysis considered non-physical values of $\lambda>0$.

The dark-energy contribution also modifies the phase structure depending on its equation-of-state parameter. In the full solution, the quintessence regime strengthens the first-order SBH/LBH transition, leading to a larger Gibbs construction area and increased latent heat, while simultaneously reducing the region of criticality. In contrast, the phantom regime exhibits spinodal behavior without phase coexistence or a genuine phase transition. Although the AdS pressure associated with the cosmological constant generally permits vdW-like criticality, the phantom contribution suppresses the SBH/LBH coexistence curve in the full configuration. Nevertheless, critical behavior can still occur in the limiting case of a Schwarzschild--AdS solution with phantom dark energy, as shown in Fig.~\ref{temperature_function2}(b) and Table~\ref{tab:critical_TS}. As a natural extension, a full Bayesian analysis using Markov chain Monte Carlo (MCMC) methods, following the approach developed in Ref.~\cite{romero2026geometric}, could be performed to quantify parameter uncertainties and assess the robustness of these trends across the allowed parameter space. Beyond the SBH/LBH phase transition, we also find that the dark sector modifies
the Hawking--Page transition between thermal AdS and the black hole
phase. In the presence of the dark-energy field ($N\neq0$), increasing $|\lambda|$
and $Q_m$ shifts the Hawking--Page temperature toward lower values,
whereas the trend is reversed in the absence of the dark-energy field
($N=0$). Thus, the dark-energy contribution changes the way PFDM and
the magnetic charge affect the global thermodynamic preference between
thermal AdS and the black hole phase. Finally, the GTD analysis provides an independent geometric characterization of the phase structure. The singularities of the thermodynamic equilibrium manifold reproduce the critical point and separate the stable and unstable branches. More importantly, the critical scaling of the GTD curvature is sensitive to the dark-energy intensity: for $N=0$, the extracted exponents differ markedly between the SBH and LBH branches, $\zeta_{\rm SBH}\simeq0.99$ and $\zeta_{\rm LBH}\simeq1.46$, whereas increasing $N$ modifies both exponents and brings them closer together. The amplitude ratio $A_{\rm LBH}/A_{\rm SBH}$ also increases substantially with $N$, indicating a progressive change in the relative strength of the thermodynamic curvature on the two branches. A Bayesian analysis based on MCMC methods is left for future
work to test the robustness of these dark-energy-induced modifications and assess the statistical
significance of the observed changes in the critical exponents and amplitude ratio.  Despite these quantitative modifications, the power-law behavior remains consistent with previous results for black holes, cosmological horizons, and real fluids, suggesting a broader universality of thermodynamic critical behavior.

\section*{Acknowledgments}
CRF acknowledge support from Conahcyt-Mexico, grant No. 4003366.
JRV is partially supported by the Centro de F\'isica Teórica de Valpara\'iso (CeFiTeV).

\bibliographystyle{unsrt}
\bibliography{referencias}

%\begin{thebibliography}

%%%%%%

%%%%

%\bibitem{SC1984} S. Chandrasekhar, {\it The Mathematical Theory of Black Holes} (Oxford University Press, Oxford, 1984).

%\bibitem{COV} N. Cruz, M. Olivares and JR. Villanueva, {\it The Geodesic structure of the Schwarzschild anti-de Sitter black hole}, Class. Quant. Grav. {\bf 22}, 1167-1190  (2005).

%\bibitem{RMW1984} R. M. Wald, \textit{General Relativity}, University of Chicago Press, Chicago (1984).

%\bibitem{EHTL1} K. Akiyama et al. [Event Horizon Telescope], Astrophys. J. Lett. {\bf 875}, L1 (2019). 

%\bibitem{EHTL4} K. Akiyama et al. [Event Horizon Telescope], Astrophys. J. Lett. {\bf 875}, L4 (2019). 

%\bibitem{EHTL6} K. Akiyama et al. [Event Horizon Telescope], Astrophys. J. Lett. {\bf 875}, L6 (2019). 

%\bibitem{EHTL12} K. Akiyama et al. [Event Horizon Telescope], Astrophys. J. Lett. {\bf 930}, L12 (2022). 

%\bibitem{EHTL14} K. Akiyama et al. [Event Horizon Telescope], Astrophys. J. Lett. {\bf 930}, L14 (2022). 

%\bibitem{EHTL15} K. Akiyama et al. [Event Horizon Telescope], Astrophys. J. Lett. {\bf 930}, L15 (2022). 

%\bibitem{EHTL16} K. Akiyama et al. [Event Horizon Telescope], Astrophys. J. Lett. {\bf 930}, L16 (2022). 

%\bibitem{EHTL17} K. Akiyama et al. [Event Horizon Telescope], Astrophys. J. Lett. {\bf 930}, L17 (2022). 

%\end{thebibliography}

\end{document}